\documentclass[aps,pra,preprint,groupedaddress]{revtex4}

\usepackage{verbatim}
\usepackage{amssymb,amsmath,epsfig,natbib}

\usepackage{graphicx}
\usepackage{dcolumn}
\usepackage{bm}
\usepackage{float}
\usepackage{enumerate}
\usepackage{tabularx} 

\usepackage{color}

\usepackage{txfonts}
\usepackage{morefloats}

\newcommand{\beq}{\begin{equation}}
\newcommand{\eeq}{\end{equation}}
\newcommand{\bea}{\begin{eqnarray}}
\newcommand{\eea}{\end{eqnarray}}

\usepackage{psfrag}
\usepackage{url}
\usepackage{fancybox}
\usepackage{afterpage}
\usepackage{color}
\usepackage{pifont}
\usepackage{exscale}
\usepackage{relsize}

\usepackage{hyperref}
\usepackage{cleveref}

\usepackage[caption=false]{subfig}
\usepackage{epstopdf}
\newcommand{\pa}{\partial}
\newcommand{\dd}{{\rm d}}

\usepackage{tikz}
\usetikzlibrary{patterns,arrows,calc,decorations.pathmorphing}

\begin{document}


\title{Vortices and magnetic impurities}



\author{Jennifer Ashcroft}
\email[]{jea29@kentforlife.net}

\affiliation{School of Mathematics, Statistics and Actuarial Science, University of Kent, Canterbury, CT2 7NF, UK}

\author{Steffen Krusch}
\email[]{S.Krusch@kent.ac.uk}
\affiliation{School of Mathematics, Statistics and Actuarial Science, University of Kent, Canterbury, CT2 7NF, UK}


\date{December 21, 2019 \\ \vspace*{3cm}}


\begin{abstract}
Ginzburg-Landau vortices in superconductors attract or repel depending on whether the value of the coupling constant $\lambda$ is less than $1$ or larger than $1.$ At critical coupling $\lambda =1,$ it was previously observed that a strongly localised magnetic impurity behaves very  similarly to a vortex. This remains true for axially symmetric configurations away from critical coupling. In particular, a delta function impurity of a suitable strength is related to a vortex configuration without impurity by singular gauge transformation. However, the interaction of vortices and impurities is more subtle and depends not only on the coupling constant $\lambda$ and the impurity strength, but also on how broad the impurity is. Furthermore, the interaction typically depends on the distance and may be attractive at short distances and repulsive at long distances. 
Numerical simulations confirm moduli space approximation results for the scattering of one and two vortices with an impurity. However, a double vortex will split up when scattering with an impurity, and the direction of the split depends on the sign of the impurity. Head-on collisions of a single vortex with different impurities away from critical coupling is also briefly discussed.
\end{abstract}



\maketitle



\section{Introduction}\label{sec:intro}

The Ginzburg-Landau model of superconductivity describes vortices in superconductors. In this paper, we investigate the scattering of vortices in the presence of magnetic impurities. Our aim is to provide an initial numerical study of such scattering processes for different choices of impurity. Previous studies of vortices in the presence of magnetic impurities have concentrated on the case of critical coupling. We carry out numerical simulations for critically coupled vortices, but also consider other values of the coupling constant $\lambda$.

Jaffe and Taubes showed in \cite{Jaffe:1980mj} that $N$ vortices at critical coupling can be described by a $2N$ dimensional moduli space. The relativistic dynamics of vortices can be approximated by the moduli space approximation \cite{Manton:1981mp} which can be rigorously justified \cite{Stuart:1994yc, Stuart:1994tc}. 
For small velocities, vortices move according to geodesic flow on the moduli space where the metric is induced by the kinetic energy. Samols found an implicit formula for the metric \cite{Samols:1991ne}. While no explicit solutions for vortices are known in flat space Witten noticed the vortex equation are integrable in hyperbolic space \cite{Witten:1976ck}. This allowed Strachan to evaluate the metric for two centred vortices \cite{Strachan:1992fb} and some more general metrics have been found subsequently \cite{Krusch:2009tn, Alqahtani:2014pva}.
There are also interesting non-Abelian vortices \cite{Hanany:2003hp, Auzzi:2003fs}. Baptista showed in \cite{Baptista:2010rv} how the moduli space metric can be calculated following Samols' method. The moduli space metric of non-Abelian vortices and their dynamics were discussed in \cite{Fujimori:2010fk,Eto:2011pj}.

The study of vortices in the presence of impurities has attracted interest in recent years. A notable example is the work of Tong and Wong in Ref.~\cite{Tong:2013iqa} concerning BPS vortices in the presence of electric and magnetic impurities. They argued that there still exists a moduli space of solitons after the addition of electric and magnetic impurities, and discussed the manner in which the moduli space dynamics is affected by each type of impurity. Ref.~\cite{Tong:2013iqa} has motivated several studies of vortices in product Abelian gauge theories which can be related to vortices in the presence of magnetic impurities. For example, existence theorems for solutions of vortices and anti-vortices in such models have been proven in Refs. \cite{Han:2015yua, ZhangLi:2015}, and similar ideas have been explored in an Abelian Chern-Simons-Higgs model in Ref. \cite{Han:2015tga}.

We have been particularly motivated by Ref.~\cite{Cockburn:2015huk},  which investigated the dynamics of vortices at critical coupling with magnetic impurities. The authors obtained solutions in flat space by numerically solving the Bogomolny equation for localised, axially symmetric impurities, and their numerics confirms the existence of a moduli space of vortex solutions. They also discussed vortices in hyperbolic space, where they calculated exact solutions and moduli space metrics for a delta function impurity.
We will give analytic proof that the profile function of an axially symmetric charge $n$ vortex sitting on top a charge $m$ delta function impurity is equal to the shape function of an $n+m$ vortex which is valid even for $\lambda \neq 1$.
We will extend their results for localised axially symmetric impurities in flat space by carrying out numerical simulations of the full field equations.
This enables us to consider vortices away from critical coupling and to numerically simulate vortex dynamics.  

As yet, there has not been a comprehensive numerical study of the scattering of vortices with magnetic impurities.  Previous numerical investigations into the scattering of two vortices in the Ginzburg-Landau model in Refs.~\cite{Shellard:1988zx,Rebbi:1991wk,Myers:1991yh,Thatcher:1997da} have explored the relationship between the scattering angle and impact parameter and any dependence of the scattering behaviour on the initial velocity given to the vortices. We carry out similar calculations for vortices scattering with magnetic impurities.

The paper begins with an introduction to vortices in the presence of magnetic impurities. We discuss the Bogomolny bound satisfied by vortices at critical coupling and obtain vacuum solutions for different values of the coupling constant $\lambda$ and for different impurities. The energy of static vortex solutions in the presence of magnetic impurities can be evaluated for axially symmetric configuration which allows us to calculate a binding energy for vortices and impurities. We also investigate the asymptotic interaction between vortices and impurities. A complete picture emerges when we calculate the energy as a function of the separation between vortex and impurities.
We carry out numerical simulations of the scattering of vortices at critical coupling with magnetic impurities and analyse our findings.
We also study head-on collisions of a single vortex with different impurities away from critical coupling for various initial velocities.
Finally we summarise the results and note opportunities for further work.

\section{Vortices with magnetic impurities}

The Lagrangian for gauged vortices in the Ginzburg-Landau model is
\begin{align}\label{vort_lagrangian}
L = \int \left( -\frac{1}{4}f_{\mu\nu}f^{\mu\nu} + \frac{1}{2} \overline{D_\mu\phi}D^\mu\phi - \frac{\lambda}{8}(1-\overline{\phi}\phi)^2\right)~\dd^2x.
\end{align}
Here, $\phi$ is a complex scalar field, and $a_\mu$ is the $U(1)$ gauge field. The field tensor $f_{\mu\nu}$ and the covariant derivative $D_\mu\phi$ are given by
\begin{align}
f_{\mu\nu} = \pa_\mu a_\nu - \pa_\nu a_\mu \quad {\rm and} \quad
D_\mu\phi =  \pa_\mu \phi - i a_\mu\phi.
\end{align}
To ensure finite energy, vortex solutions satisfy $D_\mu\phi\rightarrow0$ and $|\phi|\rightarrow1$ as $|x|\rightarrow\infty$. The Lagrangian \eqref{vort_lagrangian} is invariant under a gauge transformation
\begin{align}\label{gauge_transf}
\phi(x) \mapsto e^{i\alpha(x)}\phi(x), \quad a_\mu(x) \mapsto a_\mu(x) + \pa_\mu\alpha(x).
\end{align}

The coupling constant $\lambda$ is a real parameter, which distinguishes between Type I and Type II superconductivity. If $\lambda<1$, then we model Type I superconductivity when vortices attract, and if $\lambda>1$ then we model Type II superconductivity when vortices repel. The value $\lambda=1$ separating the two regimes is known as critical coupling.

We consider the deformation of the Lagrangian \eqref{vort_lagrangian} that was proposed in Ref.~\cite{Tong:2013iqa} to include magnetic impurities,
\begin{align}\label{vimp_lag}
L = \int\left( -\frac{1}{4}f_{\mu\nu}f^{\mu\nu} + \frac{1}{2}\overline{D_\mu\phi}D^\mu\phi - \frac{\lambda}{8}(1 + \sigma - |\phi|^2)^2 + \frac{\mu}{2}\sigma B \right) ~d^2x,
\end{align}
where $\sigma$ is a fixed, static source term for the magnetic field $B=f_{12}$. Here $\mu$ is an additional coupling parameter. We restrict our attention to localised, axially symmetric impurities of the form $\sigma(x,y) = c e^{-d(x^2+y^2)}$, where $c,d \in \mathbb{R}$, and $d>0$.

The topological charge is an integer $N$ giving the net number of vortices in a solution. It can be written in terms of the magnetic field as
\begin{align}\label{vortex_charge}
N = \frac{1}{2\pi} \int B ~\dd^2 x.
\end{align}

We begin by obtaining vacuum solutions in this model. The potential energy for vortices in the presence of magnetic impurities is given by
\begin{align}\label{vimp_pot_en}
V = \frac{1}{2} \int \left( B^2 + \overline{D_i\phi}D_i\phi + \frac{\lambda}{4}(1+\sigma-|\phi|^2)^2 - \mu \sigma B \right) ~d^2x.
\end{align}

At critical coupling $\lambda =1$ and $\mu =1,$ it is possible to obtain a Bogomolny bound on the energy. We first complete the square on the integrand of \eqref{vimp_pot_en}, to find
\begin{align}\label{vimp_en_sq}
 \left(B-\frac{1}{2}(1+\sigma-\overline{\phi}\phi)\right)^2 +\overline{(D_1\phi + iD_2\phi)}(D_1\phi+ iD_2\phi)
 + B - i\left(\pa_1(\overline{\phi}D_2\phi)-\pa_2(\overline{\phi}D_1\phi)\right).
\end{align}
The final term above integrates to zero. Using this fact in combination with the expression for the topological charge \eqref{vortex_charge} we see that
\begin{align}\label{vimp_en_sq_again}
E = V = \frac{1}{2} \int \left( \left(B-\frac{1}{2}(1+\sigma-\overline{\phi}\phi)\right)^2 +\overline{(D_1\phi + iD_2\phi)}(D_1\phi+iD_2\phi) \right) \dd^2x  + \pi N,
\end{align}
leading us to the Bogomolny bound
\begin{align}\label{vimp_bog}
E \geq \pi N.
\end{align}
This is saturated for solutions of the Bogomolny equations
\begin{align}\label{vimp_bog_eq}
D_1\phi + i D_2\phi & = 0, \nonumber \\
B - \frac{1}{2}\left(1+\sigma-\overline{\phi}\phi\right) & = 0.
\end{align}
Note that the same bound is satisfied by vortices in the original model \eqref{vort_lagrangian}, and the only difference in the equations is the additional $\sigma$ in the second Bogomolny equation. Other possibilities of modifying the vortex equations while keeping the BPS structure are discussed in Refs.~ \cite{Bazeia:2012uc, Bazeia:2011xw, Casana:2017qep, Almeida:2017wmw}.

We can derive a lower bound on the energy when $\lambda\neq1$ and $\mu \neq 1$ in a similar way. We rewrite the integrand of \eqref{vimp_pot_en} as
\begin{align}\label{gen_en_rewrite}
\left( B -\frac{1}{2}(1+\sigma-\overline{\phi}\phi)\right)^2 + \overline{(D_1\phi + iD_2\phi)}(D_1\phi + iD_2\phi) + B - i\left(\pa_1(\overline{\phi}D_2\phi) 
- \pa_2(\overline{\phi}D_1\phi)\right) \nonumber \\+\frac{\lambda-1}{4}\left(1+\sigma-\overline{\phi}\phi\right)^2 - (\mu-1)\sigma B.
\end{align}
Integrating \eqref{gen_en_rewrite} produces the same result as before, but with two additional terms. The energy is bounded from below by
\begin{align}\label{vimp_gen_bound}
  E  \geq \pi N + \frac{\lambda-1}{8}\int \left(1+\sigma-\overline{\phi}\phi\right)^2 ~\dd^2 x
- \frac{\mu-1}{2} \int \sigma B\, \dd^2 x.
\end{align}
When $\lambda\geq1$ and $\mu$ takes values such that the additional terms are non-negative, then we have the bound \eqref{vimp_bog} which is saturated at critical coupling $\lambda =1$ and $\mu = 1.$ For the more general bound \eqref{vimp_gen_bound} to be meaningful, we must argue
\begin{equation}
  \int(1+\sigma-\overline{\phi}\phi)^2 ~\dd^2 x\quad {\rm and} \quad
  \int \sigma B\, \dd^2 x
\end{equation}
are bounded. This is true as long as $\sigma,$  $\phi$ and $B$ are non-singular and decay sufficiently quickly at infinity. We consider impurities of the form $\sigma(x,y) = ce^{-d(x^2+y^2)}$, which all decay very fast, and the asymptotics derived in Sect. \ref{impasymp} ensures that $\phi$ and $B$ also decay fast enough. We will also discuss the limit in which $\sigma$ approaches a delta function. In this case, a square of a delta function is introduced into the Lagrangian \eqref{vimp_lag}, which is not defined. However, as noted in Ref.~\cite{Cockburn:2015huk}, it does make sense to substitute a delta function for $\sigma$ in the equations of motion, and we can consider this to be a limit of impurities for which the energy is well-defined. In the following, we restrict our attention to the case $\mu =1.$ Then, the moduli space approximation is still applicable, and 
$\lambda \neq 1$ induces a potential on the moduli space of the BPS vortices satisfying equations \eqref{vimp_bog_eq}. The case without impurities, but $\lambda \neq 1,$ has been rigorously treated in \cite{Stuart:1994yc}.


\subsection{Symmetric solutions}

We first study the effect of the impurity on the vacuum configuration. To simplify the problem, we assume circular symmetry which allows us to convert to polar coordinates $r,~\theta$, and fix the radial gauge $a_r = 0$. Then we have $\phi(r,\theta) = \phi(r)e^{iN\theta}$, and $a_\theta(r,\theta) = a_\theta(r)$, and will solve for the real profile functions $\phi(r)$ and $a_\theta(r)$. For fields of this form, the energy \eqref{vimp_pot_en} becomes
\begin{align}\label{vimp_prof_en}
V = \pi \int \left( \phi'^2 + \frac{a_\theta'^2}{r^2} + \frac{(N-a_\theta)^2}{r^2}\phi^2 + \frac{\lambda}{4}(1+\sigma-\phi^2)^2 - \frac{\sigma}{r}a_\theta' \right) r ~dr.  
\end{align}

To calculate $\phi(r)$ and $a_\theta(r)$, we solve the reduced field equations
\begin{align}\label{vimp_prof_eq}
\phi'' + \frac{\phi'}{r} - \frac{(N-a_\theta)^2}{r^2}\phi + \frac{\lambda}{2}(1+\sigma-\phi^2)\phi = 0, \nonumber \\ 
a''_\theta - \frac{a'_\theta}{r} - \frac{r}{2}\sigma' + (N-a_\theta)\phi^2 = 0,
\end{align}
via a finite difference method on grids typically of size 2001 with spacing $\Delta r = 0.01$, subject to the boundary conditions $\phi'(0)=0$, $a_\theta(0)=0$, $\phi(\infty) = 1$, $a_\theta(\infty)=N$. We consider impurities of the form $\sigma(r) = c e^{-dr^2}$, where $c,d \in \mathbb{R}$, and $d>0$.


\begin{figure}[!htb]
\begin{center}
  \subfloat[\, $\phi(r)$]{\includegraphics[totalheight=7.0cm]
    {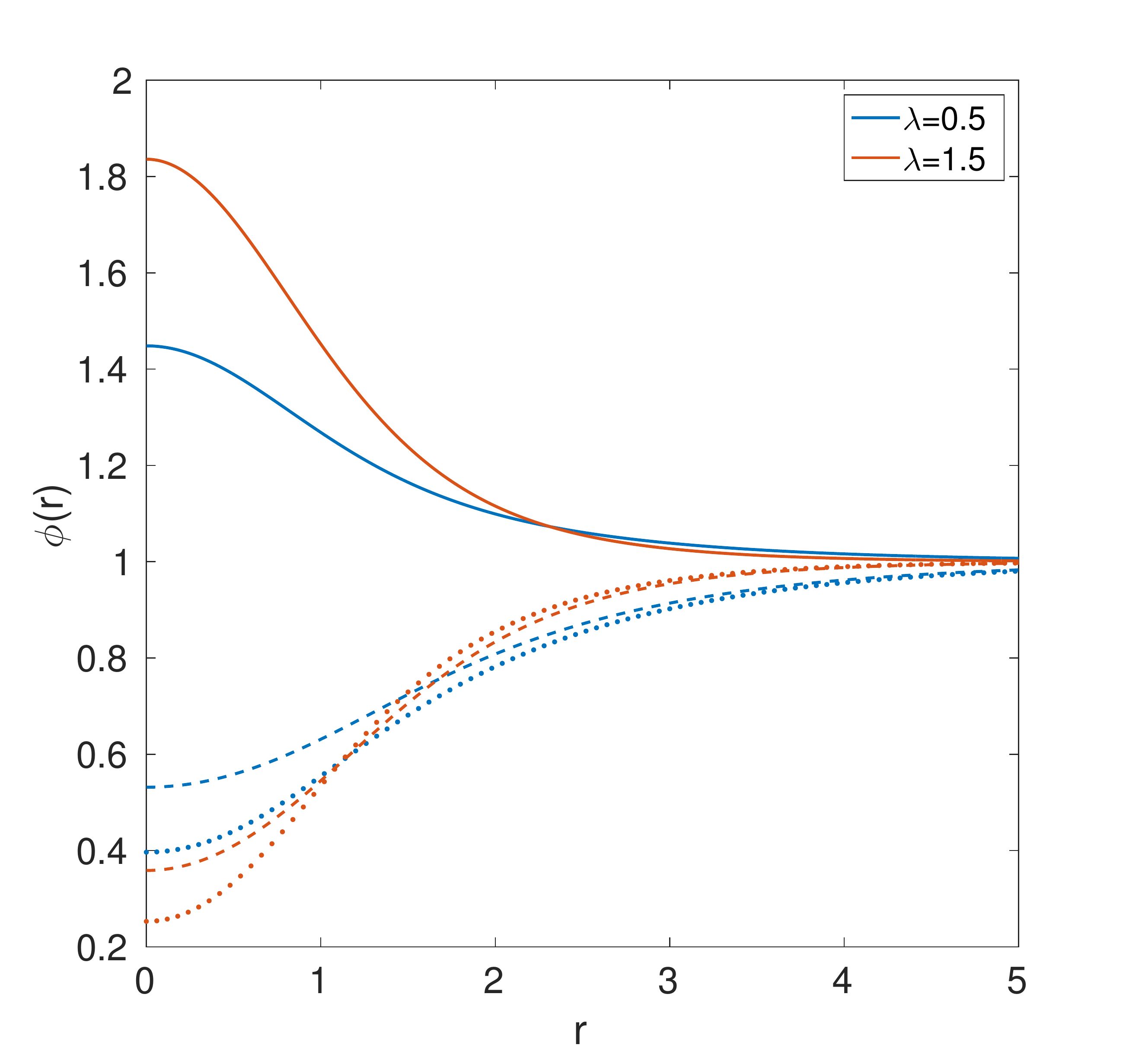}}
  \subfloat[\, $a_\theta(r)$]{\includegraphics[totalheight=7.0cm]
    {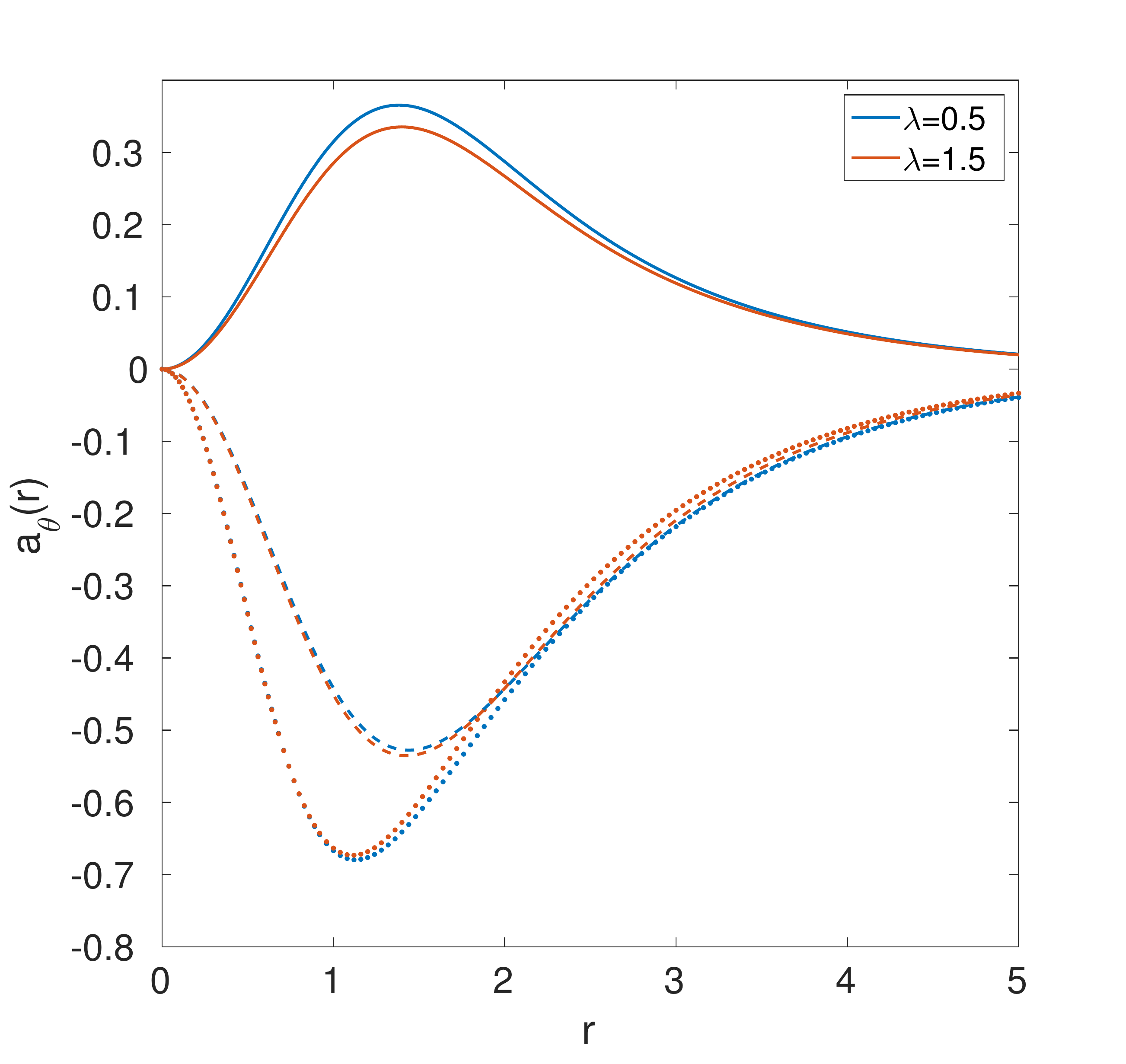}} \\
  \subfloat[\, $E(r)$]{\includegraphics[totalheight=7.0cm]
    {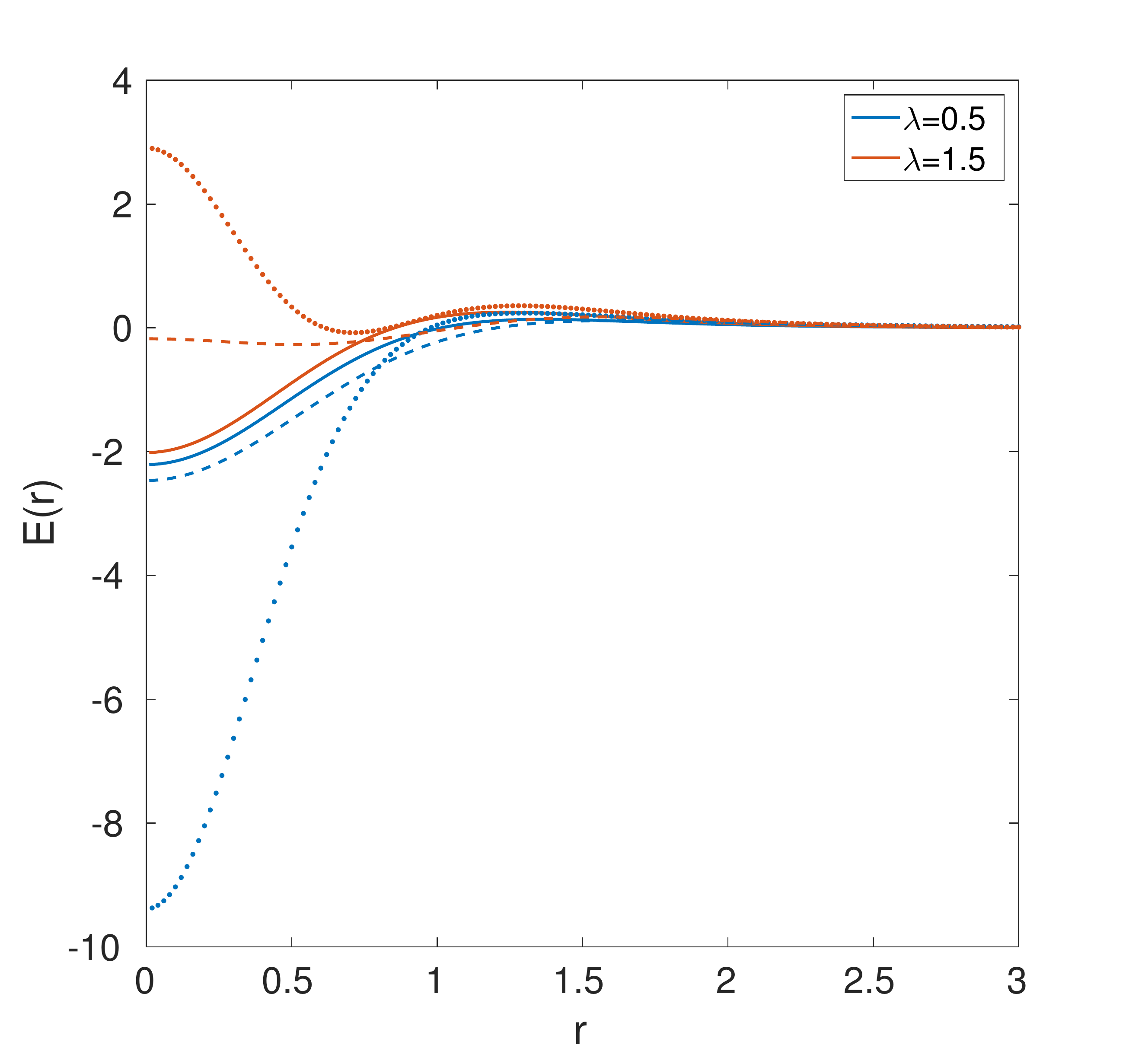}}
  \subfloat[\, $B(r)$]{\includegraphics[totalheight=7.0cm]
    {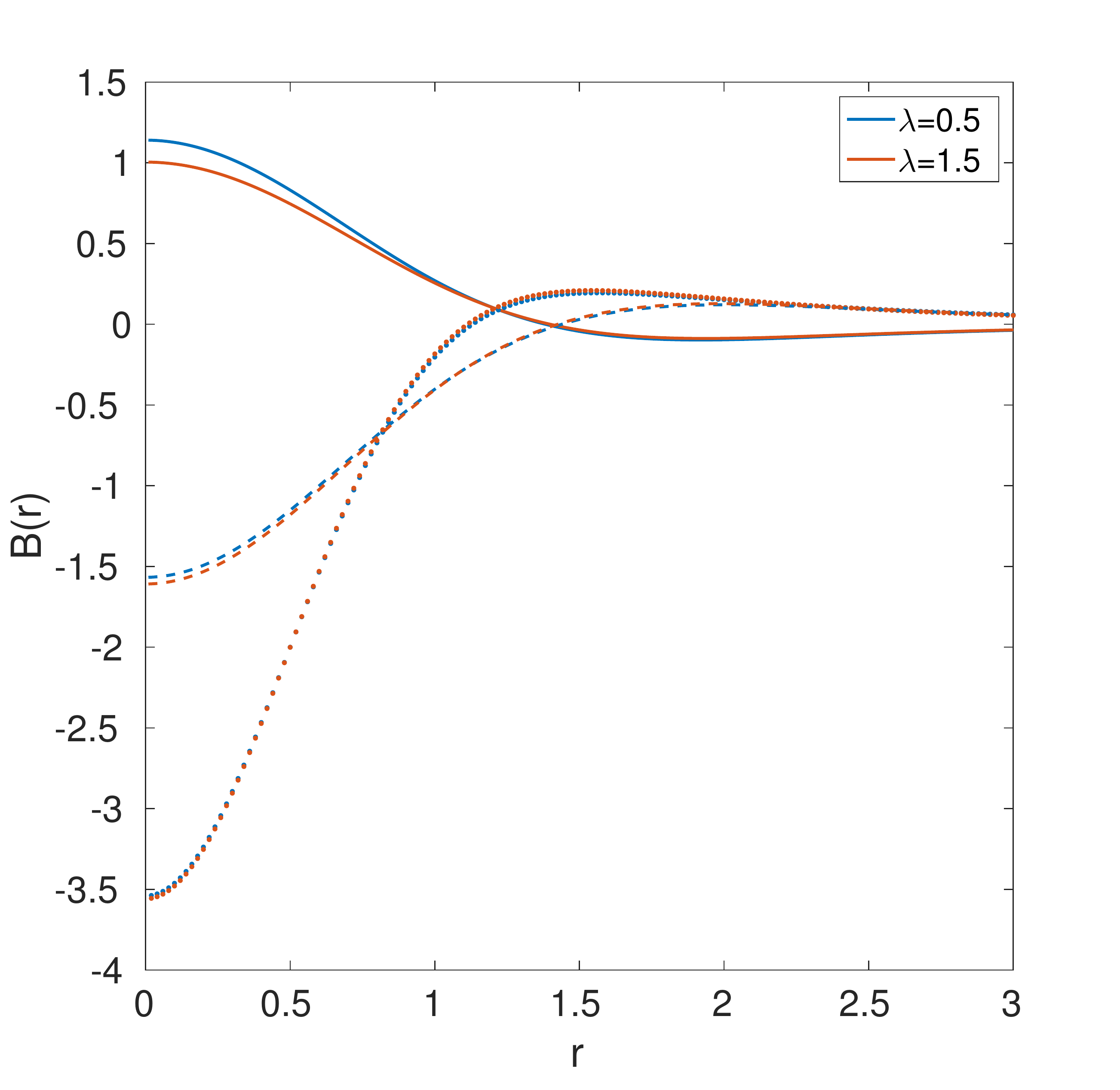}}
\caption{Vacuum solutions obtained by solving equations \eqref{vimp_prof_eq} with $\lambda=0.5$ (in blue), and $\lambda=1.5$ (in red), in the presence of the impurities $\sigma(r)=4e^{-r^2}$ (solid lines), $\sigma((r)=-4e^{-r^2}$ (dashed lines), and $\sigma(r)=-8e^{-2r^2}$ (dotted lines).  In the subfigures we display (a) profile function $\phi(r)$, (b) profile function $a_\theta(r)$, (c) energy density $E(r)$, (d) magnetic field $B(r)$.}
\label{fig:vimp_prof_N0}
\end{center}
\end{figure}

In Fig. \ref{fig:vimp_prof_N0} we display the profile functions $\phi(r),~a_\theta(r)$, the energy density $E(r)$, and the magnetic field $B(r)$ for vacuum solutions $(N=0)$ in the presence of three different magnetic impurities: $\sigma(r)=4e^{-r^2}$ in solid lines, $\sigma(r)=-4e^{-r^2}$ in dashed lines, and $\sigma(r)=-8e^{-2r^2}$ in dotted lines. Solutions for $\lambda=0.5$ are shown in blue and those for $\lambda=1.5$ are shown in red. Although the profile functions were calculated over $r\in[0,20]$, to highlight the more interesting features of the solutions we only display the ranges $r\in[0,5]$ for the profile functions and $r\in[0,3]$ for the energy density and magnetic field. We see that the effect of the impurity is localised, with the fields taking their usual vacuum values away from the impurity. In Ref.~\cite{Cockburn:2015huk} it was noted that for vortices at critical coupling $\phi(0)\rightarrow 0$ as $c\rightarrow-\infty$, and $\phi(0) \rightarrow \infty$ as $c\rightarrow+\infty$. We have observed the same behaviour away from critical coupling, and this has been tested over a much greater range of $c$ than those shown here. In Fig.~\ref{fig:vimp_prof_N0}(b) we see that $a_\theta(r)\geq0$ for $c>0$ and $a_\theta\leq0$ for $c<0$. 

The energy density plot shows that there is a region of negative energy density, which is not the case in the absence of a magnetic impurity. We previously saw that it is possible to derive a lower bound on the energy for any $\lambda$ as \eqref{vimp_gen_bound}. This bound is saturated when $\lambda=1$ and can be negative for $\lambda <1.$

For $c>0$, the energy density for $\lambda=0.5$ is similar in shape to that for $\lambda=1.5$ and the same impurity, however for $c<0$ there is a significant difference in the energy density depending on $\lambda$. For $\lambda=0.5$, the region of negative energy density becomes more significant, which is especially clear for $\sigma(r)=-8e^{-2r^2}$. By contrast, for $\lambda=1.5$, the region of negative energy density becomes much less significant. For $\sigma(r) = -8e^{-2r^2}$, the energy density is positive at the origin for $\lambda = 1.5$, whereas all other solutions have negative energy density at the origin. We see from Fig.~\ref{fig:vimp_prof_N0}(d) that the value of $\lambda$ does not significantly alter the magnetic field $B(r)$. Furthermore, changing the sign of $c$ roughly reverses the sign of $B(r)$.

\begin{figure}[!htb]
\begin{center}
\subfloat[\, $\phi(r)$]{\includegraphics[totalheight=7.0cm]{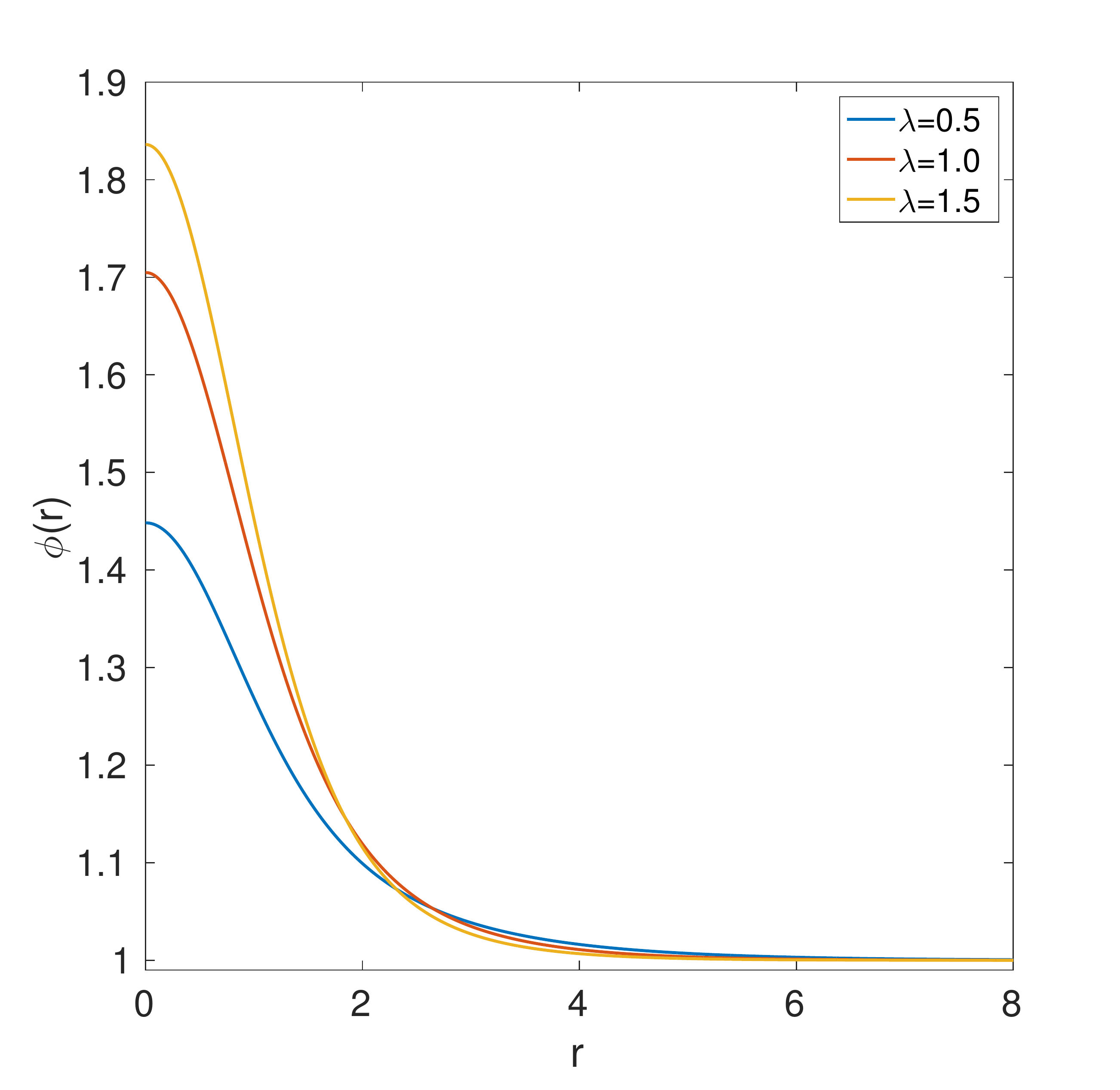}}
\subfloat[\, $a_\theta(r)$]{\includegraphics[totalheight=7.0cm]{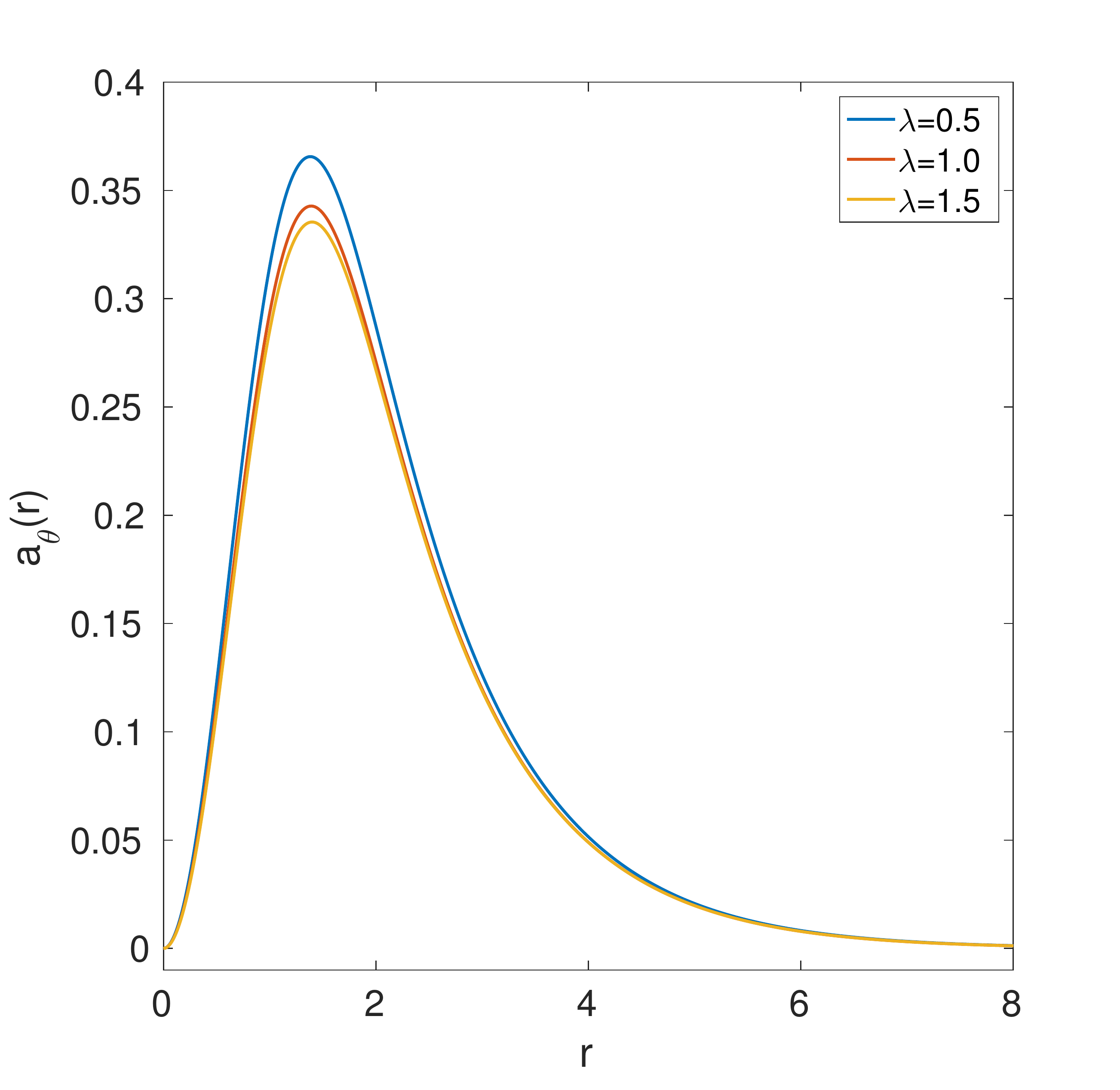}} \\
\subfloat[\, $E(r)$]{\includegraphics[totalheight=7.0cm]{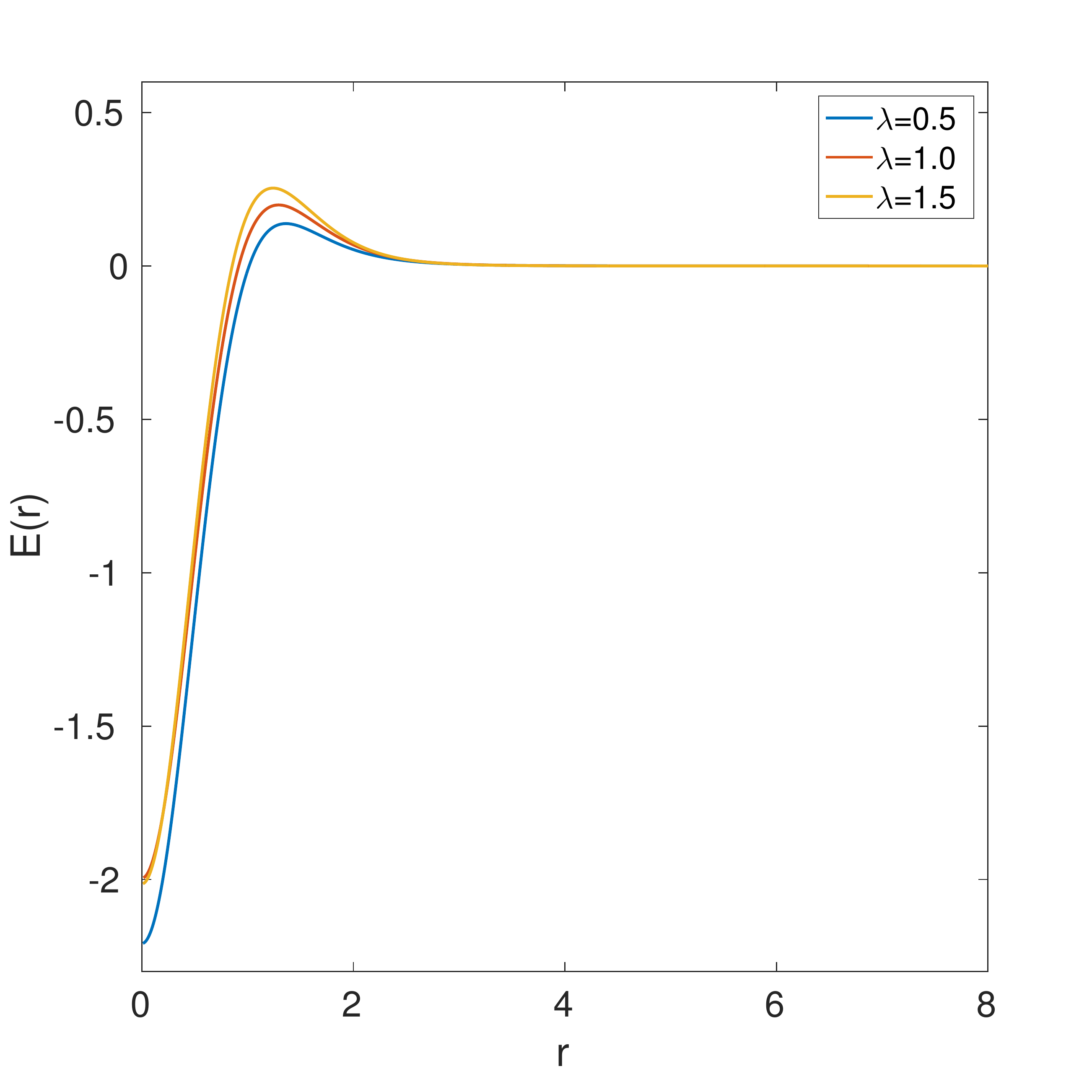}}
\subfloat[\, $B(r)$]{\includegraphics[totalheight=7.0cm]{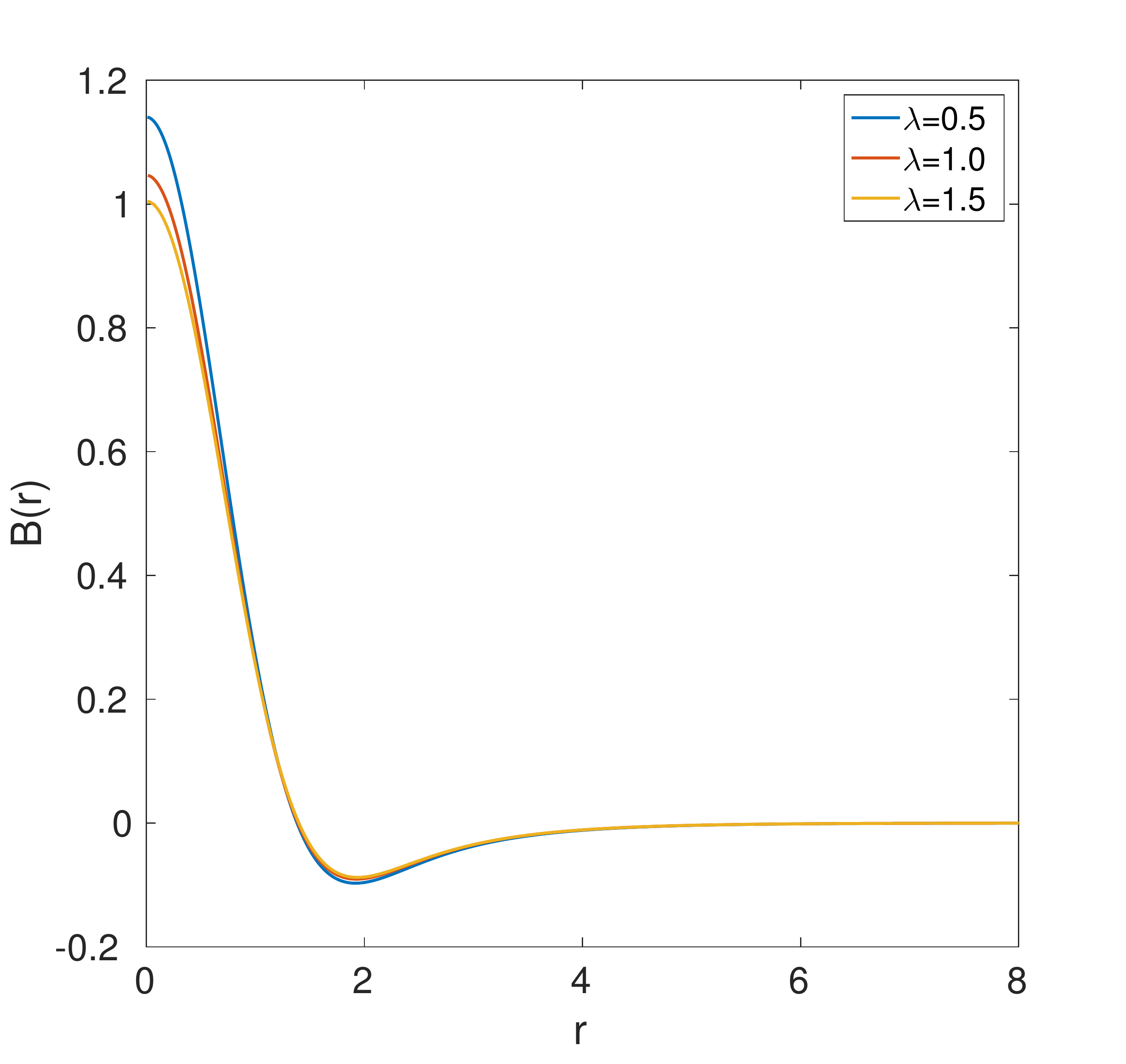}}
\caption{Vacuum solutions obtained by solving equations \eqref{vimp_prof_eq} with three different values of $\lambda$ in the presence of the impurity $\sigma(r)=4e^{-r^2}$. In each subfigure, we display (a)~profile function $\phi(r)$, (b)~profile function $a_\theta(r)$, (c)~energy density $E(r)$, (d)~magnetic field $B(r)$.}
\label{fig:vimp_prof_N0_lambda}
\end{center}
\end{figure}

We examine more carefully the effect of varying the coupling constant $\lambda$ on the vacuum solutions in Fig.~\ref{fig:vimp_prof_N0_lambda}. This displays the profile functions, energy density and magnetic field for the vacuum solutions with $\sigma(r)=4e^{-r^2}$, and $\lambda=0.5,~1.0,~1.5$. We see in  Fig.~\ref{fig:vimp_prof_N0_lambda}(a) that the value of $\phi(0)$ increases with increasing $\lambda$. Similarly, Fig.~\ref{fig:vimp_prof_N0_lambda}(b) indicates that the maximum of $a_\theta(r)$ decreases with increasing $\lambda$. The energy density, as shown in Fig.~\ref{fig:vimp_prof_N0_lambda}(c), has a more significant region of negative energy for $\lambda<1$, and for such values of the coupling constant, the total vacuum energy is negative. For $\lambda>1$, the positive region of the energy density is more significant, and the total vacuum energy is positive. At critical coupling our numerical methods evaluate the energy as zero to five decimal places. Although we see from Fig.~\ref{fig:vimp_prof_N0_lambda}(d) that there are some differences in the magnetic field depending on $\lambda$, the value of the topological charge is unaffected, as we would expect.

\begin{figure}[!htb]
\begin{center}
\subfloat[\, $\lambda=0.5,~\phi(r)$]{\includegraphics[totalheight=5.cm]{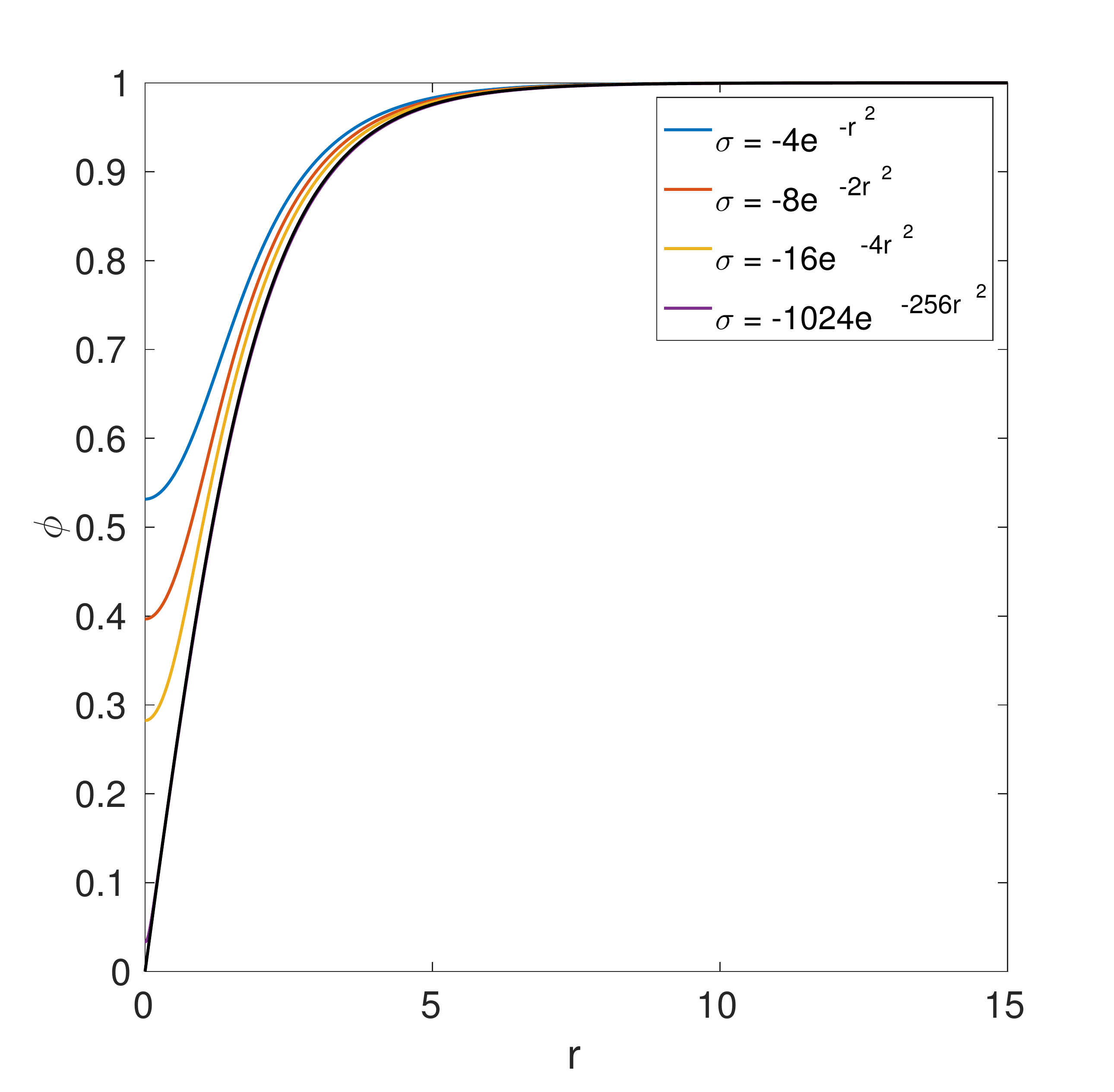}}
\subfloat[\, $\lambda=0.5,~a_\theta(r)$]{\includegraphics[totalheight=5.cm]{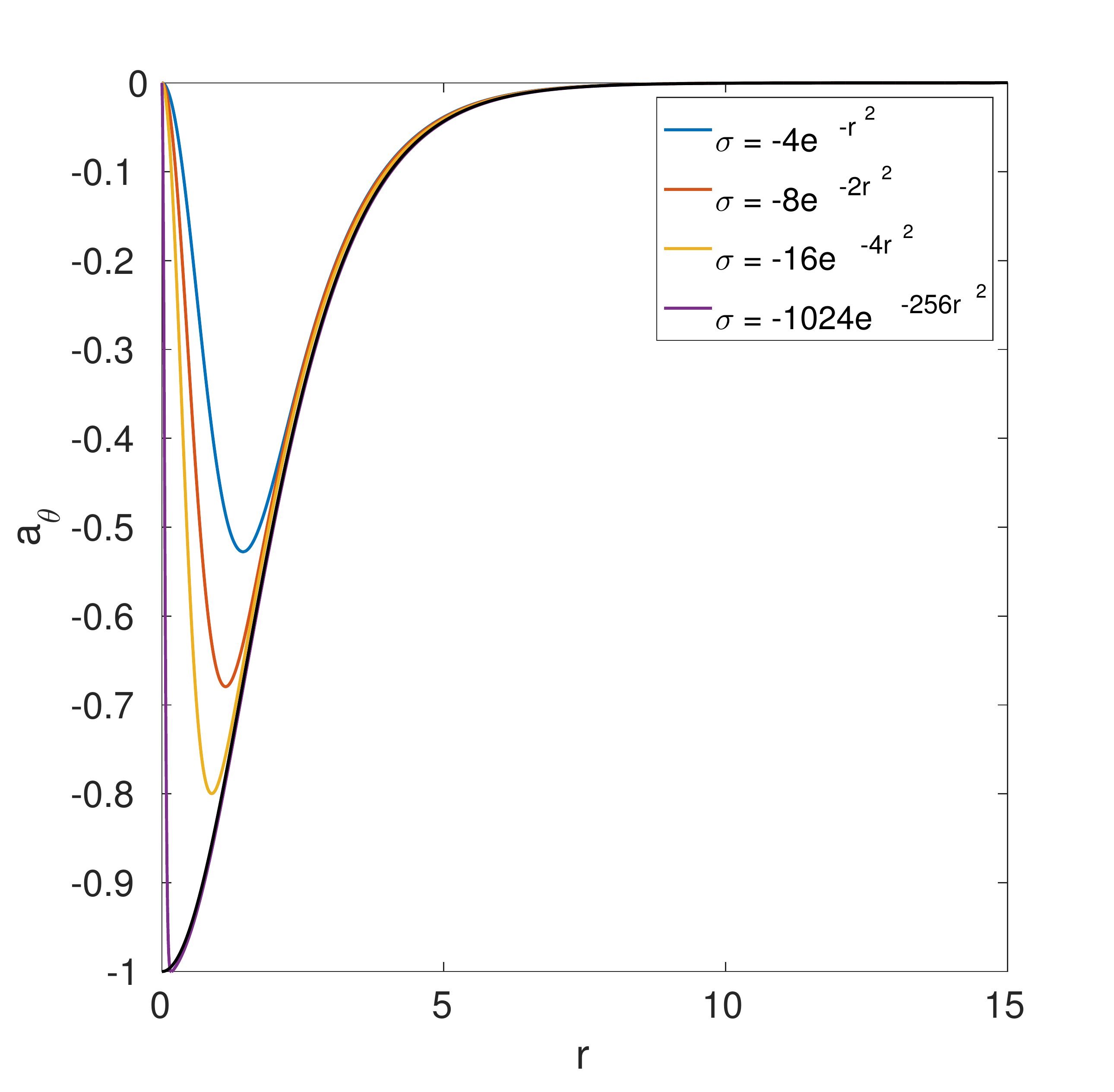}}\\
\subfloat[\, $\lambda=1.0,~\phi(r)$]{\includegraphics[totalheight=5.cm]{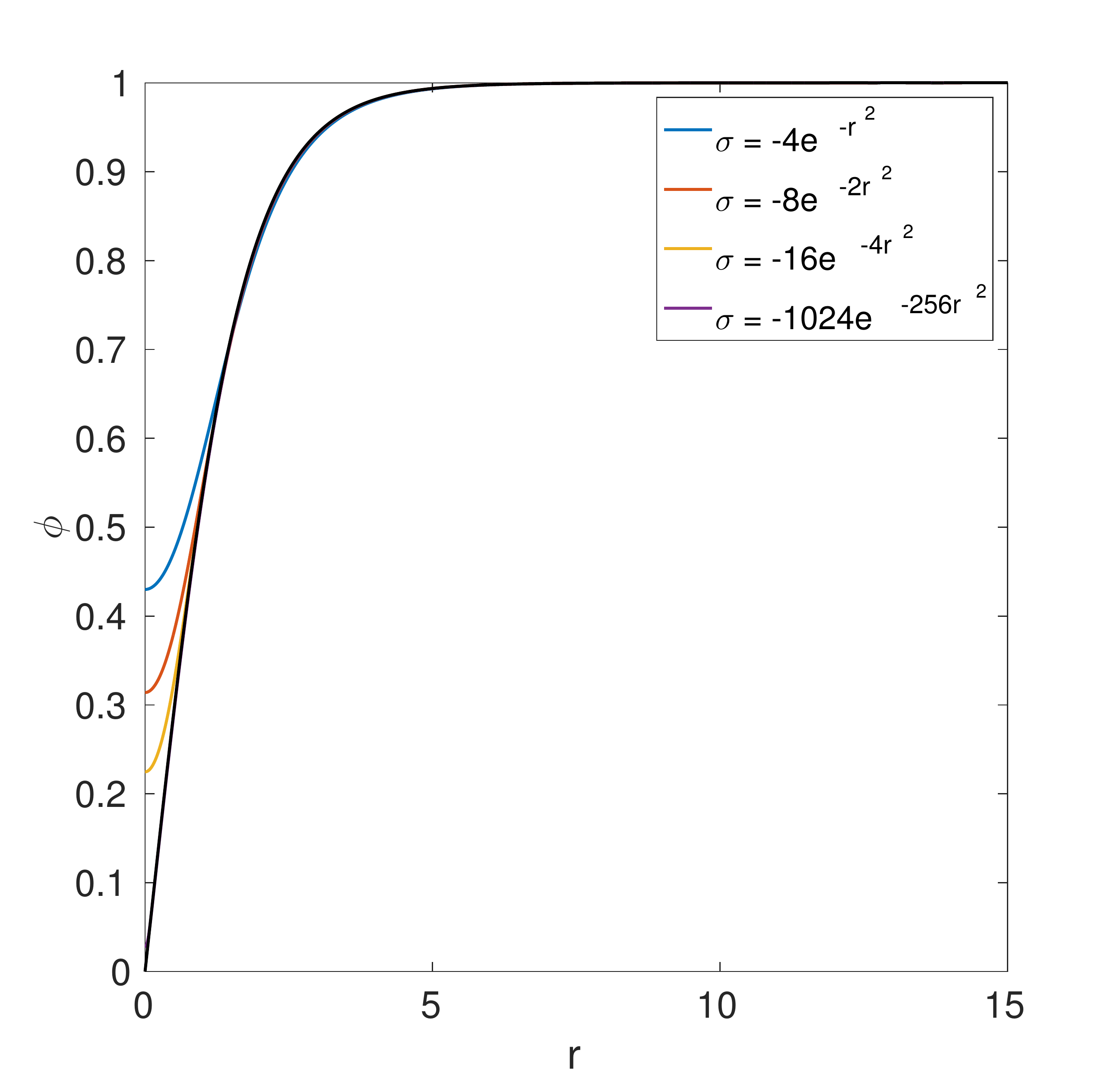}} 
\subfloat[\, $\lambda=1.0,~a_\theta(r)$]{\includegraphics[totalheight=5.cm]{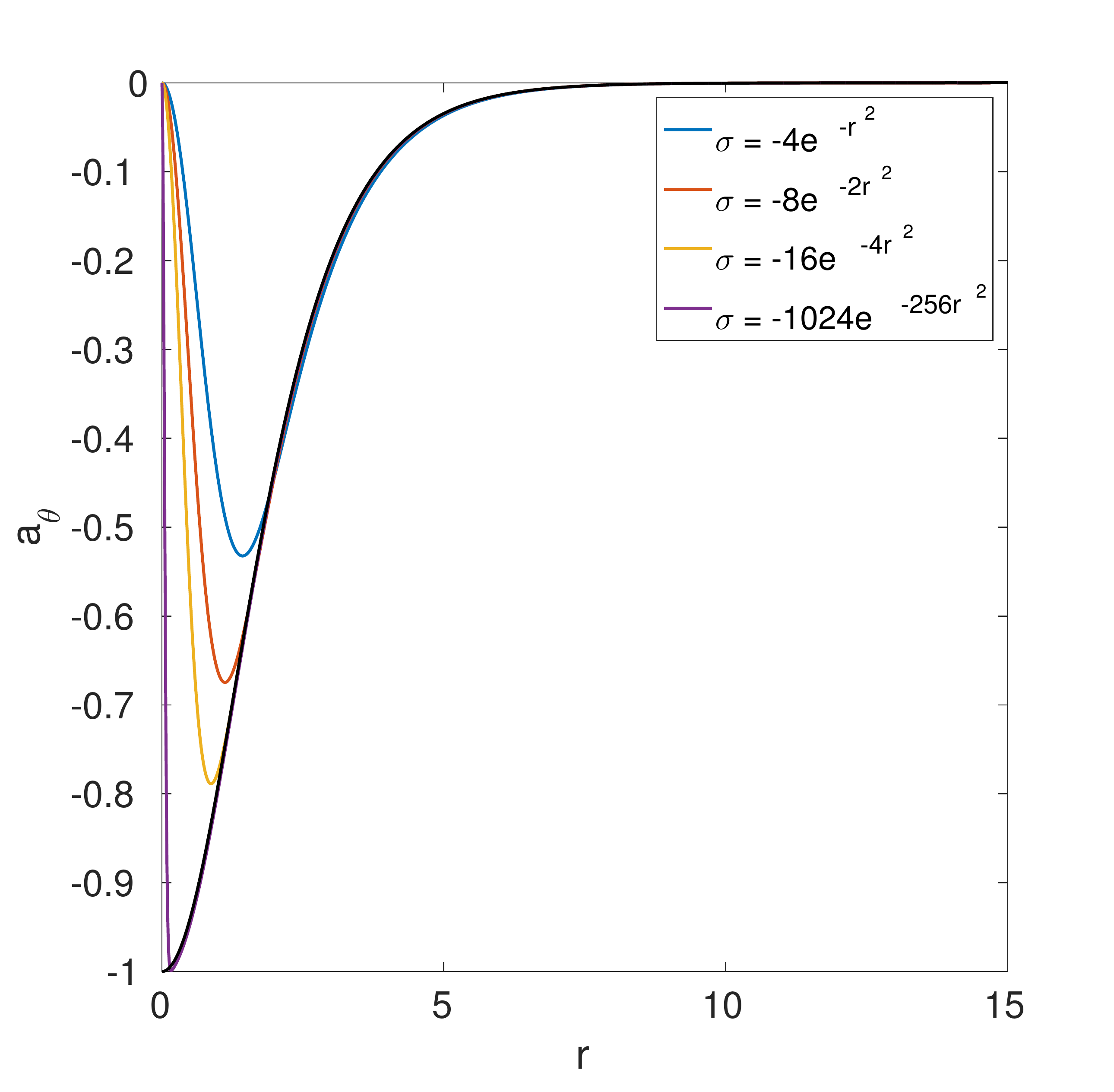}} \\
\subfloat[\, $\lambda=1.5,~\phi(r)$]{\includegraphics[totalheight=5.cm]{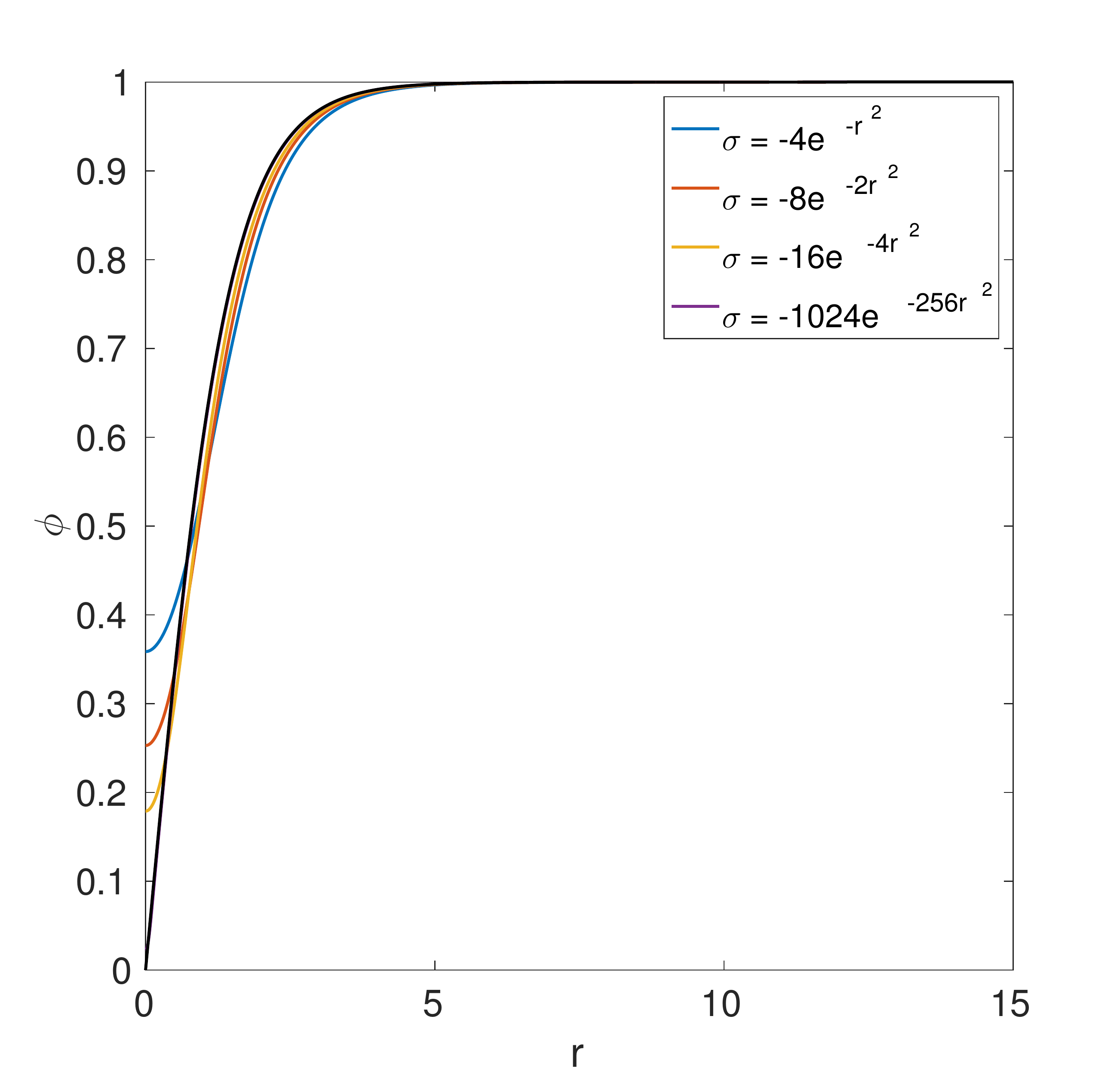}}
\subfloat[\, $\lambda=1.5,~a_\theta(r)$]{\includegraphics[totalheight=5.cm]{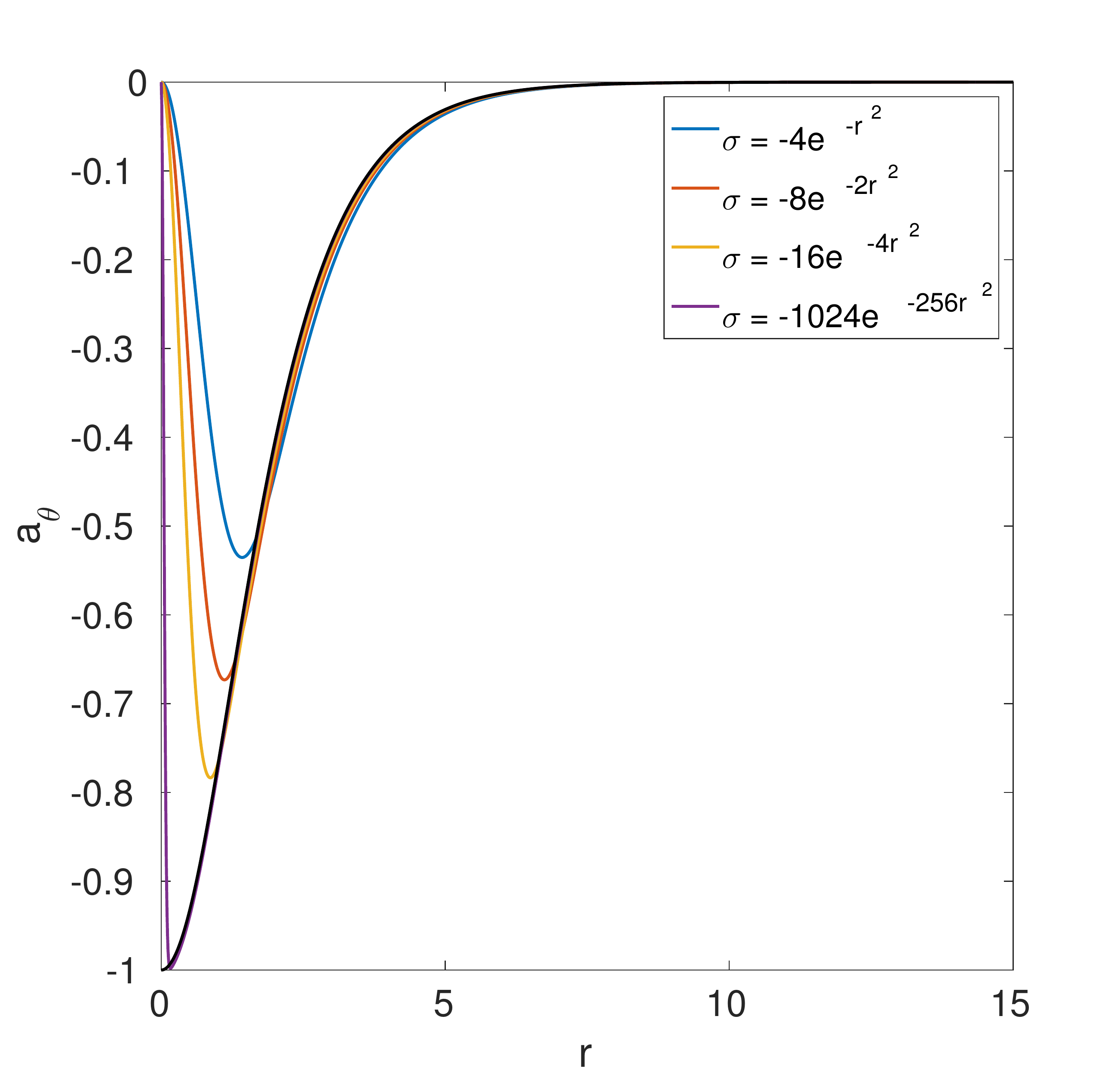}}
\caption{Vacuum profile functions $\phi(r),~a_\theta(r)$ for impurities $\sigma(r) = -4de^{-dr^2}$ and coupling constants (a)~$\lambda=0.5$, (b)~$\lambda=1.0$, (c)~$\lambda=1.5$. In each subfigure we plot an $N=1$ vortex profile function for that coupling constant as a solid black line. Note that for $a_\theta$ the profile function is shifted down to enable comparison with the impurity vacuum solutions.}
\label{fig:vimp_approach_delta}
\end{center} 
\end{figure}

\subsection{Delta function impurities}

For $c<0$, the authors of Ref.~\cite{Cockburn:2015huk} applied a singular gauge transformation to show in the case of critical coupling that a delta function impurity of the form $-4\pi\alpha\delta(z)$ for $\alpha\in\mathbb{N}$ ``behaves" like an $(N+\alpha)$-vortex solution. By ``behaves" we mean that the vortex and impurity solution looks identical to an $(N+\alpha)$-vortex solution with the gauge field shifted down by $\alpha$. In Fig.~\ref{fig:vimp_approach_delta}, we plot vacuum profile functions $\phi(r),~a_\theta(r)$ for impurities of the form $\sigma(r) = -4de^{-dr^2}$. As $d$ increases, these impurities approach the delta function with $\alpha=1$. We display vacuum solutions for $\lambda=0.5,~1.0,~1.5$, and in each subfigure we also plot an $N=1$ vortex profile function for the same $\lambda$ as a solid black line. For $a_\theta(r)$, the $N=1$ profile function is shifted down so that it can be compared with the impurity vacuum solutions. As $d$ increases, regardless of the value of $\lambda$, the vacuum solutions approach the vortex profile functions, with a singularity at the origin for $a_\theta(r)$. This suggests that, for negative $c$, a delta function impurity behaves like a vortex in the Type I and II regimes, as well as at critical coupling. The numerical evidence is compelling and encourages us to generalise the argument of Ref.~\cite{Cockburn:2015huk} to any value of $\lambda$.


The key idea is to formally relate an axial vortex of degree $N=n+m$ with an axial vortex of degree $n$ in the presence of an impurity of strength $m$ via a singular gauge transformation. As in the argument at critical coupling, we define 
$h = 2\log\phi,$
so $e^h = |\phi|^2$. The quantity $h$ is gauge invariant, and it is finite everywhere except for at the zeros of $\phi$. We will rewrite the profile function equations \eqref{vimp_prof_eq} for an axial vortex of degree $N$ in terms of $h$. First we consider the equation for $\phi$ from \eqref{vimp_prof_eq} with no impurity ($\sigma \equiv 0$). This can be written in terms of $h$ as
\begin{equation}
\label{heq}
h^{\prime\prime} + \frac{1}{r}h^\prime + \frac{1}{2} \left(h^\prime\right)^2 -\frac{2\left(N-a_\theta\right)^2}{r^2} + \lambda \left(1 - {\rm e}^h\right) =
4\pi N \delta(r).
\end{equation}
This is well-defined for $r>0$, but the first four terms become singular as $r \to 0$. For $r\approx 0,$ the Higgs field satisfies $\phi \approx r^N,$ so $h\approx 2N \log r.$ The terms $h^{\prime\prime} + \frac{1}{r}h^\prime$  are the radial part of the Laplacian. Since $\log r$ is the Green's function of the two dimensional Laplacian, we have regularised these terms  by adding $4\pi N \delta(r)$ on the right-hand-side of equation \eqref{heq}.

The remaining singular terms as $r \to 0$ are
\begin{align}
\frac{\left(h'\right)^2}{2} - \frac{2(N-a_\theta)^2}{r^2}. 
\end{align}
Recall that the gauge field $a_\theta$ satisfies the boundary conditions $a_\theta(0) = 0$ and $a_\theta(\infty) = N$. Using the boundary condition $a_\theta(0)=0$ and differentiating $h$ for $r\approx0$ as $h'\approx 2N/r$, we see that these two terms cancel.

We have now rewritten the profile function equation \eqref{vimp_prof_eq} for the Higgs field $\phi$ of an axial vortex of degree $N=n+m$ in terms of $h$. To relate this to an axial vortex of degree $n$ in the presence of an impurity of strength $m$, we perform the singular gauge transformation $a_\theta(r) \mapsto {\tilde a}_\theta(r) = a_\theta(r) -m,$ for $r> 0.$ Then equation \eqref{heq} becomes 
\begin{equation}
\label{heqs}
h^{\prime\prime} + \frac{1}{r}h^\prime + \frac{1}{2} \left(h^\prime\right)^2 -\frac{2\left(n-{\tilde a}_\theta\right)^2}{r^2} + \lambda \left(1 + \sigma(r) - {\rm e}^h\right) = 4\pi n \delta(r),
\end{equation}
where we have moved $4\pi m \delta(r)$ to the left-hand-side and defined
$\sigma(r) = -\frac{4\pi m}{\lambda} \delta(r).$ This is the radial equation of a vortex of charge $n$ in the presence of an impurity $\sigma(r)$ at the origin. 

Equation \eqref{vimp_prof_eq} for the gauge field is
\begin{align}
a''_\theta - \frac{a'_\theta}{r} + (N-a_\theta) {\rm e}^h = 0.
\end{align}
Note that for fixed $h$ this is a linear equation in $a_\theta.$ The singular gauge transformation  $a_\theta(r) \mapsto {\tilde a}_\theta(r) = a_\theta(r) -m$ results in
\begin{align}\label{a_h2}
{\tilde a}_\theta^{\prime\prime} - \frac{{\tilde a}_\theta^\prime}{r} - \frac{r}{2} \sigma^\prime + (n-{\tilde a}_\theta) {\rm e}^h= 0,
\end{align}
where we added the appropriate impurity term. This is justified for $r>0$ when $\sigma = -\frac{4\pi m}{\lambda} \delta(r)$ is a delta function. Hence, we have shown that for $r>0$ the gauge field ${\tilde a}_\theta$ satisfies the equation for an $n$ vortex with an impurity at the origin. 
As $r\to \infty$, the gauge potential ${\tilde a}_\theta\to n,$ which is the topological charge. As $a_\theta(0)=0$, we have ${\tilde a}_\theta(r)\to -m$ as $r\to 0$. However, to be well-defined at the origin, the gauge field ${\tilde a}_\theta$ would need to satisfy ${\tilde a}_\theta(0)=0$. So, ${\tilde a}_\theta$ is singular at the origin. In summary, if we impose axial symmetry then an $N=n+m$ vortex configuration is related to an $n$ vortex configuration in the presence of an impurity of the form $\sigma(r) = -\frac{4\pi m}{\lambda} \delta(r)$ by a singular gauge transformation.


\begin{figure}[!htb]
\begin{center}
\subfloat[\, $N=0$]{\includegraphics[totalheight=7cm]{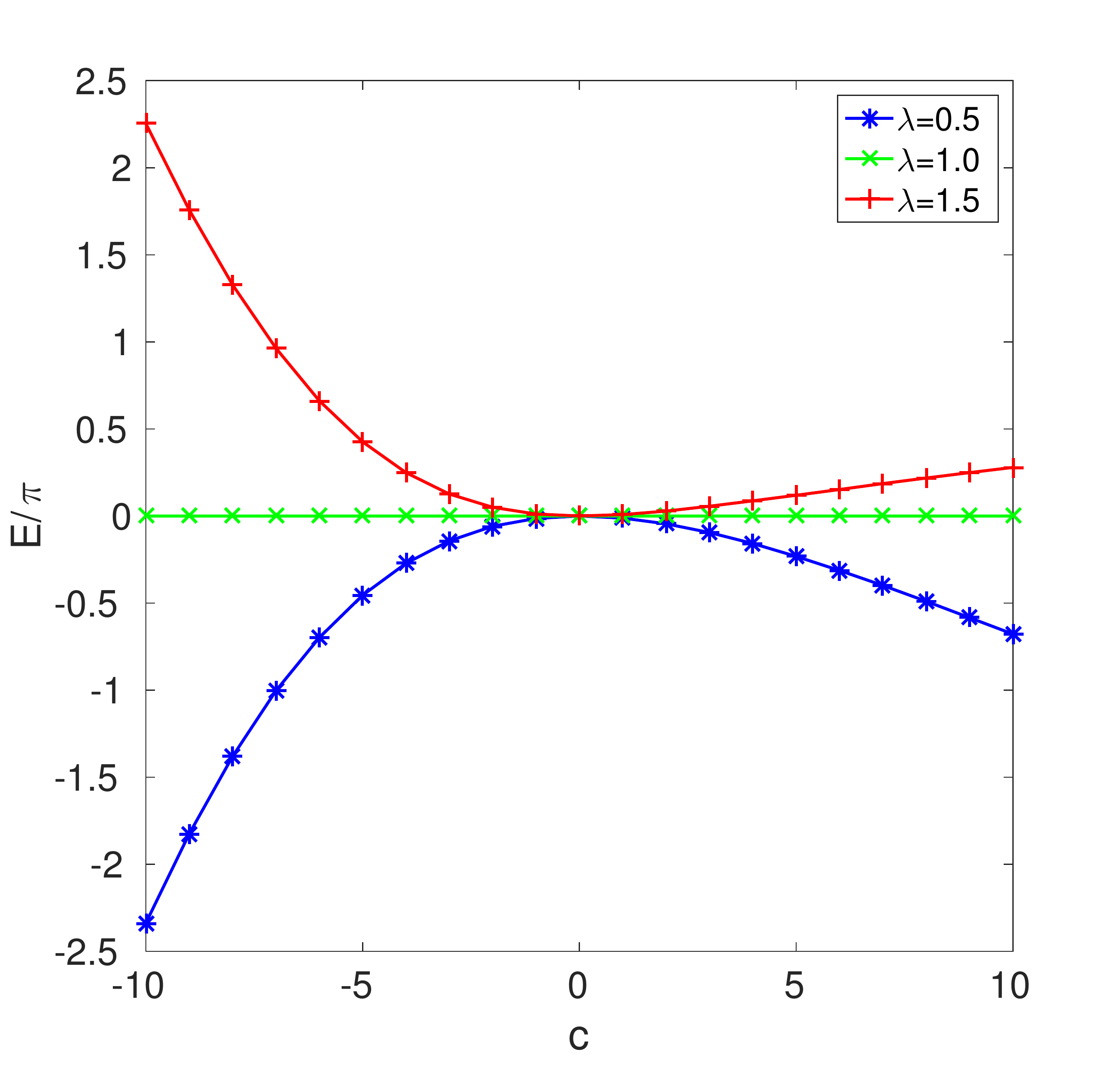}} 
\subfloat[\, $N=1$]{\includegraphics[totalheight=7cm]{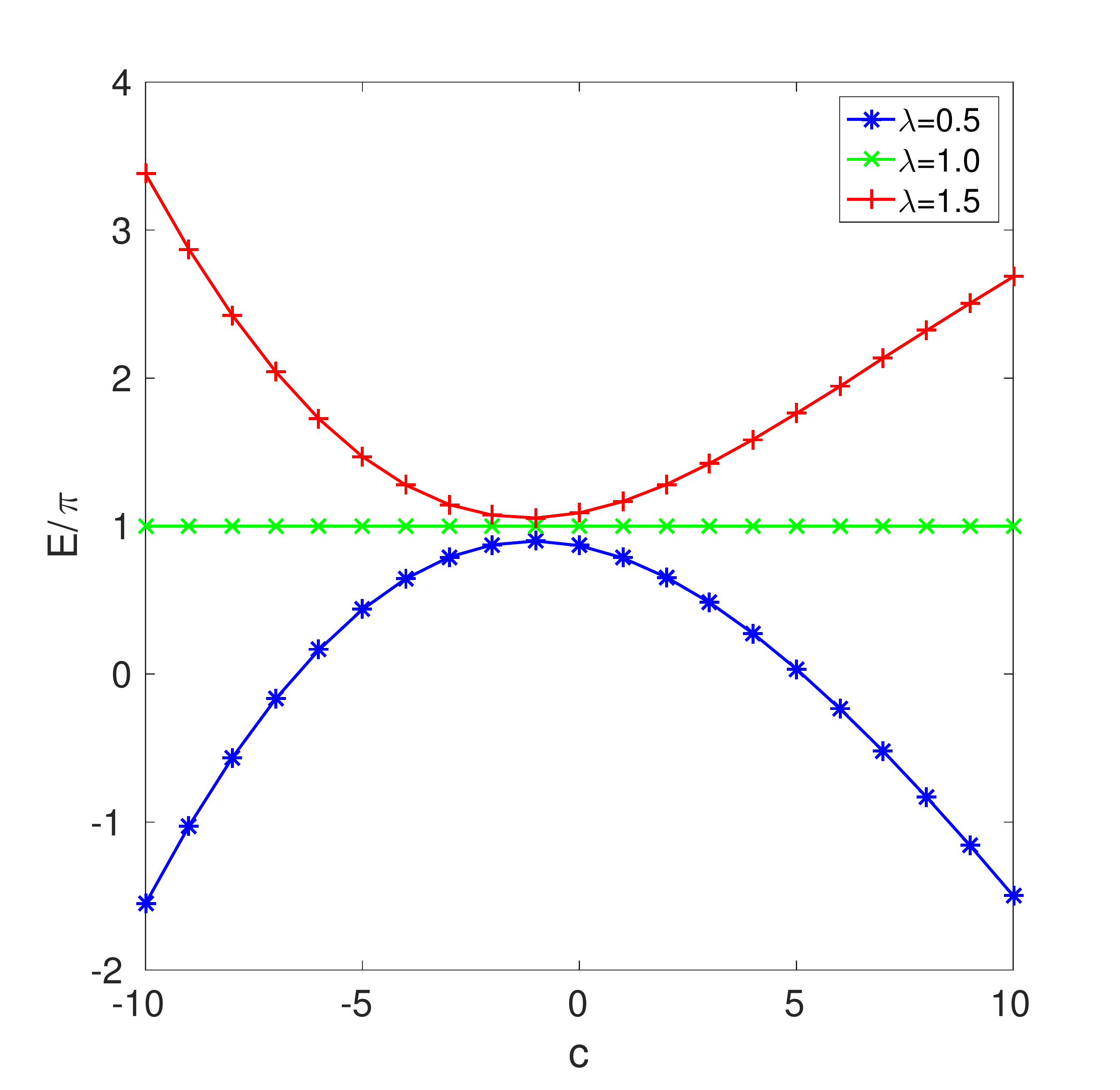}}
\caption{Energy $E/\pi$ of profile function solutions obtained by solving equations \eqref{vimp_prof_eq}. In each subfigure we display energy against $c$ for impurities of the form $\sigma(r)=ce^{-r^2}$ with topological charge (a)~$N=0$, and (b)~$N=1$.}
\label{fig:vimp_prof_energy}
\end{center} 
\end{figure}

\subsection{Binding energy of vortices and impurities}

In Fig.~\ref{fig:vimp_prof_energy} we compare the energy of profile function solutions to equations \eqref{vimp_prof_eq} for $\lambda=0.5,~1.0,~1.5$ and different values of the topological charge $N$. Fig.~\ref{fig:vimp_prof_energy}(a) gives the vacuum energy as a function of $c$ for impurities of the form $\sigma(r)=ce^{-r^2}$. For any $\lambda$, the vacuum energy is zero at $c=0$, since this describes the case in which there is no impurity. For $\lambda=0.5$, the vacuum energy is negative and decreases as $|c|$ increases, whilst for $\lambda=1.5$ the energy is positive and increasing with $|c|$. Note that for $c<0$, the energy increases or decreases more steeply than for $c>0$, and at roughly the same rate for both $\lambda=0.5$ and $\lambda=1.5$. However for $c>0$, the decrease in energy for $\lambda=0.5$ is steeper than the increase in energy for $\lambda=1.5$.

Fig.~\ref{fig:vimp_prof_energy}(b) displays energy as a function of $c$ for topological charge $N=1$. For both $\lambda=0.5$ and $\lambda=1.5$, the energy curves away from the constant energy $E_1\big|_{\lambda=1}/\pi=1$, with the point of closest approach to this line being at $c\approx-1.3$, though the specific value is slightly different for each $\lambda$. For $c>0$, the energy for $\lambda=1.5$ is strictly increasing with increasing $c$, and for $\lambda=0.5$ it is strictly decreasing. However for $c<0$ there is a region in which the energy decreases with decreasing $c$ for $\lambda=1.5$, and increases for $\lambda=0.5$. This is interesting as it indicates that for $\lambda>1$ we can lower the energy of a vortex by including an impurity. As is the case in the absence of an impurity, the energy for $\lambda<1$ is strictly less than $E_1\big|_{\lambda=1}/\pi=1$ for any $c$, whilst the energy for $\lambda>1$ is strictly greater than 1.

\begin{figure}[!htb]
\begin{center}
\includegraphics[totalheight=10.0cm]{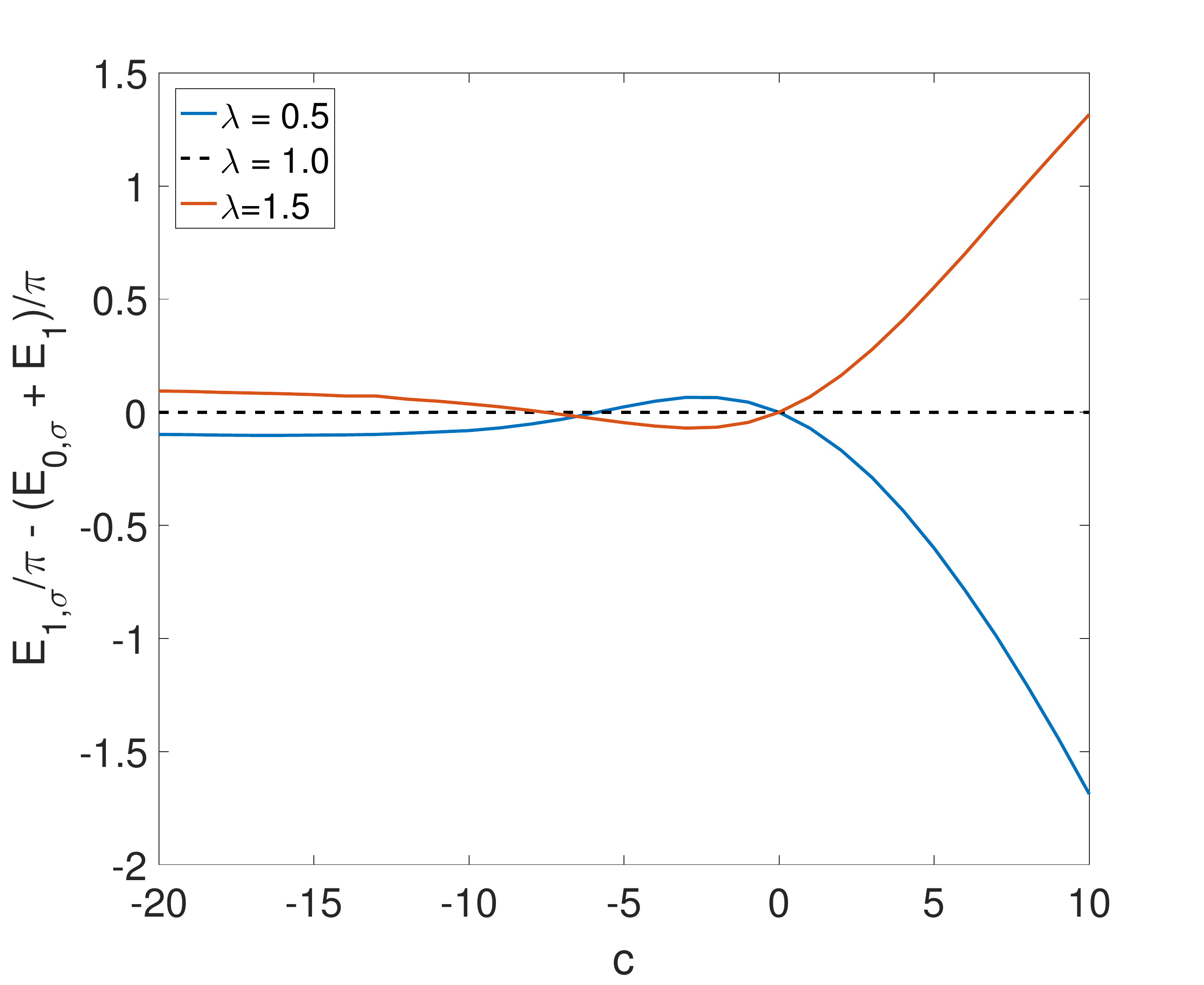}
\caption{Energy difference $\left(E_{1,\sigma}-(E_{0,\sigma}+E_1)\right)/\pi$ as a function of $c$ for impurities of the form $\sigma(r) = ce^{-r^2}$, and $\lambda=0.5,~1.0,~1.5$. Where the energy difference is positive, the impurity will repel a vortex, and where it is negative the impurity and vortex will attract. }
\label{fig:energy_difference}
\end{center}
\end{figure}

To determine whether a vortex and impurity will attract or repel, we consider
the binding energy of a vortex and an impurity that is
the difference between the energy of an impurity coincident to a vortex and one which is well-separated from the vortex. Let $E_{N,\sigma}$ denote the energy of an $N$-vortex profile function coincident to an impurity $\sigma$, and $E_N$ the energy of an $N$-vortex profile function with no impurity. We approximate the energy of a well-separated vortex and impurity by $(E_1+E_{0,\sigma})$: the sum of the energy of a single vortex with no impurity and the vacuum energy for the impurity $\sigma$. If $E_{1,\sigma}-(E_1+E_{0,\sigma})>0$, then the energy is lower when the vortex and impurity are well-separated and so they will repel. If $E_{1,\sigma}-(E_1+E_{0,\sigma})<0$ then the energy is lower when the vortex and impurity are coincident and so they will attract. 

In Fig.~\ref{fig:energy_difference}, we plot the energy difference $E_{1,\sigma}-(E_1+E_{0,\sigma})$ against $c$ for impurities of the form  $\sigma(r)=ce^{-r^2}$, and $\lambda=0.5,~1.0,~1.5$. At critical coupling, the energy difference is always zero since the vortex and impurity neither attract nor repel. The energy difference is also zero at $c=0$ for any $\lambda$ because there is no impurity in this case. For $c>0$, the energy difference is positive for $\lambda=1.5$, and negative for $\lambda=0.5$, indicating that a vortex and impurity will repel for $\lambda=1.5$ and attract for $\lambda=0.5$. For $c<0$, there are two regimes in which the impurity can either attract or repel a vortex, and the range of $c$ for which each behaviour occurs depends on $\lambda$. For $\lambda=1.5$, if $c\in(-7.5,0)$ then the energy difference is negative so the impurity attracts a vortex, and for $c<-7.5$ the energy difference is positive so the vortex and impurity repel. Similarly, for $\lambda=0.5$, the energy difference is positive for $c\in(-5.84,0)$, and negative for $c<-5.84$. At the critical values of $c$ between each regime, the energy difference is zero, and so the vortex and impurity neither attract nor repel.

\begin{figure}[!htb]
\begin{center}
  \includegraphics[totalheight=10.0cm]
                  {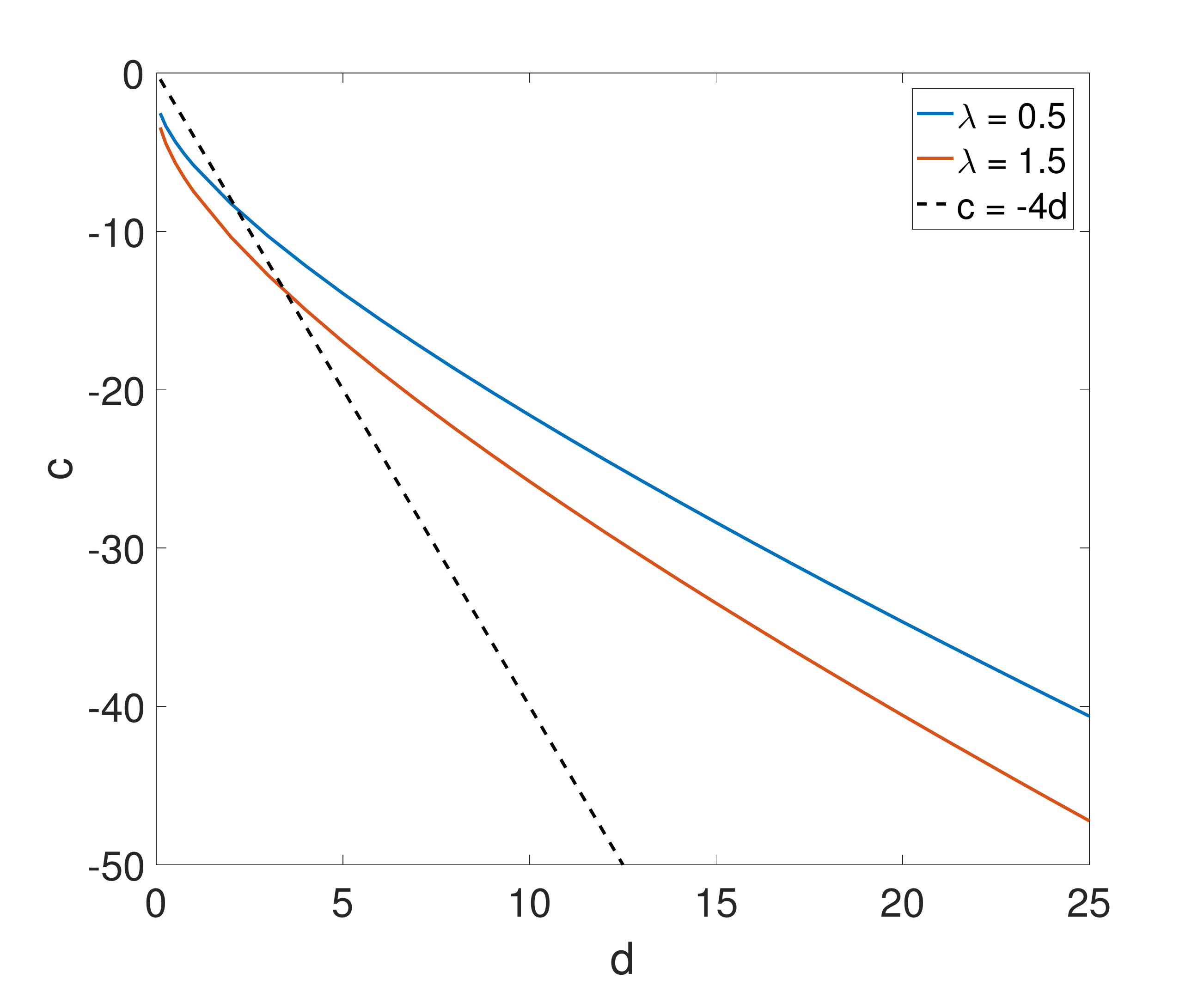}
\caption{We plot the critical values of $c$ at which the impurity changes from attracting to repelling a vortex against $d$ for impurities of the form $\sigma(r)=ce^{-dr^2}$, and $\lambda=0.5,~1.5$. The line $c=-4d$, where the impurities approach a delta function of strength $\alpha=1$ as $d$ increases, is shown as a dashed line for comparison.}
\label{fig:varying_d}
\end{center}
\end{figure}

Since for $c<0$ a delta function impurity behaves like another vortex, we would expect it to attract a vortex for $\lambda<1$ and repel it for $\lambda>1$. However we have found two different regimes of behaviour for $c<0$ in which the impurity can either attract or repel a vortex. The critical value of $c$ separating the two regimes depends on the coupling constant $\lambda$ and the impurity parameter $d$. For each $\lambda$, we calculate these values by fixing $d$ and evaluating the energy difference $E_{1,\sigma}-(E_1+E_{0,\sigma})$  for a range of $c$. The value of $c$ at which the energy difference is zero is the critical value separating the two regimes for the chosen $d$. In Fig.~\ref{fig:varying_d}, we plot the critical values of $c$ against $d$ for $\lambda=0.5,~1.5$ in blue and red respectively. We will refer to these as the \emph{critical lines} separating the two regimes of behaviour. Impurities for which $(d,c)$ is located below the critical line will attract a vortex for $\lambda=0.5$ and repel it for $\lambda=1.5$. Conversely if $(d,c)$ is located above the critical line, then the impurity repels a vortex for $\lambda=0.5$ and attracts it for $\lambda=1.5$.  

We previously approximated a delta function impurity by  $\sigma(r)=-4de^{-dr^2}$, which approaches a delta function of strength $\alpha=1$ as $d$ increases. We also plot the line $c=-4d$ corresponding to these impurities in Fig.~\ref{fig:varying_d}. For small $d$, our approximation to a delta function impurity does not behave like a vortex: the impurities are located above the critical line and will repel a vortex for $\lambda=0.5$ and attract it for $\lambda=1.5$. However as $d$ increases, the line $c=-4d$ crosses the critical $c$ line, and the impurities do behave like vortices. For $\lambda=1.5$ the intersection between the two lines occurs at $d\approx3.5$, and for $\lambda=0.5$ at $d\approx2.2$. Since an impurity of this form only becomes more like a delta function as $d$ increases, this supports our  conclusion that a delta function impurity behaves like a vortex for $c<0$.

\subsection{Impurity asymptotics} \label{impasymp}
As a final comment on the profile function equations, we discuss the impurity strength. This is calculated in Ref.~\cite{Cockburn:2015huk} by linearising the Bogomolny equation for large $r$, but we will calculate the same quantities by linearising the profile function equations \eqref{vimp_prof_eq}, following a similar argument for vortices given in Ref.~\cite{Speight:1996px}. We consider a vacuum solution in the presence of an impurity $\sigma$. As $r\rightarrow\infty$, we have the boundary conditions $\phi(\infty)=1$, and $a_\theta(\infty)=0$. We linearise the profile functions at infinity by taking
\begin{align}\label{vimp_lin_field}
\phi(r) = 1 + \alpha(r), \qquad a_\theta(r) = \beta(r),
\end{align}
where $\alpha(r)$ and $\beta(r)$ are small. Assuming that the impurity decays sufficiently rapidly as $r\rightarrow\infty$, we obtain the linearised profile function equations
\begin{align}\label{vimp_lin_eqs}
\alpha'' +\frac{1}{r}\alpha' -\lambda\alpha = 0, \nonumber \\
\beta'' - \frac{1}{r}\beta' -\beta =0,
\end{align}
or equivalently
\begin{align}\label{vimp_lin_eqs_Bessel}
\lambda\left( \frac{d^2\alpha(\sqrt{\lambda}r)}{d(\sqrt{\lambda}r)^2} + \frac{1}{\sqrt{\lambda}r}\frac{d\alpha(\sqrt{\lambda}r)}{d(\sqrt{\lambda}r)} -\alpha \right) = 0, \nonumber \\
\frac{d}{dr}\left(\frac{\beta}{r}\right) + \frac{1}{r}\frac{d}{dr}\left(\frac{\beta}{r}\right) - \left(1+\frac{1}{r^2}\right)\frac{\beta}{r}=0.
\end{align}
These are identical to the equations that were obtained for a vortex in Ref.~\cite{Speight:1996px} and their solutions are
\begin{align}\label{vimp_lin_sol}
\alpha(r) = qK_0\left(\sqrt{\lambda}r\right), \qquad \beta(r) = mrK_1(r),
\end{align}
where $K_0(r)$ and $K_1(r)$ are Bessel functions. The interpretation is that for large $r$, the vortex, or in this case the impurity, can be considered to be made up of a scalar monopole of charge $q$ and a magnetic dipole of moment $m$ \cite{Speight:1996px}. At critical coupling $q=m$, and this is what we call the point charge of the impurity. To show that $q=m$ in this case, we substitute the profile function expressions for $N=0$ into the Bogomolny equations \eqref{vimp_bog_eq} to obtain
\begin{align}
r\frac{d\phi}{dr} + a_\theta\phi = 0,\nonumber \\
\frac{1}{r}\frac{da_\theta}{dr} - \frac{1}{2}(1 + \sigma -\phi^2 ) = 0.
\end{align}
We substitute the asymptotic forms \eqref{vimp_lin_field} into the first equation above and find to leading order
\begin{align}
\beta(r) = -r \frac{d\alpha}{dr}.
\end{align}
Since $\frac{dK_0}{dr} = -K_1(r)$, and we have already found that $\alpha(r) = qK_0(\sqrt{\lambda}r)$, we can write $\beta(r)$ as
\begin{align}
\beta(r) = q rK_1(r).
\end{align}
Comparing this with the expression for $\beta(r)$ in  \eqref{vimp_lin_sol}, we find $q=m$.

For any $\lambda$, we can calculate $q$ and $m$ by fitting the solutions \eqref{vimp_lin_sol} to numerical vacuum solutions of \eqref{vimp_prof_eq}. In Table~\ref{tab:imp_strength}, we give the values of $q$ and $m$ for different impurities and $\lambda=0.5,~1.0,~1.5$. Notice that the values of $q$ and $m$ are positive for $c>0$ and negative for $c<0$. We can see this by considering how the approach to the vacuum values of the profile functions in Fig.~\ref{fig:vimp_prof_N0} changes with the sign of $c$ and comparing this with the shape of the Bessel functions $K_0$ and $K_1$. For $c>0$ the profile functions approach the vacuum from above, asymptotically matching the shape of a Bessel function, but for $c<0$ they approach from below, matching the shapes of $-K_0$ and $-K_1$. Earlier we discussed the conjecture that impurities $\sigma(r)$ approaching a delta function will behave like a vortex. In Table~\ref{tab:imp_strength} we illustrate this at critical coupling with impurities $\sigma(r) = -4e^{-r^2},~-8e^{-2r^2},~-16e^{-4r^2}$ approaching a delta function. The point charge of a single critically coupled vortex was numerically calculated in  Ref.~\cite{deVega:1976xbp} as 1.7079. We see in the table that the strength of the impurity approaches this value as the impurity approaches a delta function, with the values already identical to two decimal places for $\sigma(r)=-16e^{-4r^2}$.

\def\arraystretch{1.2}
\begin{table}[htb!]
  \begin{small}
\caption{Some different impurities and the corresponding values of $q$ and $m$.}
\begin{center}
\begin{tabular}{ c | c | c c }
$\lambda$ & Impurity & $q$ & $m$ \\\hline
  &~$e^{-r^2}$ & 0.12  & 0.30   \\
 & $-e^{-r^2}$ & $-0.16$ & $-0.35$ \\
0.5 & ~$2e^{-r^2}$ & 0.22  & 0.57 \\
 & $-2e^{-r^2}$ & $-0.35$ & $-0.76$ \\
 & ~$4e^{-r^2}$ &  0.37 & 1.02      \\
 & $-4e^{-r^2}$ &$-0.89$ & $-1.91$ \\
\hline
  &~$e^{-r^2}$ &    ~0.29  &    ~0.29\\
 & $-e^{-r^2}$ & $-0.35$  & $-0.35$ \\
1.0 & ~$2e^{-r^2}$ &   0.56 & 0.56 \\
 & $-2e^{-r^2}$ &  $-0.74$ & $-0.74$ \\
 & ~$4e^{-r^2}$ &  ~0.99 &    ~0.99 \\
 & $-4e^{-r^2}$ & $-1.82$ & $-1.82$ \\
 & $-8e^{-2r^2}$ & $-1.73$ & $-1.73$\\
& $-16e^{-4r^2}$ &   $-1.71$ & $-1.71$\\
\hline
  &~$e^{-r^2}$ & 0.50  & 0.30    \\
 & $-e^{-r^2}$ & $-0.59$ & $-0.35$  \\
1.5 & ~$2e^{-r^2}$ &  0.94  & 0.55 \\
 & $-2e^{-r^2}$ & $-1.28$ & $-0.75$ \\
 & ~$4e^{-r^2}$ &  1.64 &   0.98  \\
 & $-4e^{-r^2}$ & $-2.99$ &  $-1.77$ \\

\end{tabular}
\end{center}
\label{tab:imp_strength}
\end{small} 
\end{table}

\begin{figure}[!htb]
\begin{center}
\subfloat[\, $\sigma = 4e^{-r^2}$]{\includegraphics[totalheight=4.1cm]{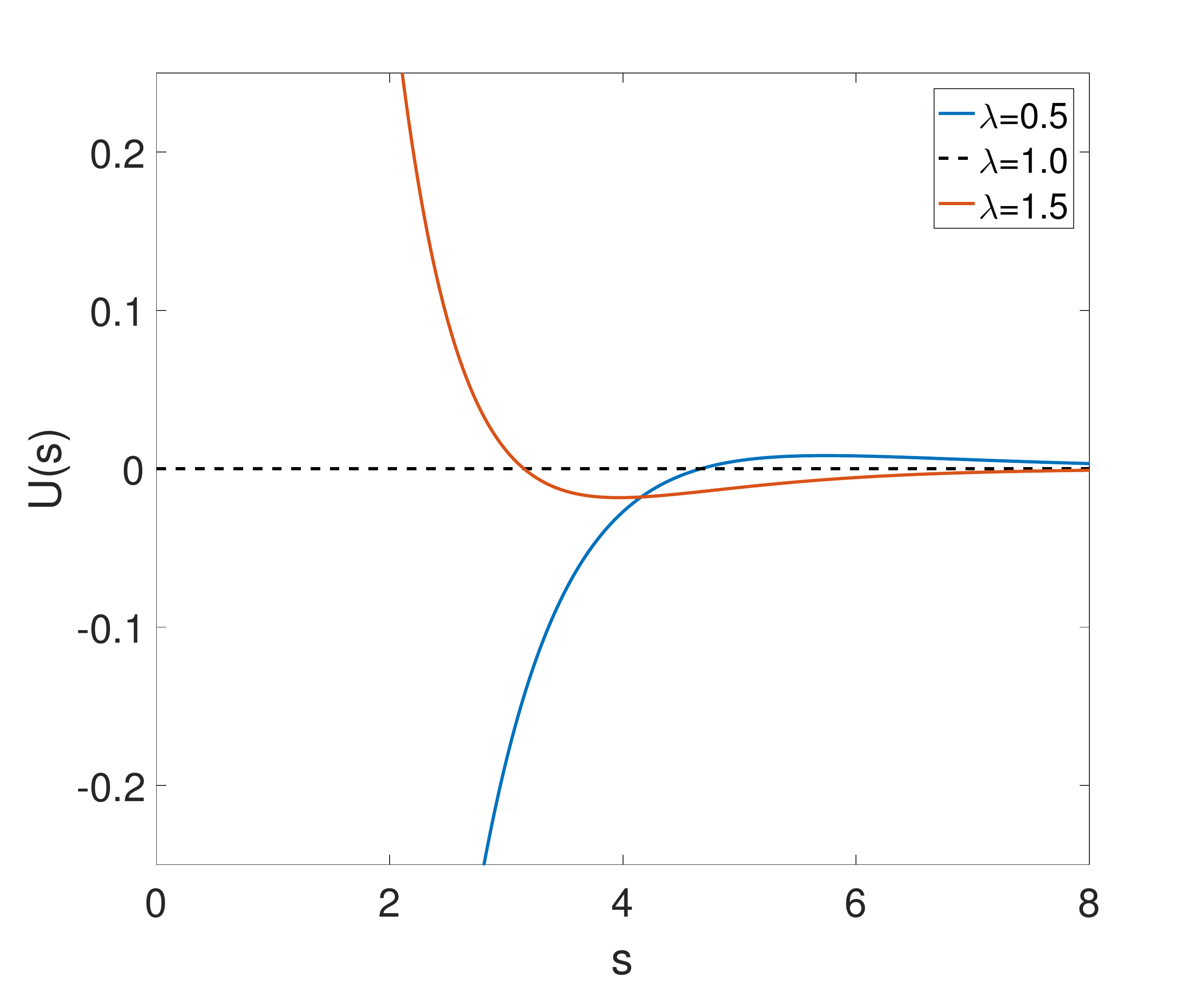}}
\subfloat[\, $\sigma = -4e^{-r^2}$]{\includegraphics[totalheight=4.1cm]{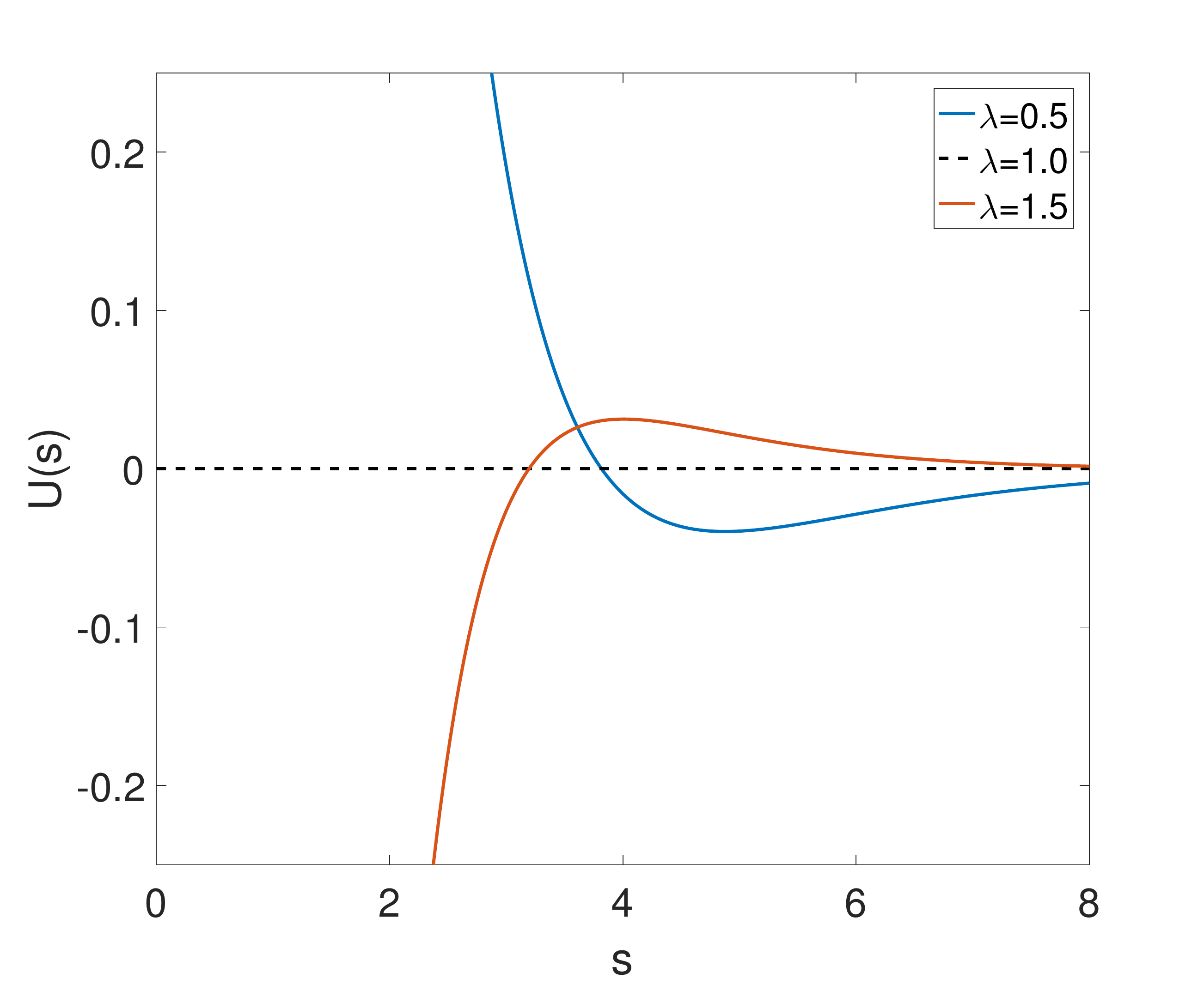}} 
\subfloat[\, $\sigma = -16e^{-4r^2}$]{\includegraphics[totalheight=4.1cm]{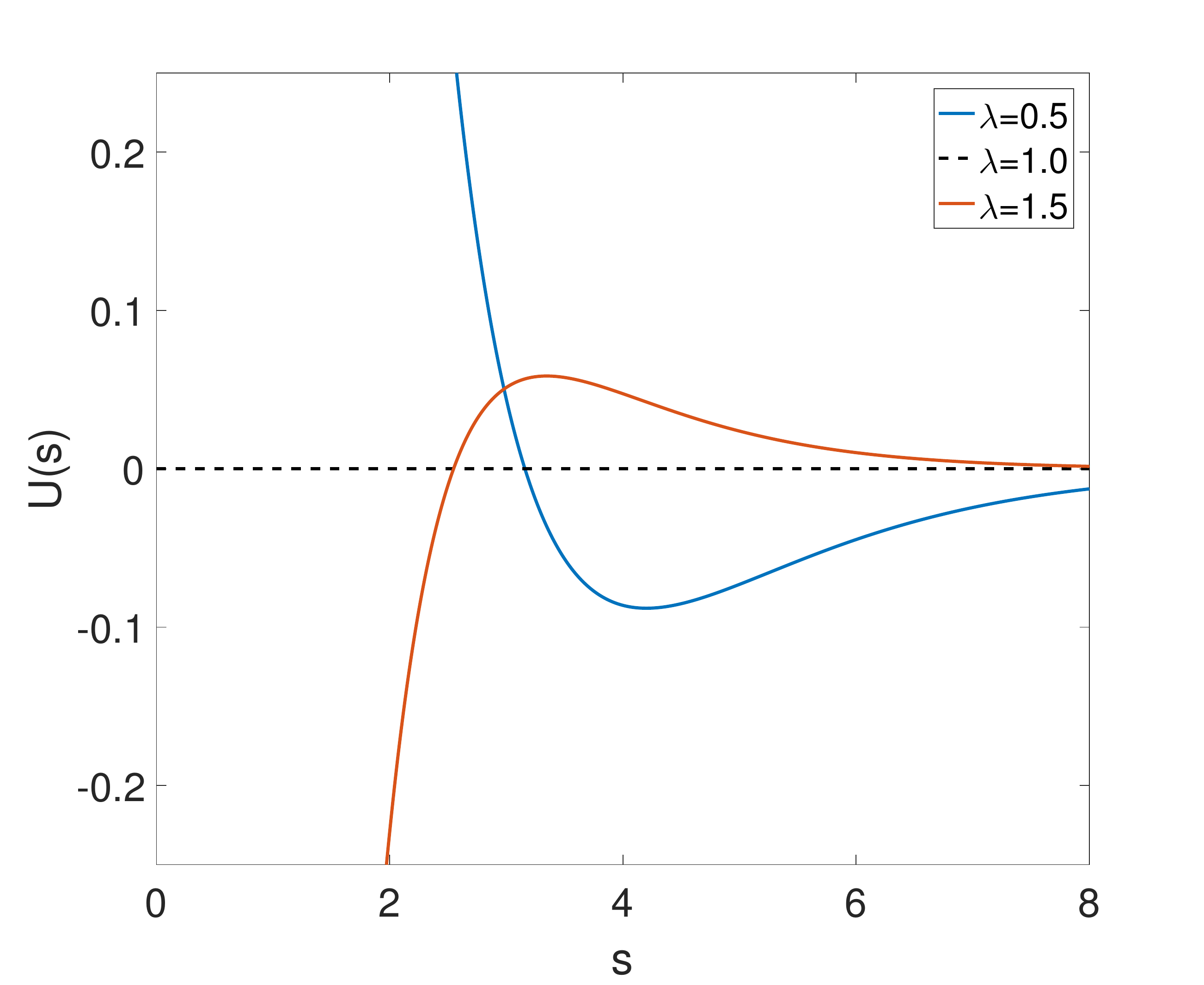}} 
\caption{The potential \eqref{vimp_intervort_pot} for a vortex and impurity with $\lambda=0.5,~1.0,~1.5$ and impurities (a) $\sigma(r)=4e^{-r^2}$, (b) $\sigma(r)=-4e^{-r^2}$, and (c) $\sigma(r) = -16e^{-4r^2}$. }
\label{fig:vimp_intervort_pot}
\end{center}
\end{figure}

In Ref.~\cite{Speight:1996px}, the vortex asymptotics was used to understand how the attraction or repulsion between two vortices depends on the value of $\lambda$. Since we have found the large $r$ behaviour of an impurity to be of the same form as that of a vortex, we adapt the expression for the intervortex potential of two vortices (see for example Ref.~\cite{Speight:1996px}) to obtain that for a vortex and an impurity. Let $q_\sigma,~m_\sigma$ be the charge and moment of an impurity $\sigma$, and $q_V,~m_V$ correspond to those of a vortex. Then the static potential is given by
\begin{align}\label{vimp_intervort_pot}
U(s) = 2\pi\left(m_\sigma m_V K_0(s) - q_\sigma q_V K_0(\sqrt{\lambda}s)\right).
\end{align}
Here $s$ denotes the distance between the vortex and impurity. We plot the potential \eqref{vimp_intervort_pot} for three different impurities in Fig.~\ref{fig:vimp_intervort_pot}. In each figure, the black dashed line gives the potential for $\lambda=1$, which, since $q=m$ here, is always zero. We plot the potential for $\lambda=0.5$ in blue, and for $\lambda=1.5$ in red. Each figure represents a different regime of impurity behaviour as observed in the previous section: (a)~$c>0$, (b)~$c<0$ above the critical line separating the two behaviours (as shown in Fig.~\ref{fig:varying_d}), and (c)~$c<0$ below the critical line. The potential for $c>0$ is the most clearly different from the other two, as the signs of the Bessel functions have been reversed. Figs.~\ref{fig:vimp_intervort_pot}(b) and (c) are a similar shape to one another, though in (c) the critical points of the functions are more exaggerated. We note that this approximation is only accurate past some critical value of the separation between the vortex and impurity, $s_c$ which depends on the values of $q$ and $m$, and we expect it to break down for $s\leq s_c$. 

The force due to $U(s)$ is given by
\begin{align}\label{vimp_force}
-U'(s) = 2\pi\left( m_\sigma m_V K_1(s) - \sqrt{\lambda}q_\sigma q_V K_1(\sqrt{\lambda}s) \right).
\end{align}
When $\lambda=1$, we know that $q=m$ for both a vortex and an impurity, so these terms cancel and there is no force. If $\lambda<1$, then $K_1(s)$ decays faster than $K_1(\sqrt{\lambda}s)$, and so the behaviour is determined by the sign of $q_\sigma$. For a vortex $q_v<0$, and for an impurity we saw that $q_\sigma<0$ if $c<0$ and $q_\sigma>0$ if $c>0$. Combining these, we find that for $\lambda<1$ the net force is negative if $c<0$ and positive if $c>0$. So this argument suggests that a vortex and impurity will attract if $c<0$ and repel if $c>0$. Similarly, if $\lambda>1$, then the $K_1(\sqrt{\lambda}s)$ term decays faster and so the $K_1(s)$ term dominates. The type of force depends on the sign of $m_\sigma$. We know that $m_V<0$, and we have that $m_\sigma<0$ if $c<0$ and $m_\sigma>0$ if $c>0$. The net force is positive if $c<0$ and negative if $c>0$, indicating that a vortex and impurity will repel for $c<0$ and attract for $c>0$. Comparing these predictions with the three regimes that we have found by considering the energy difference, we see that they are only accurate in one case -- for impurities $\sigma$ found below the critical line (shown in Fig.~\ref{fig:varying_d} for $\lambda=0.5,~1.5$). This argument predicts a different behaviour for both $c>0$ and impurities with $c<0$ found above the critical line. However the case in which an impurity approaches a delta function is correctly predicted to behave like a vortex. The next subsection will provide an explanation why energy differences and asymptotics appear to lead to different results.

\subsection{Static vortices and impurities}

Next we obtain static vortex solutions in the presence of magnetic impurities on a 2D grid. We solve the gradient flow equations
\begin{align}\label{vimp_gflow}
\pa_0\phi &= D_{ii}\phi + \frac{\lambda}{2}\left(1 + \sigma - |\phi|^2\right)\phi, \nonumber \\
\pa_0a_i &= -\epsilon_{ij}\left(\pa_jB - \frac{1}{2}\pa_j\sigma \right) - \frac{i}{2}\left(\overline{\phi}D_i\phi - \phi \overline{D_i\phi}\right),
\end{align}
using a finite difference method that is first order in time, and fourth order in space. We use timestep $\Delta t = 0.001$, and grid spacing $\Delta x = \Delta y = 0.1$, and typically solve on grids of size $401\times401$. 

To generate initial conditions, we first use the vacuum profile functions obtained by solving \eqref{vimp_prof_eq} to create a vacuum solution on the 2D grid. We can combine vortex solutions by using the Abrikosov ansatz.
This states that, given a vortex solution $(\phi(\mathbf{x}),a_\mu(\mathbf{x}))$, we can obtain an approximate multi-vortex solution as
\begin{align}\label{Abrikosov}
\hat{\phi}(\mathbf{x}) = \prod_i \phi(\mathbf{x}-\mathbf{x}_i), \quad \hat{a}_\mu(\mathbf{x}) = \sum_i a_\mu(\mathbf{x}-\mathbf{x}_i),
\end{align}
where the $\{\mathbf{x}_i\}$ are the positions of the vortex centres. This ansatz is very accurate if all vortices are widely separated.

Choosing in \eqref{Abrikosov} a solution describing a vortex at a given position in the absence of an impurity and the vacuum solution in the presence of a impurity creates an approximate solution of a vortex at the specified position in the presence of a magnetic impurity located at the origin. Where the vortex and impurity are well-separated this is already very accurate, but even for vortices close to the origin it provides us with a useful initial condition from which to begin solving \eqref{vimp_gflow}.

\begin{figure}[!htb]
\begin{center}
\subfloat[\, $\lambda=1.0$,~ $\sigma = -e^{-r^2}$]{\includegraphics[totalheight=7.0cm]{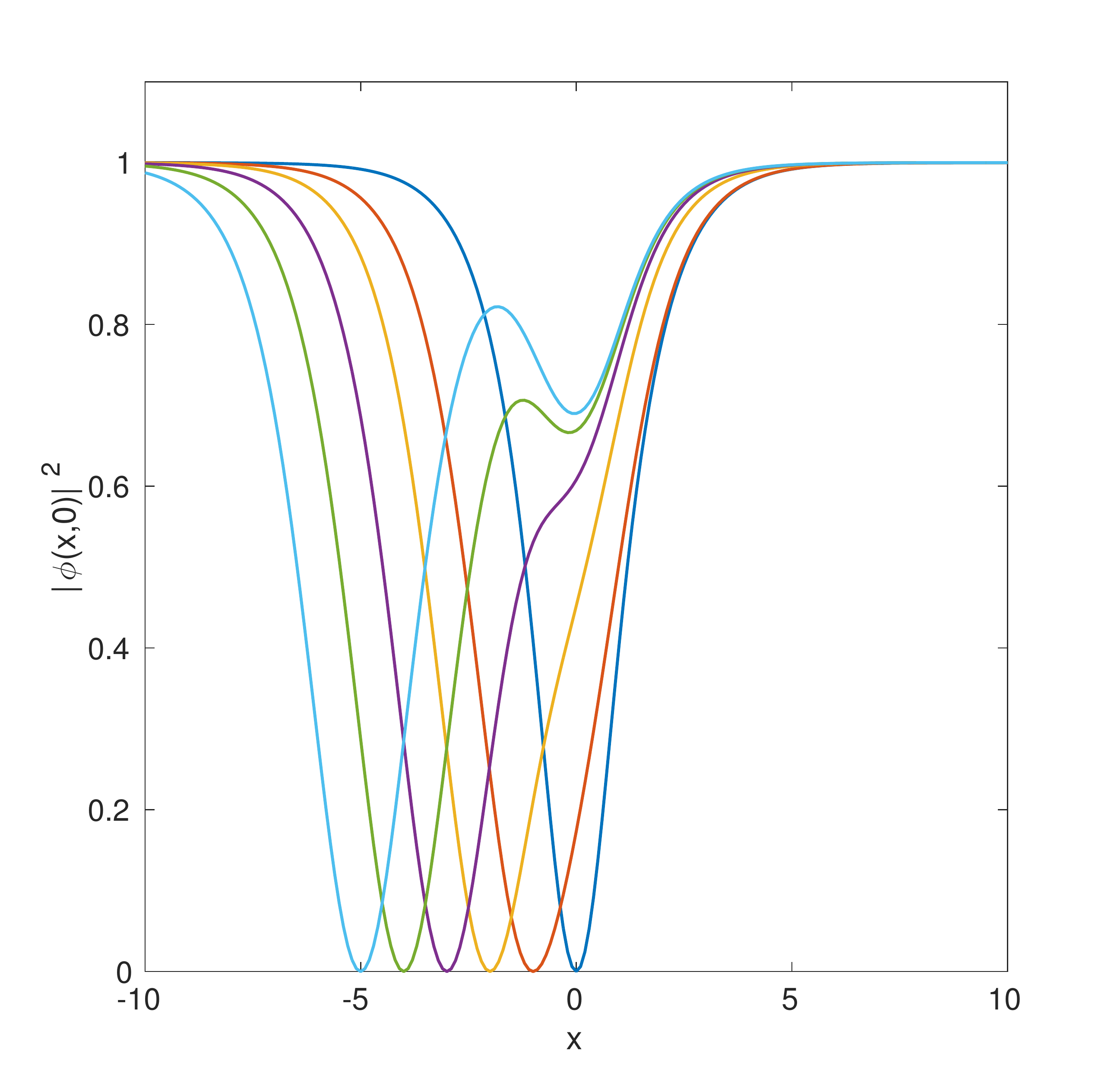}}
\subfloat[\, $\lambda=1.0$,~ $\sigma = e^{-r^2}$]{\includegraphics[totalheight=7.0cm]{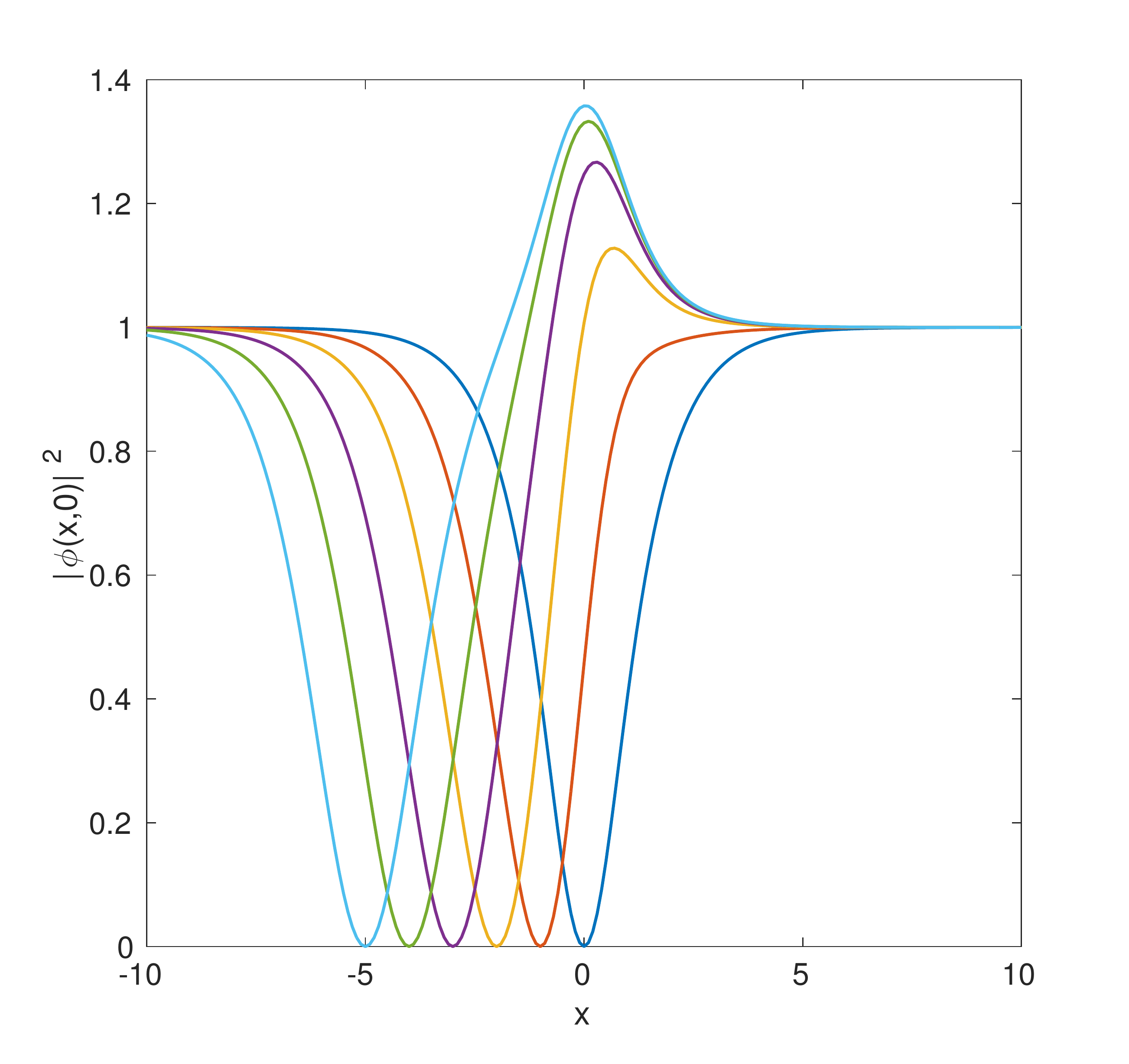}} 
\caption{Plot of $|\phi(x,0)|^2$ against $x$ for one vortex at critical coupling placed at different positions, and two different impurities $\sigma$.}
\label{fig:vimp_prof_N1_move}
\end{center}
\end{figure}

In Fig. \ref{fig:vimp_prof_N1_move}, we plot $|\phi(x,0)|^2$ against $x$ for critically coupled vortices positioned at a range of initial locations in the presence of the impurities (a) $\sigma(r) = -e^{-r^2},$ and (b) $\sigma(r) = e^{-r^2}$. We note that these solutions were also calculated in Ref.~\cite{Cockburn:2015huk} using the Bogomolny equation \eqref{vimp_bog_eq}. We have obtained the same solutions by solving the gradient flow equations \eqref{vimp_gflow}. For vortices positioned far away from the impurity, the solution looks like a superposition of a single vortex in the absence of an impurity with the vacuum solution in the presence of a magnetic impurity. When the vortex and impurity are closer together, the effect of the impurity on the vortex becomes more apparent. We see in both of the subfigures that a vortex positioned at the origin effectively ``screens" the impurity.


\begin{figure}[!htb]
\begin{center}
\subfloat[\, $\sigma = 4e^{-r^2}$]{\includegraphics[totalheight=4.9cm]{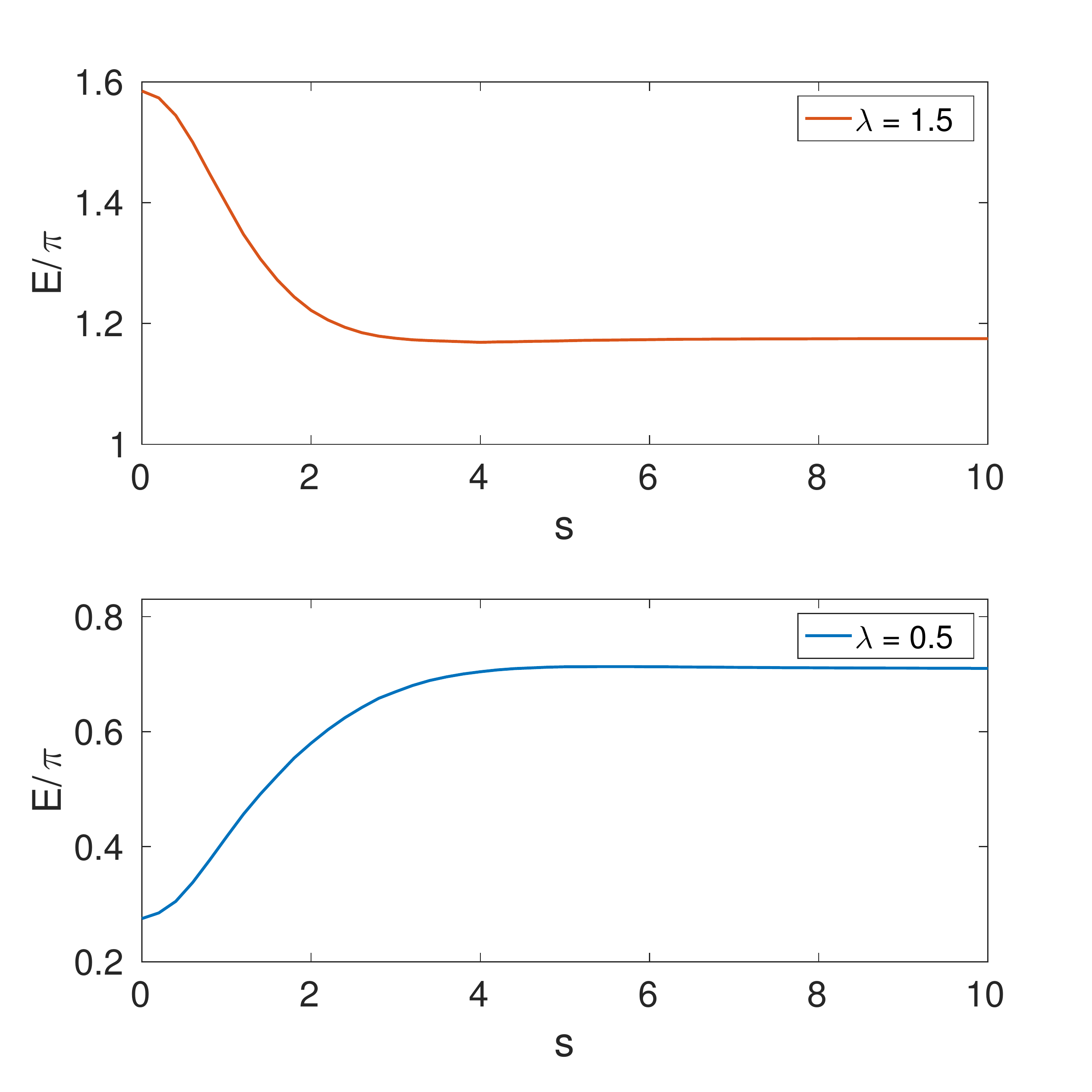}} 
\subfloat[\, $\sigma = -4e^{-r^2}$]{\includegraphics[totalheight=4.9cm]{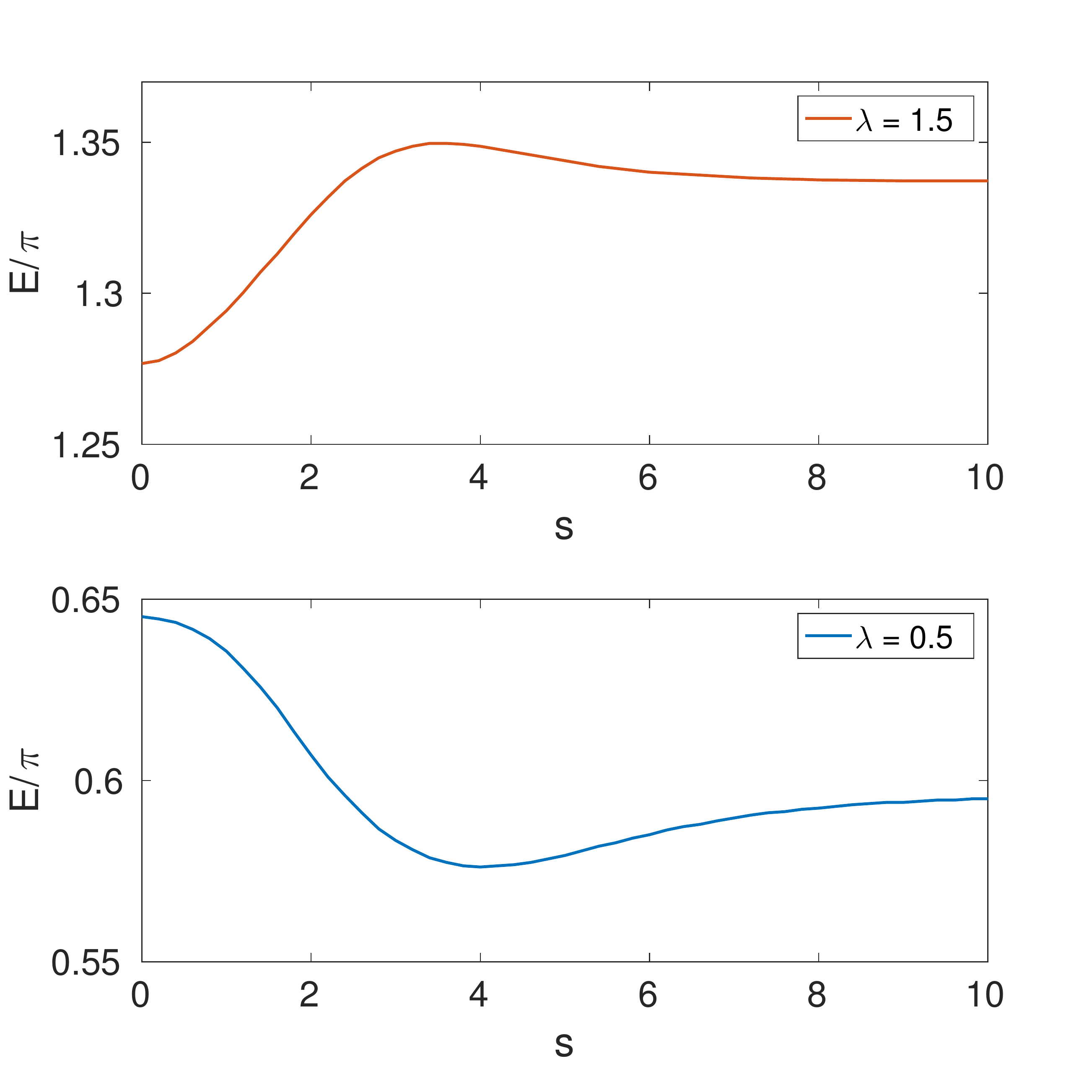}}
\subfloat[\, $\sigma = -16e^{-4r^2}$]{\includegraphics[totalheight=4.9cm]{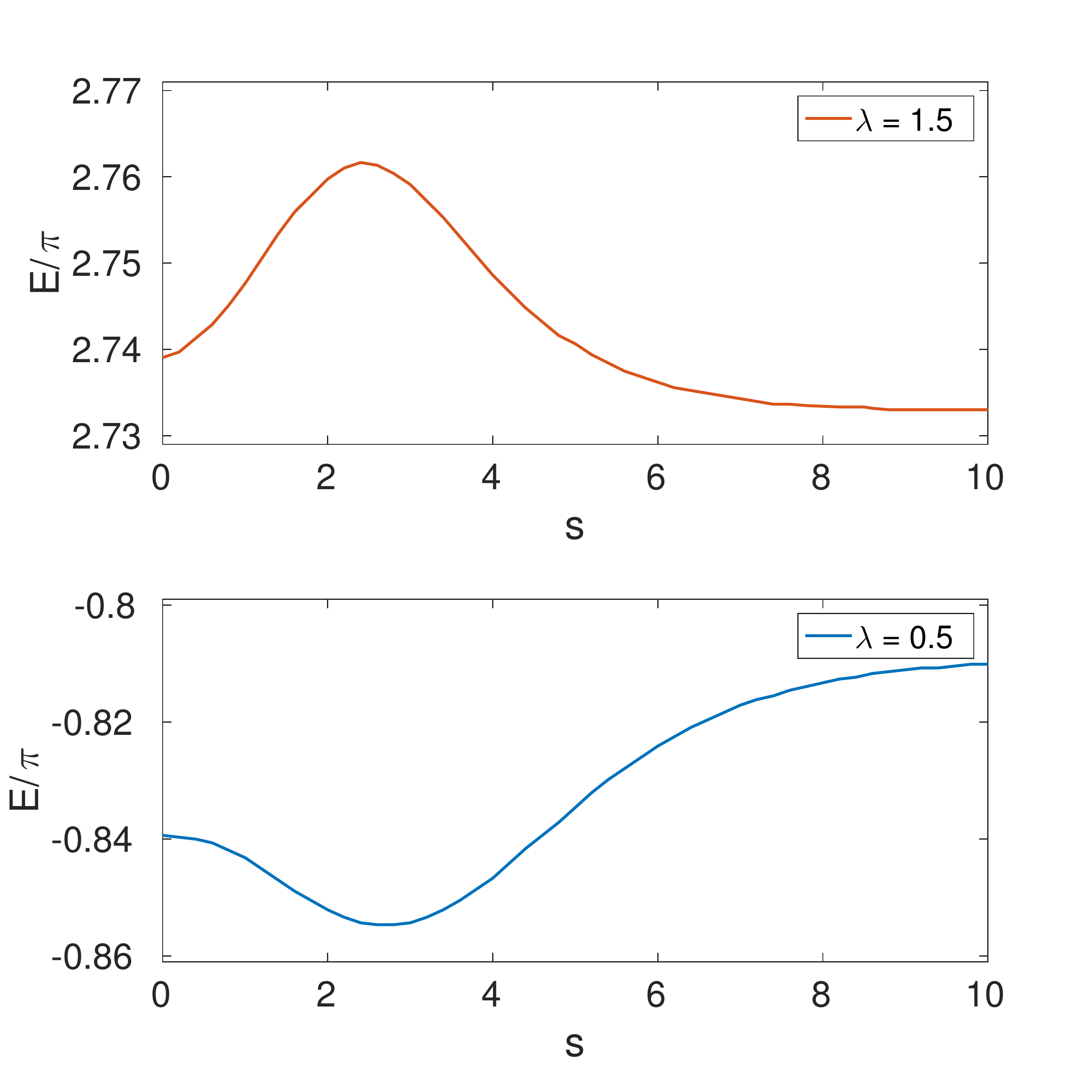}} \\
\subfloat[\, $\sigma = 4e^{-r^2}$]{\includegraphics[totalheight=4.9cm]{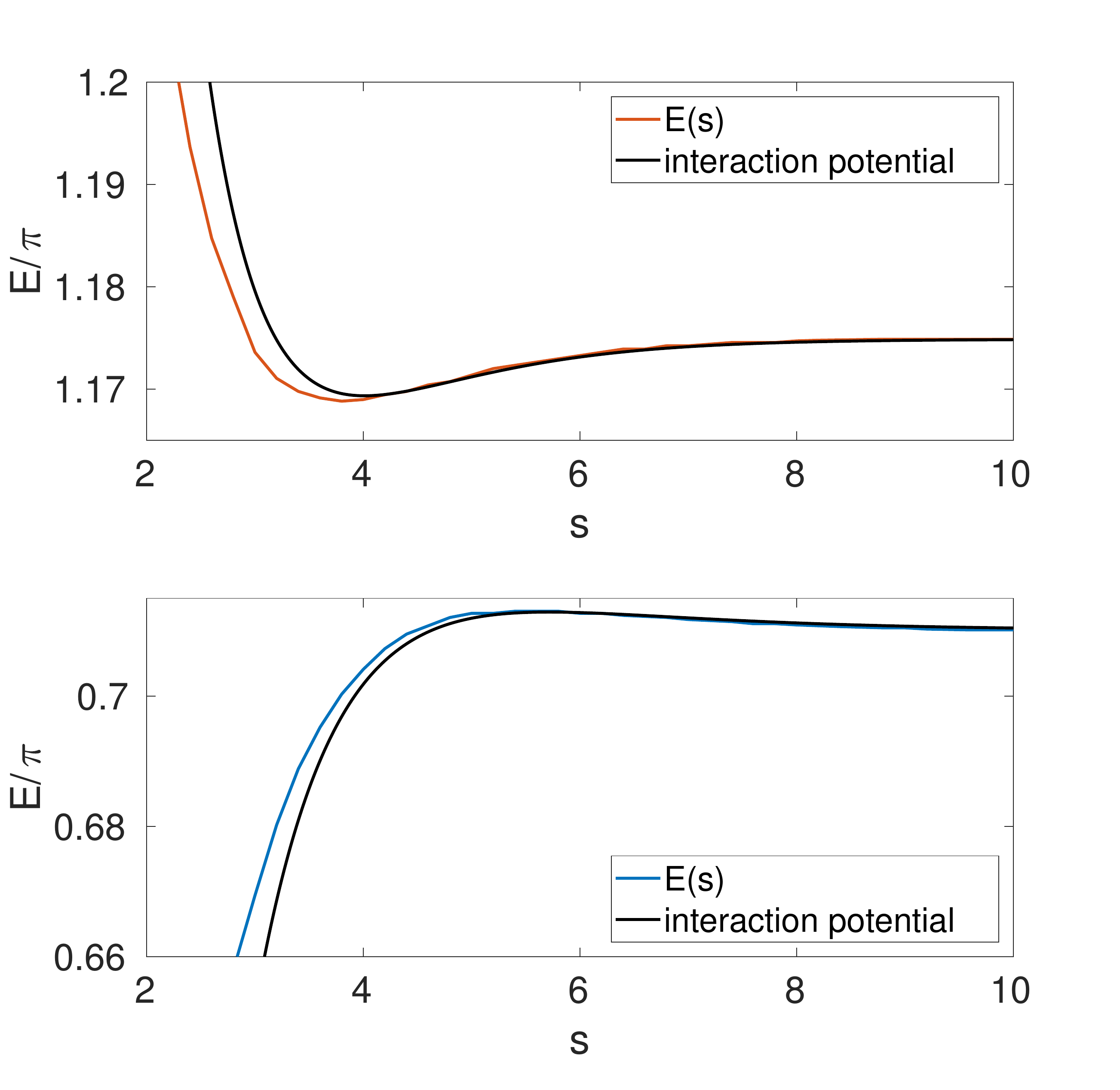}} 
\subfloat[\, $\sigma = -4e^{-r^2}$]{\includegraphics[totalheight=4.9cm]{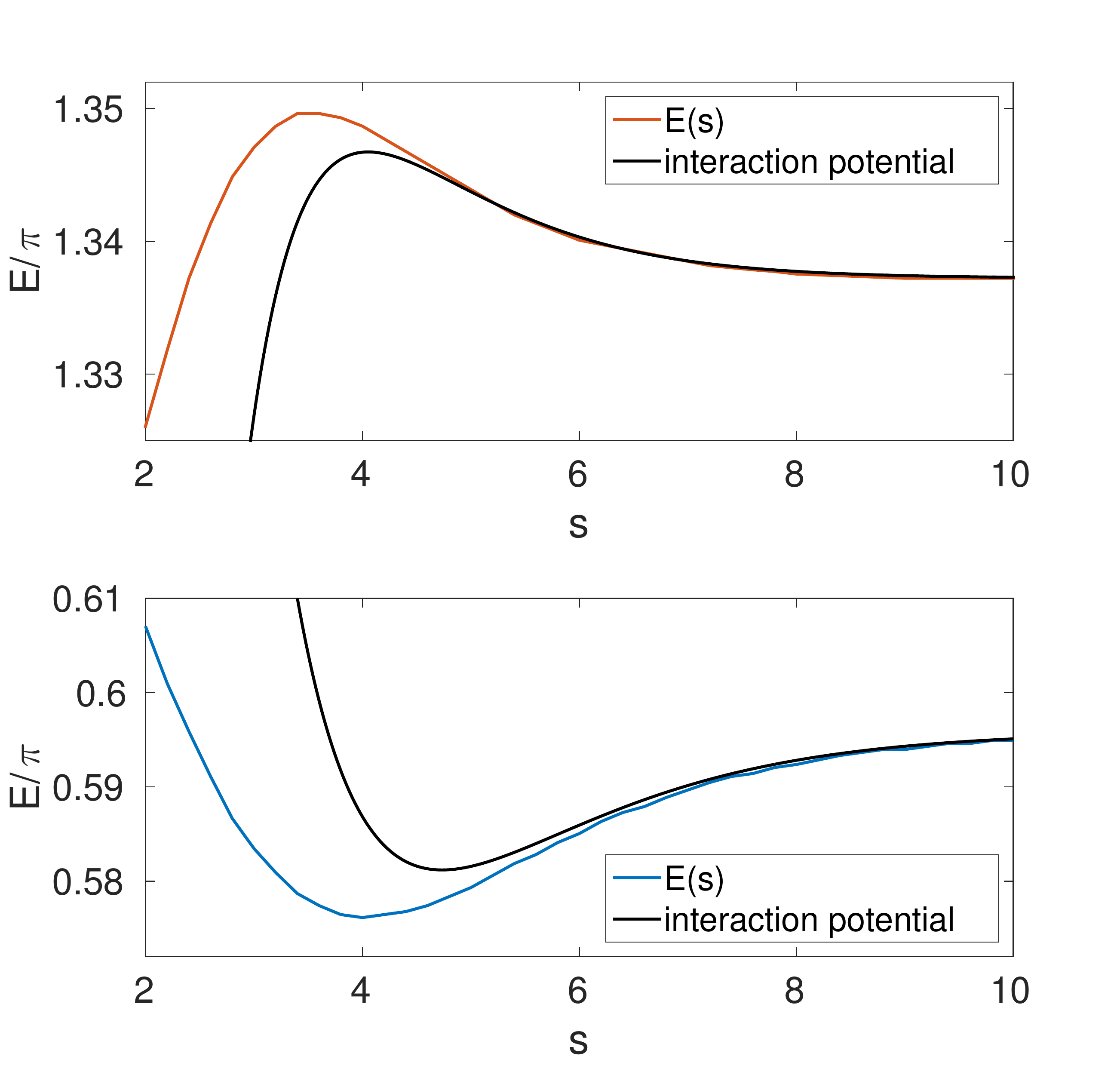}}
\subfloat[\, $\sigma = -16e^{-4r^2}$]{\includegraphics[totalheight=4.9cm]{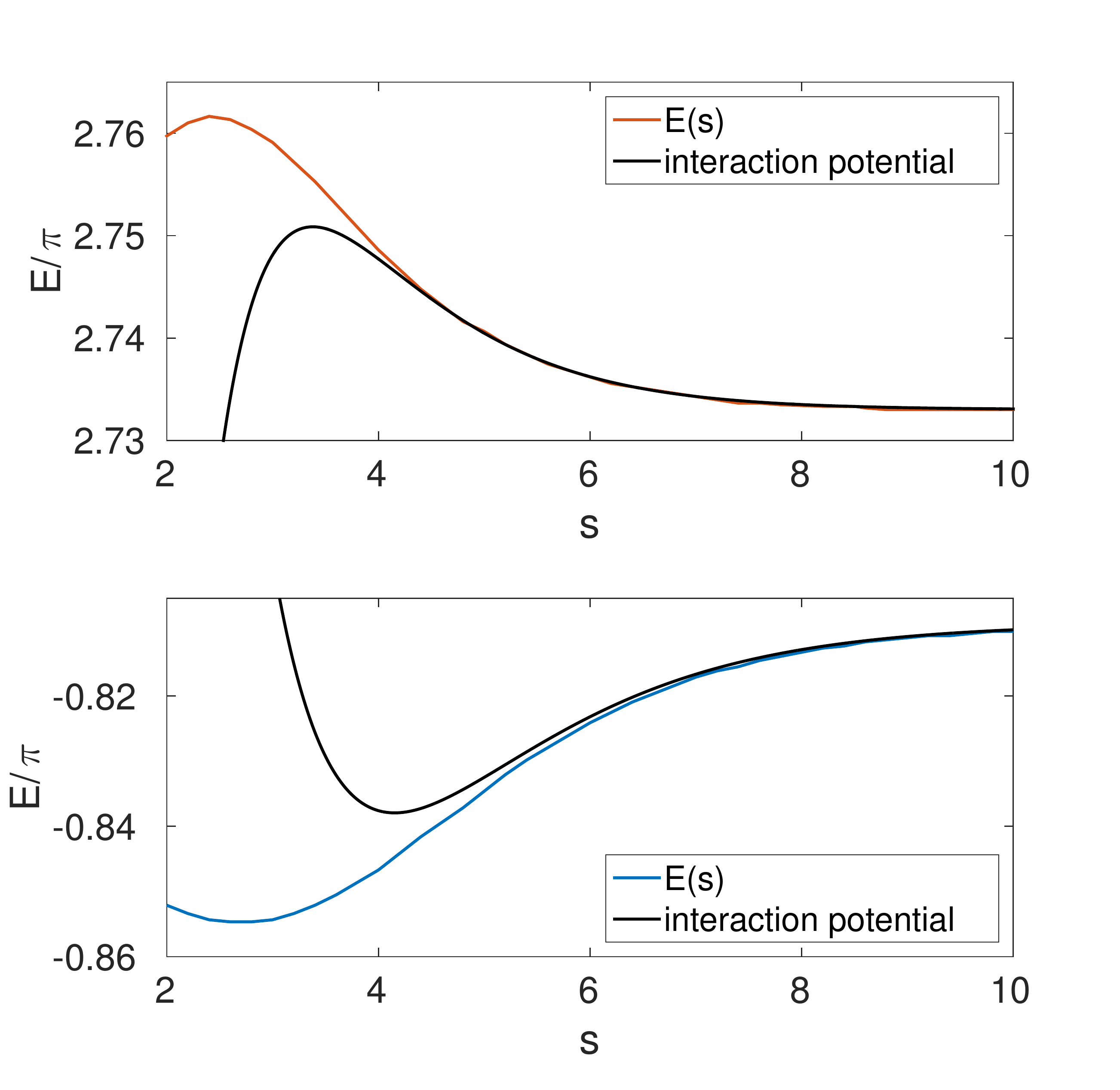}} 
\caption{(a)-(c) Energy $E/\pi$ as a function of the separation $s$ between a single vortex and an impurity $\sigma$ for $\lambda=0.5$ (in blue) and $\lambda=1.5$ (in red). (d)-(f) We zoom into the figures (a)-(c) for $s\geq2$ and compare them with the corresponding interaction potential \eqref{vimp_intervort_pot} shifted by the energy of a well-separated vortex and impurity (shown in black). }
\label{fig:vimp_en_vs_sep}
\end{center}
\end{figure}

To numerically confirm the existence of a moduli space of solutions at critical coupling we evaluate the energy of vortex and impurity solutions for different values of the separation~$s$ between the vortex and the impurity. We use gradient flow with various initial conditions, then calculate the separation and the corresponding minimal energy. This process works well because the gradient flow quickly settles into a valley keeping the separation approximately fixed. On a larger time-scale it then flows towards to a local minimum energy configuration by reducing or increasing the separation as appropriate.
Similar calculations for two vortices have been carried out in \cite{Jacobs:1978ch}. We calculate the energy $E/\pi$ for a single vortex and impurity at critical coupling to be $1$ to four decimal places regardless of the separation $s$. To illustrate the possible cases away from critical coupling, in Fig.~\ref{fig:vimp_en_vs_sep} we display the energy $E/\pi$ of a single vortex in the presence of an impurity $\sigma$ as a function of $s$ for $\lambda=0.5,~1.5$ and three different impurities. This reveals a more complicated relationship between the impurity and vortex than that predicted by considering the sign of $E_{1,\sigma}-(E_1+E_{0,\sigma})$ and explains why the asymptotics disagreed with these results. 

Figs.~\ref{fig:vimp_en_vs_sep}(a) and (d) correspond to the impurity $\sigma(r)=4e^{-r^2}$. Here the energy calculation $E_{1,\sigma}-(E_1+E_{0,\sigma})$ predicted that the impurity would repel the vortex for $\lambda>1$ and attract it for $\lambda<1$, though the asymptotics suggested the opposite effect. Initially the energy decreases with increasing $s$ for $\lambda=1.5$ and increases with increasing $s$ for $\lambda=0.5$. However a close examination of the energy at larger values of $s$ (see Fig.~\ref{fig:vimp_en_vs_sep}(d)) reveals that this behaviour reverses for $s>4$: the energy begins to slightly increase with $s$ for $\lambda=1.5$ and to very slightly decrease with $s$ for $\lambda=0.5$, before settling on a constant value $E|_{s\rightarrow\infty}$. This is the behaviour predicted by the asymptotics, and in Fig.~\ref{fig:vimp_en_vs_sep}(d) we also plot the corresponding interaction potential \eqref{vimp_intervort_pot} shifted by the value $E|_{s\rightarrow\infty}$ as a black line. The asymptotic prediction agrees very well with $E(s)/\pi$ for $s>5$. The energy calculation disagreed with the asymptotic prediction because it only compared the energy of a well-separated vortex and impurity $E|_{s\rightarrow\infty}$ to that of a coincident vortex and impurity $E(0)$, and so it missed the subtle change in behaviour at larger values of $s$. The true behaviour is more complicated than that indicated by either prediction: when vortex and impurity are close, the impurity will attract the vortex for $\lambda<1$ and repel it for $\lambda>1$, but when they are further apart the impurity will repel the vortex for $\lambda<1$ and attract it for $\lambda>1$

In Figs.~\ref{fig:vimp_en_vs_sep}(b) and (e) we show similar plots for the impurity $\sigma(r)=-4e^{-r^2}$.  In this case the energy calculation predicted that the impurity would attract the vortex for $\lambda>1$ and repel it for $\lambda<1$, and the asymptotic prediction disagreed. As before, we see that the energy prediction is correct for smaller values of $s$, but that the behaviour changes when $s>4$, at which point the shifted interaction potential becomes a good fit for $E(s)/\pi$. 

Figs.~\ref{fig:vimp_en_vs_sep}(c) and (f) correspond to the impurity $\sigma(r)=-16e^{-4r^2}$, which was one of the impurities for which the predictions of both the energy calculation and the asymptotics agreed. In this case they both predicted that the impurity should repel the vortex when $\lambda>1$ and attract it when $\lambda<1$. However, as shown in the figures, the opposite is true for small $s$, with the behaviour changing for $s>2.5$. Let $E_{\text{stat}}$ denote the energy at the stationary point of $E(s)$. Unlike the previous cases, the change in energy of the behaviour at small $s$ is less significant than that of the behaviour at large $s$, i.e. $|E_{\text{stat}}-E(0)|<|E_{\text{stat}}-E|_{s\rightarrow\infty}|$. So it is the large $s$ behaviour which determines whether $E|_{s\rightarrow\infty}$ is greater than or less than $E(0)$. Since the large $s$ behaviour has the most significant change in energy, the energy calculation agrees with the asymptotics. In the other cases, the small $s$ behaviour had the most significant change in energy and so the predictions disagreed.  

Recall that we earlier discussed a critical line separating the two different regions of behaviour for impurities with $c<0$. For impurities on this line, the difference in energy between a coincident vortex and impurity and a well-separated vortex and impurity was zero. However this does not mean the vortex and impurity can be placed anywhere without affecting the energy. There is still a region where, using the example of $\lambda>1$, for small $s$ the energy increases with increasing $s$, and for larger $s$ the energy decreases with increasing $s$ until settling on a constant value $E|_{s\rightarrow\infty}$. The energy calculation gives zero in this case because $E(0)=E|_{s\rightarrow\infty}$, but $E(s)$ is not a constant function.


\section{Vortex dynamics with magnetic impurities}

To study vortex dynamics in the presence of magnetic impurities, we return to the Lagrangian \eqref{vimp_lag}. The corresponding equations of motion are
\begin{align}\label{vimp_eqm}
 D_\mu D^\mu\phi - \frac{\lambda}{2}\left(1 + \sigma - |\phi|^2\right)\phi = 0, \nonumber \\
\pa_\mu f^{\mu 0} + \frac{i}{2}\left(\overline{\phi}D^0\phi - \phi \overline{D^0\phi}\right) = 0, \nonumber \\
\pa_\mu f^{\mu 1} +\frac{1}{2}\pa_2\sigma + \frac{i}{2}\left(\overline{\phi}D^1\phi - \phi \overline{D^1\phi}\right) = 0,\nonumber \\
\pa_\mu f^{\mu 2} -\frac{1}{2}\pa_1\sigma  + \frac{i}{2}\left(\overline{\phi}D^2\phi - \phi \overline{D^2\phi}\right) = 0.
\end{align}
We solve these equations using a leapfrog method, with derivatives that are second order accurate in time and fourth order in space. The timestep used is $\Delta t = 0.01$, with grid spacing $\Delta x = \Delta y = 0.1$ over grids that are typically of size $401\times401$ or $801\times801$. We work in the temporal gauge $a_0=0$ and generate initial conditions by boosting an ordinary vortex solution at a given initial position with velocity $v$ and using the Abrikosov ansatz \eqref{Abrikosov} to combine this with a static vacuum solution for the chosen magnetic impurity. 

\begin{figure}[!htb]
\begin{center}
\begin{tikzpicture}

\draw [dashed] (-6,0) -- (6,0);
\filldraw [black] (0,0) circle (6pt);
\node[label={ Impurity}] () at (0,0) {};

\draw [dashed] (-6,-3) -- (6,-3);
\filldraw [black] (-5,-3) circle (6pt);
\node[label={Vortex}] () at (-5,-3) {};

\draw [<->, thick] (-3,-2.8) -- (-3, -0.2);
\draw (-3,-1.5)  node [right] {\large $b$};

\draw [thick] plot [smooth] coordinates {(-6, -3) (-0.5,-3) (1,-2.8) (2,-2) (5, 2)};
\draw [ultra thick, ->] (-5,-3) -- (-4,-3);

\draw (1.3,-3) -- (5,2);
\draw (1.3,-3) -- (3,-3);
\filldraw [black] (1.3,-3) circle (1pt); 
\draw  (1.6,-2.595) to  [bend left]  (1.8,-3) ;
\node () at (2,-2.7) {$\Theta$}; 

\end{tikzpicture}
\caption{Diagram to illustrate the definition of the scattering angle $\Theta$ and impact parameter $b$ in our simulations.}
\label{fig:vimp_angle_diag}
\end{center}
\end{figure}
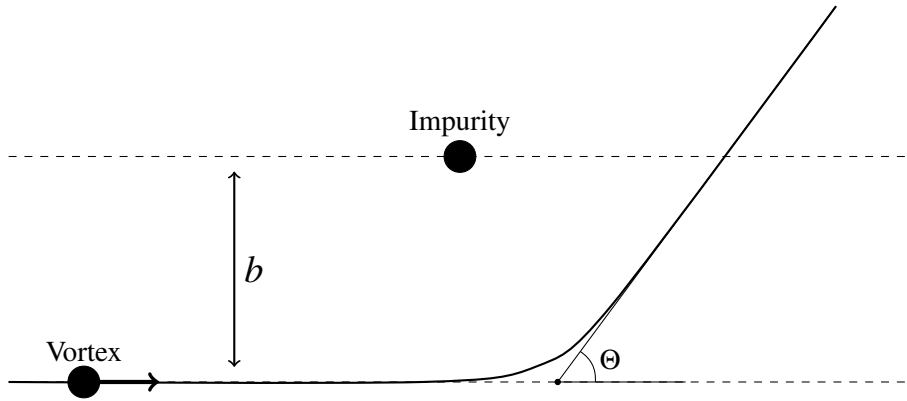

Two quantities of interest in the study of vortex scattering are the scattering angle~$\Theta$ and impact parameter~$b$. Fig.~\ref{fig:vimp_angle_diag} indicates how each of these quantities are defined within our vortex and impurity scattering simulations. The scattering angle $\Theta$ is the the angle between the trajectory of the vortex after scattering and its initial trajectory. The impact parameter $b$ is the vertical distance between the initial trajectory of the vortex and the impurity.

\subsection{Vortex scattering at critical coupling}

We first consider the scattering of a single vortex at critical coupling with an impurity of the form $\sigma(r)=ce^{-r^2}$. It has already been noted in Ref.~\cite{Cockburn:2015huk} that an impurity of this form with $c>0$ will attract a slow moving vortex, and with $c<0$ will repel the vortex. In the figures presented below, we take two different impurities $\sigma(r)=e^{-r^2}$ and $\sigma(r) = -e^{-r^2}$, but we have verified that the general behaviour is the same for other impurities also. In our simulations, the initial vortex position is $(x,y)~=~(-4,-b)$ for a range of $b\in[0,8]$, and the impurity is fixed at the origin.  

\begin{figure}[!htb]
\begin{center}
\subfloat[\, ]{\includegraphics[totalheight=7.4cm]{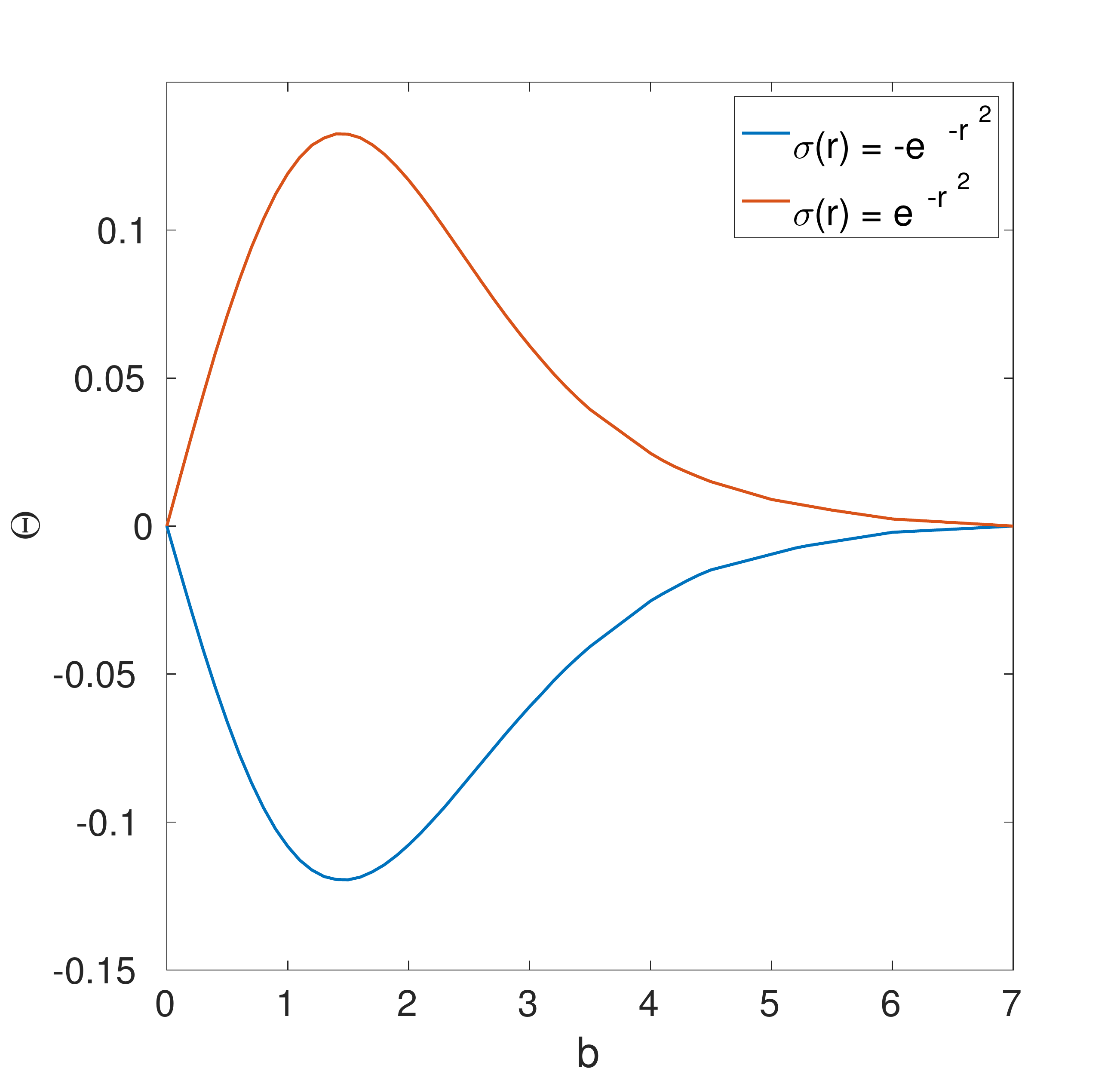}}
\subfloat[\, ]{\includegraphics[totalheight=7.4cm]{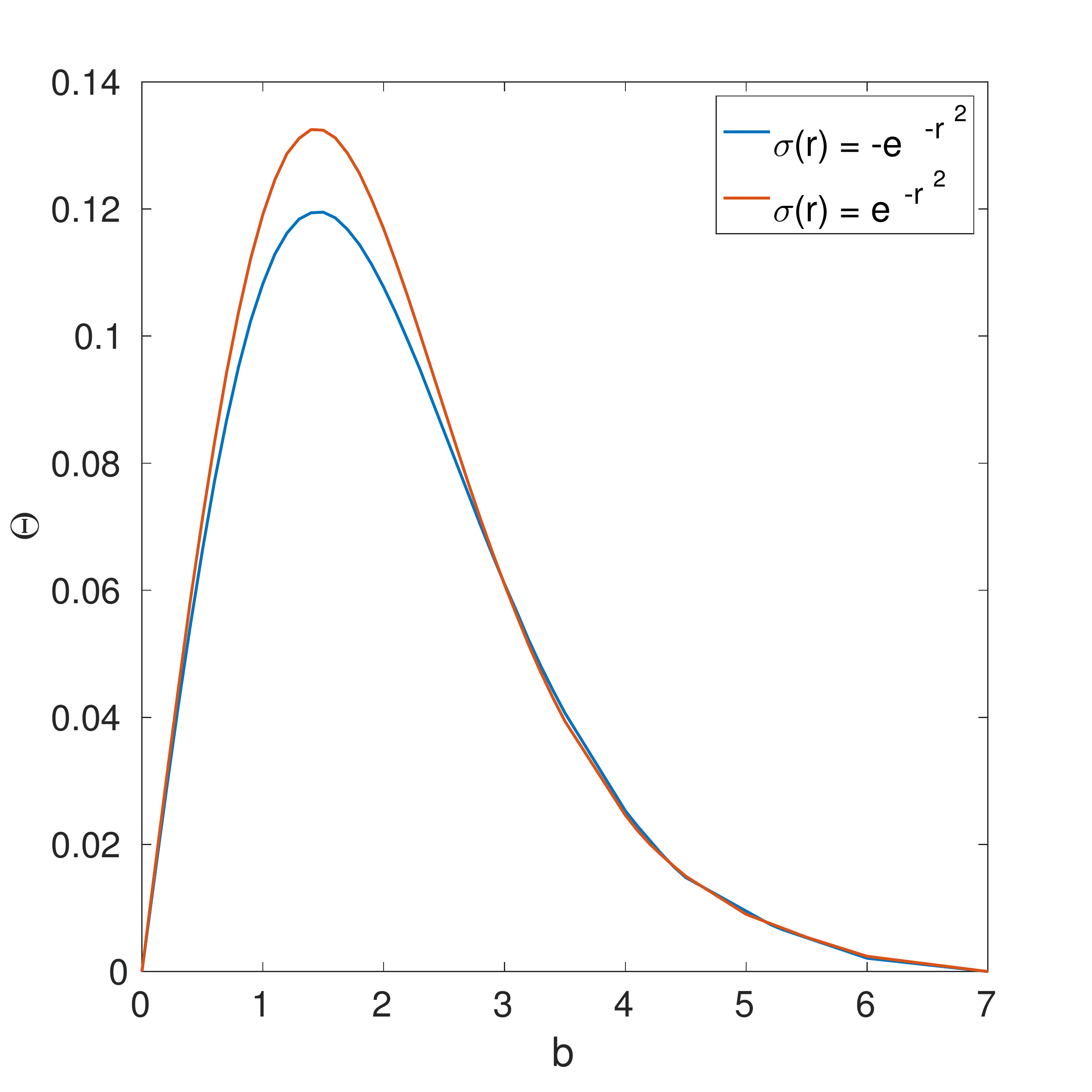}} 
\caption{(a) Scattering angle $\Theta$ in radians against impact parameter $b$ for the scattering of a single vortex at critical coupling with impurities $\sigma(r)=e^{-r^2}$ (in red), and $\sigma(r)=-e^{-r^2}$ (in blue). The initial vortex speed is $v=0.3$. (b) We display the same data, but for $\sigma(r)=-e^{-r^2}$ we plot $-\Theta$ for easier comparison with the scattering angle for $\sigma(r)=e^{-r^2}$.}
\label{fig:vimp_angle}
\end{center}
\end{figure}

In Fig.~\ref{fig:vimp_angle}, we plot the scattering angle~$\Theta$ in radians as a function of the impact parameter~$b$ for a single vortex at critical coupling scattering with the magnetic impurities $\sigma(r)=e^{-r^2}$ and $\sigma(r) = -e^{-r^2}$.
Note that for $\lambda = 1$ the vortex motion can be approximated by geodesic motion on the moduli space, and the trajectories are essentially independent on the initial velocity. This approximation breaks down for higher velocities and in particular when relativistic effects become relevant. 
In the following figures the initial velocity given to the vortex is $v=0.3$. The sign of the scattering angle for $\sigma=-e^{-r^2}$ is reversed in (b) for easier comparison with the other impurity. For both impurities, when impact parameter $b=0$, the scattering angle $\Theta=0$: the vortex passes through the impurity in a head-on collision. As $b$ increases, $\Theta$ also increases up to a maximum value occurring at $b\approx 1.5$. Then the scattering angle steadily decreases, returning to zero when the impurity is so far from the vortex as to no longer influence its trajectory.  We note that the general shape of this plot is consistent with results obtained in Refs.~\cite{Cockburn:2015huk, Krusch} for vortices and impurities in hyperbolic space.

\begin{figure}[!htb]
\begin{center}
\subfloat[\, $\sigma(r) = e^{-r^2}$]{\includegraphics[totalheight=7.0cm]{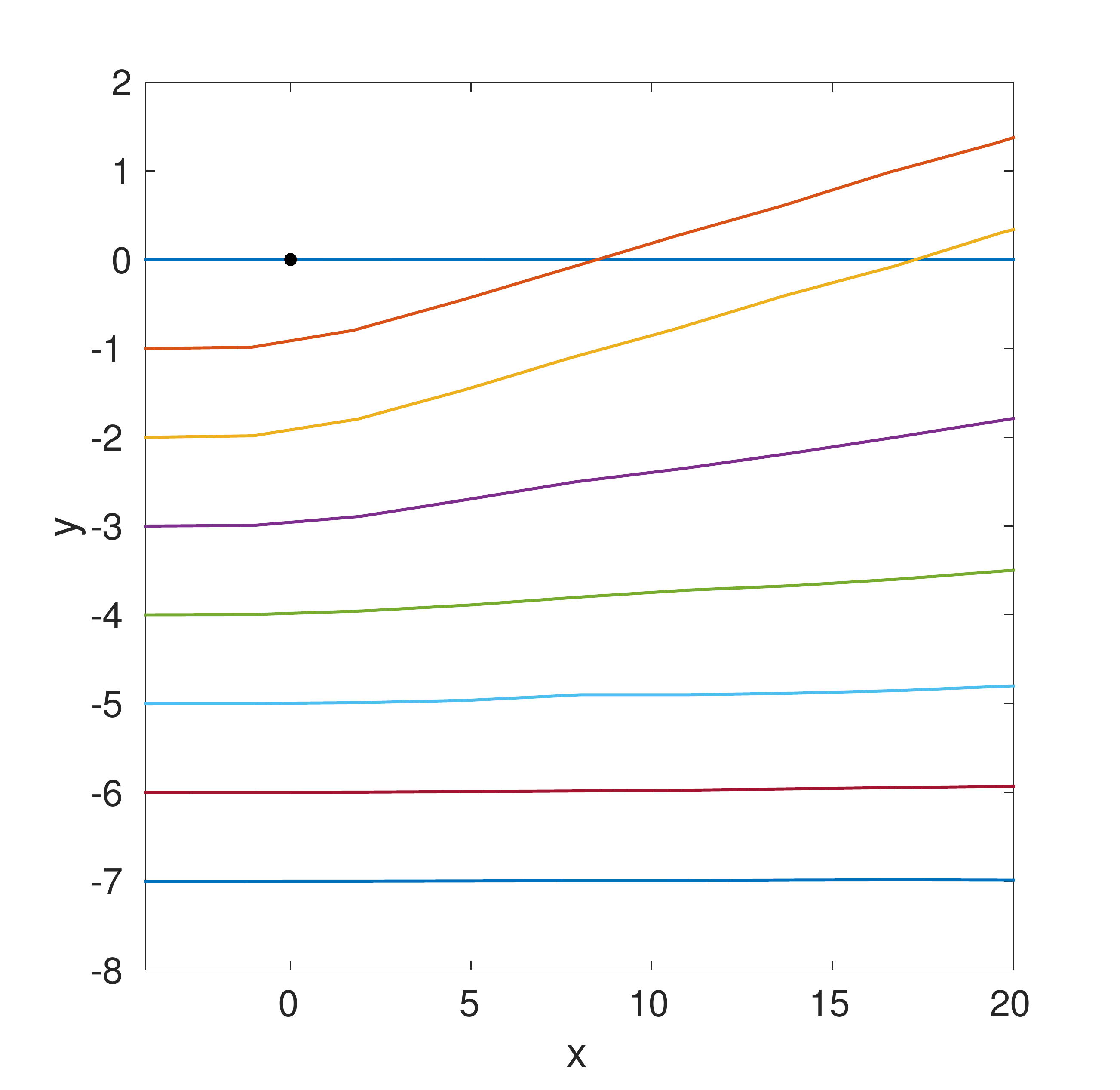}}
\subfloat[\, $\sigma(r) = -e^{-r^2}$]{\includegraphics[totalheight=7.0cm]{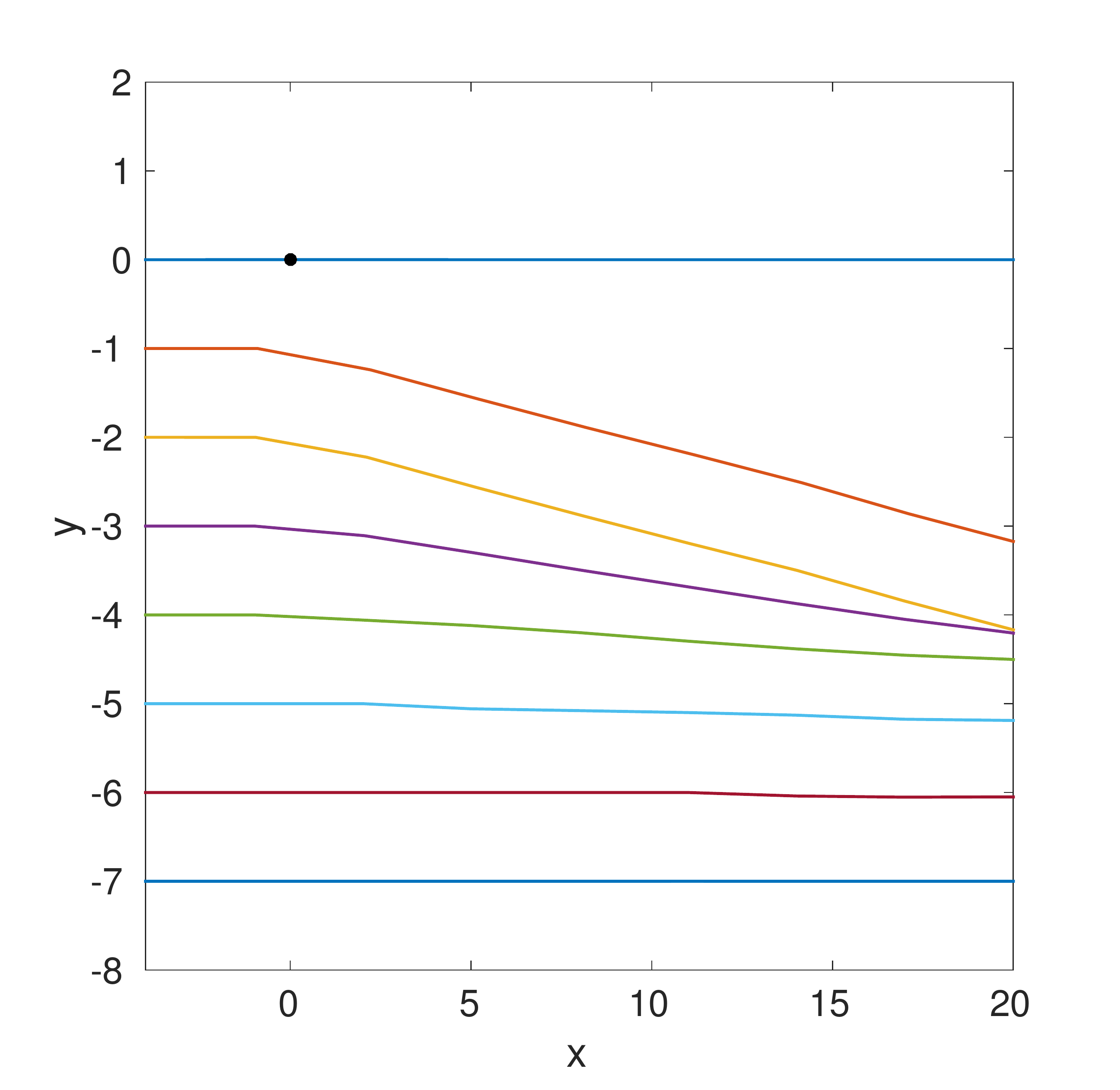}} 
\caption{Vortex trajectories during the scattering of a single vortex at critical coupling with impurities (a)~$\sigma(r) = e^{-r^2}$ and (b)~$\sigma(r) = -e^{-r^2}$ for the impact parameter values $b=0,1,2,3,4,5,6,7$ and initial velocity $v=0.3$. The location of the impurity is indicated with a black dot. 
}
\label{fig:vimp_N1pos}
\end{center}
\end{figure}

The scattering angle plots can be compared with Fig.~\ref{fig:vimp_N1pos} which shows the corresponding vortex trajectories for a selection of impact parameter values. The location of the impurity is indicated with a black dot. We see that  in a head-on collision with either impurity, the vortex will pass through the impurity and continue on its original trajectory. Similarly in both cases, for impact parameter $b=7$, the vortex is far enough from the impurity that its trajectory is unaffected, and it continues to travel along the $x$-axis as though the impurity was not there. In between these values, the vortex trajectories are altered by the presence of the impurity, bending towards the impurity for $c>0$ and away from it for $c<0$. We saw in Fig.~\ref{fig:vimp_angle} that the maximum scattering angle is found when $b\approx1.5$, after which the angle decreases with increasing impact parameter. In Fig.~\ref{fig:vimp_N1pos}, we see that the most altered trajectories are those for $b=1,2$, and after this the trajectories start to flatten out.

\begin{figure}[!htb]
\begin{center}
\subfloat[\, ]{\includegraphics[width=4.5cm,totalheight=4.8cm]{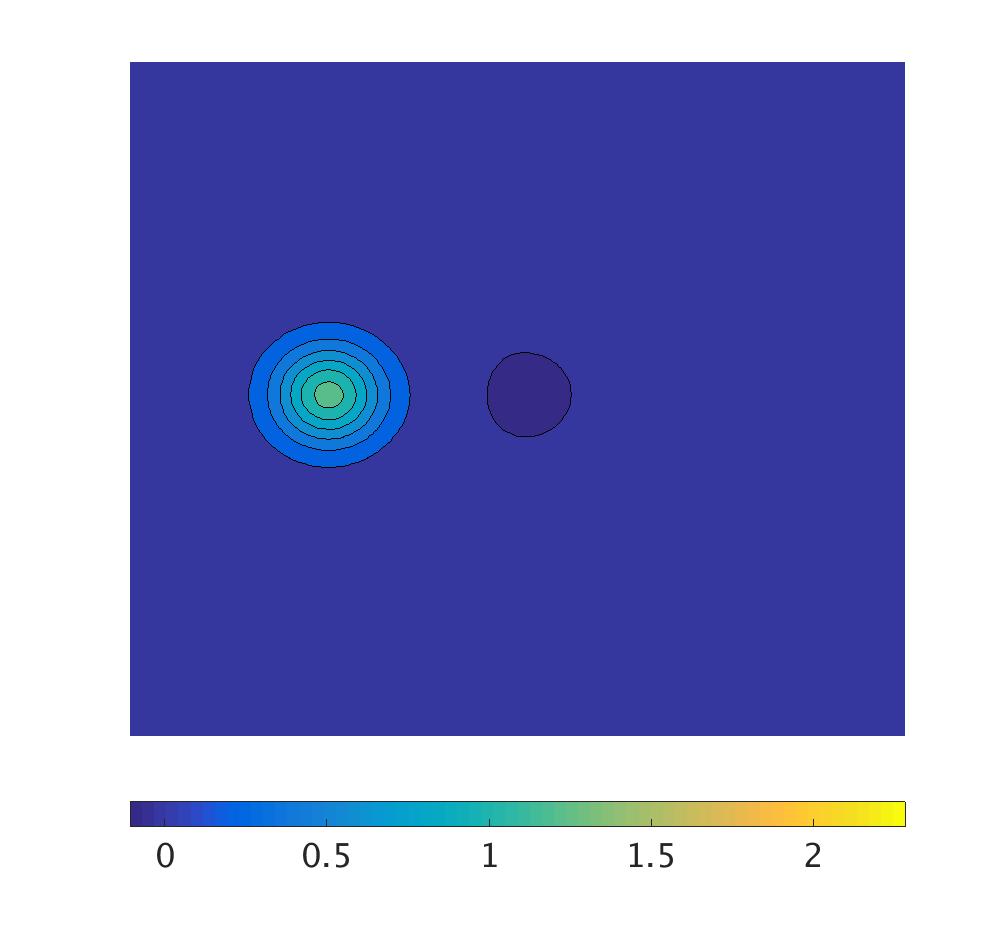}}
\subfloat[\, ]{\includegraphics[width=4.5cm,totalheight=4.8cm]{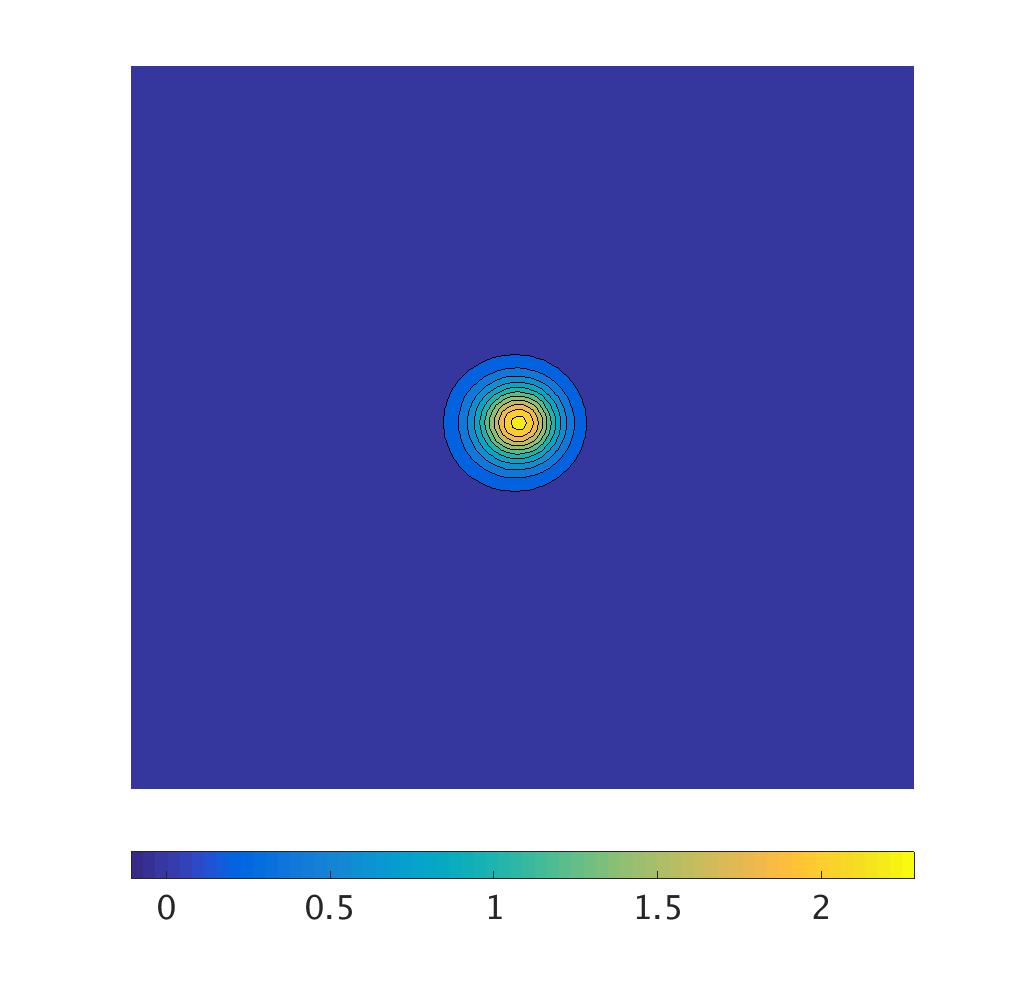}} 
\subfloat[\, ]{\includegraphics[width=4.5cm,totalheight=4.8cm]{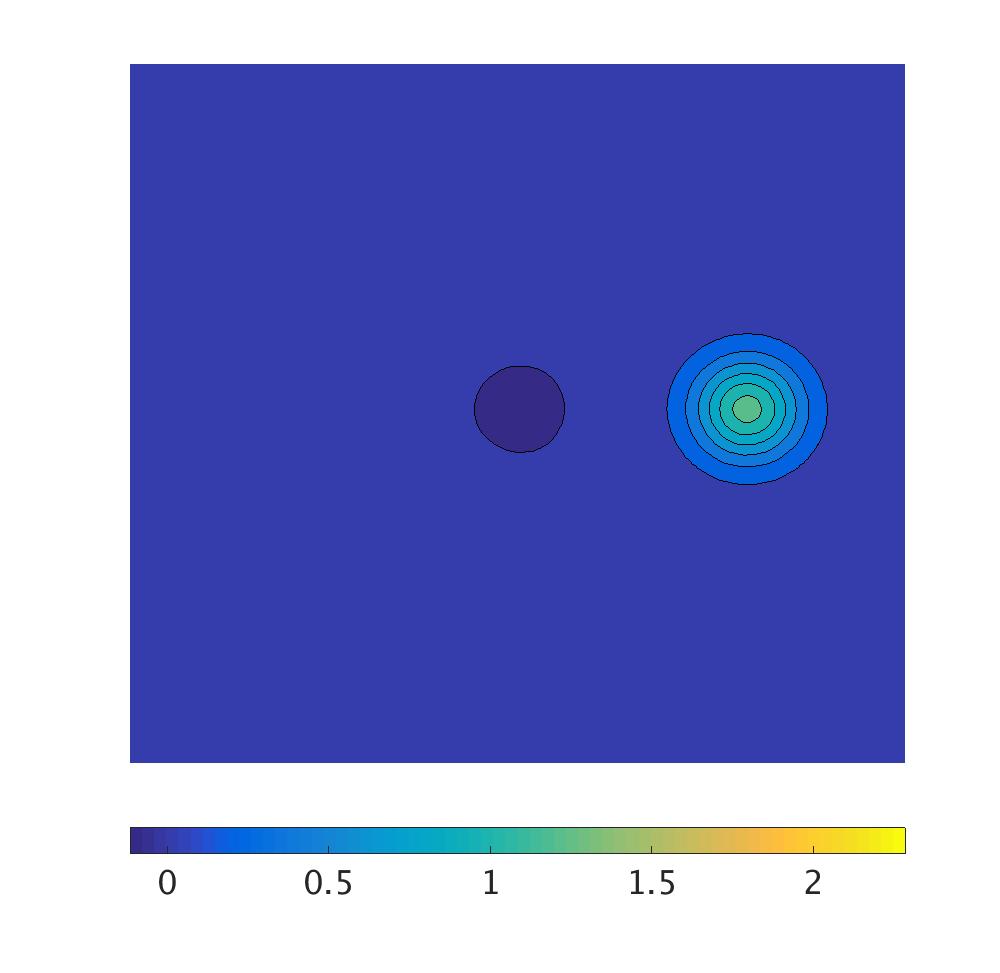}} \\
\subfloat[\, ]{\includegraphics[width=4.5cm,totalheight=4.8cm]{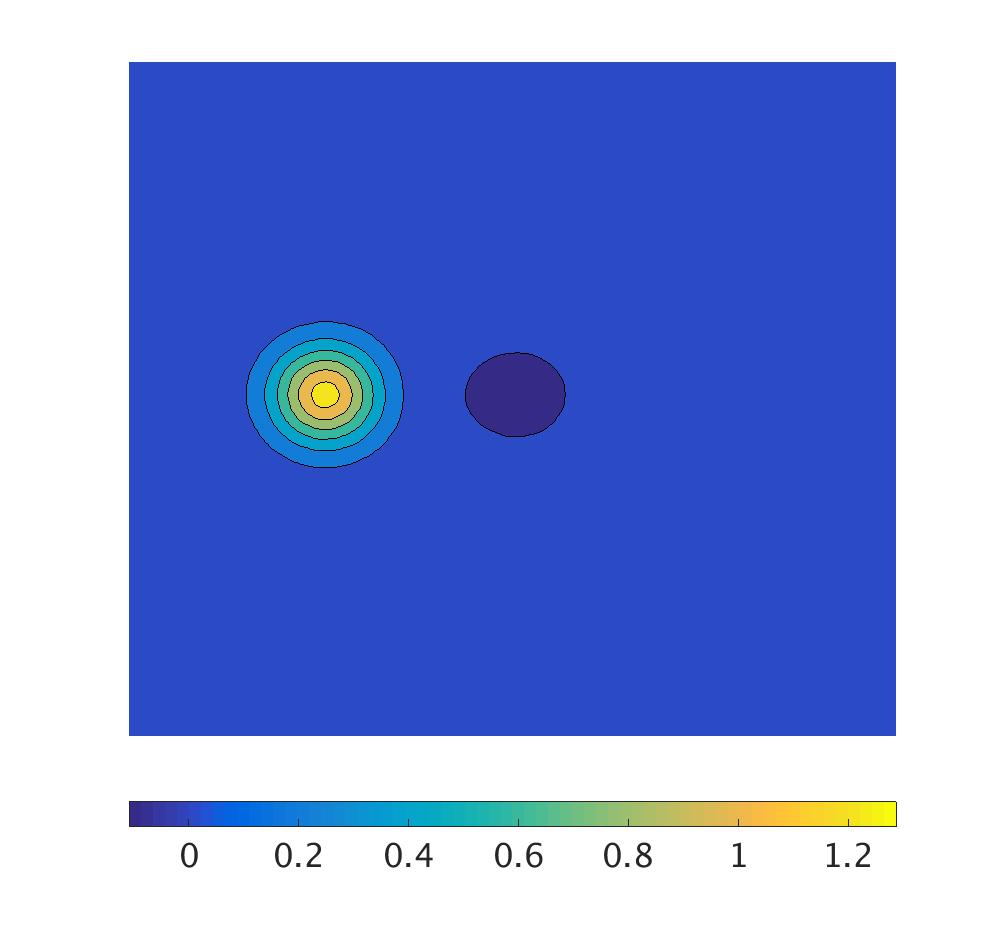}}
\subfloat[\, ]{\includegraphics[width=4.5cm,totalheight=4.8cm]{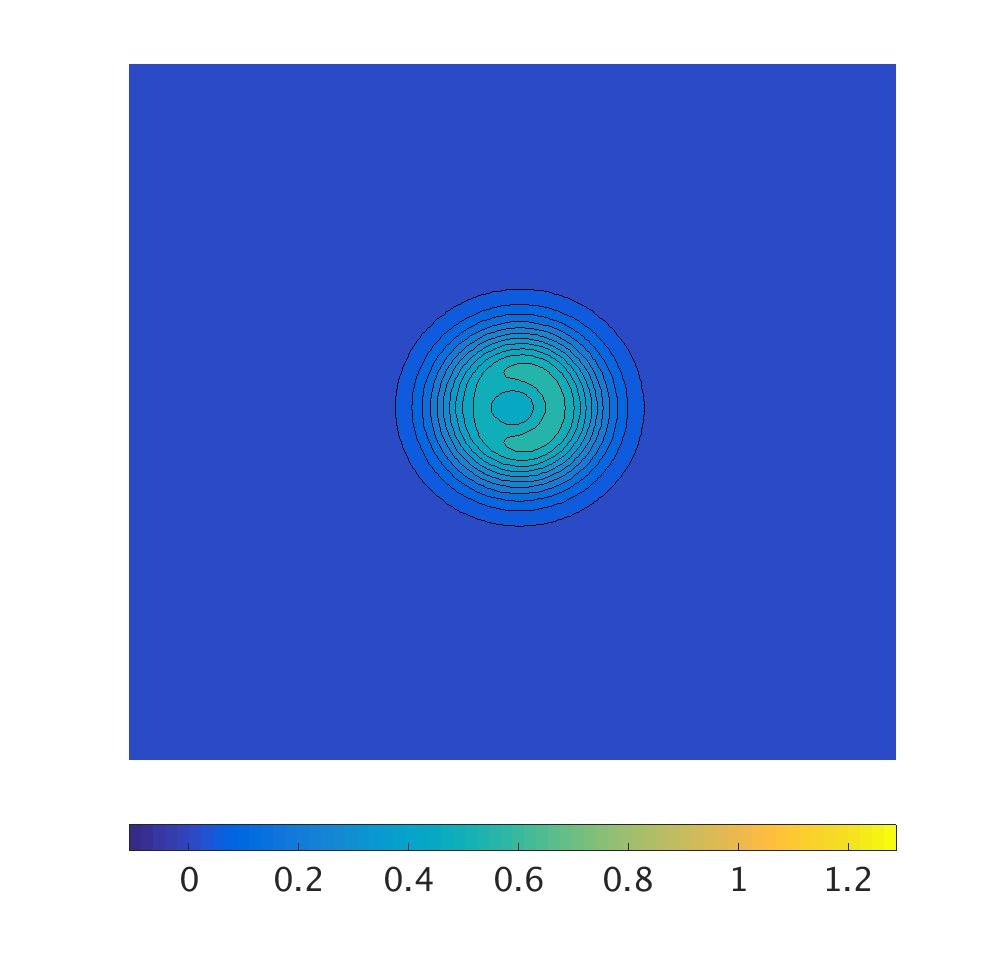}} 
\subfloat[\, ]{\includegraphics[width=4.5cm,totalheight=4.8cm]{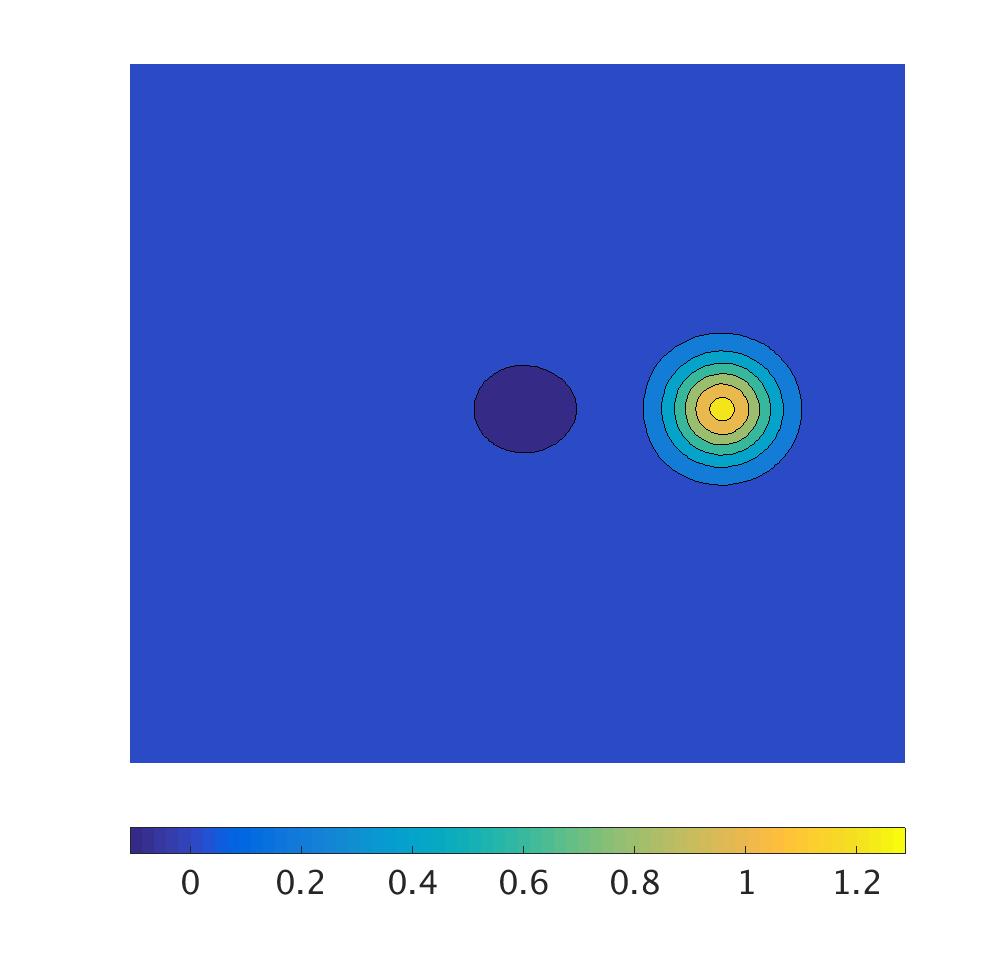}} 
\caption{Snapshots of the energy density during the scattering of a single vortex at critical coupling with impurities (a)-(c):~$\sigma(r) = e^{-r^2}$ and (d)-(f):~$\sigma(r) = -e^{-r^2}$ for initial velocity $v=0.3$.
}
\label{fig:vimp_N1snapshot}
\end{center}
\end{figure}

Although we have seen that the vortex will travel through both impurities in a head-on collision without altering its initial trajectory, the details are different depending on the sign of $c$. Fig.~\ref{fig:vimp_N1snapshot} displays snapshots of the energy density of the configurations at different times during the head-on scattering of a single vortex at critical coupling with the impurities (a)-(c):~$\sigma(r) = e^{-r^2}$, and (d)-(f):~$\sigma(r) = -e^{-r^2}$. In both cases, we begin with the vortex and impurity well separated. The effect of the impurity on the energy density can be seen in the contour plot as a region of negative energy density, and the vortex is a localised lump of energy. When the vortex crosses the impurity with $c>0$, as seen in Fig.~\ref{fig:vimp_N1snapshot}(b), there is a localised lump of energy at the origin which is taller than the energy of the vortex alone. By contrast, for $c<0$, as seen in Fig.~\ref{fig:vimp_N1snapshot}(e), the energy forms a less localised ring, resembling an $N=2$ vortex, at the origin, which is smaller than the energy of the single vortex was originally. In both cases, after scattering the vortex and impurity appear unchanged by their interaction, and the vortex continues on its original path.

\begin{figure}[!htb]
\begin{center}
\subfloat[\, ]{\includegraphics[width=4.5cm,totalheight=4.8cm]{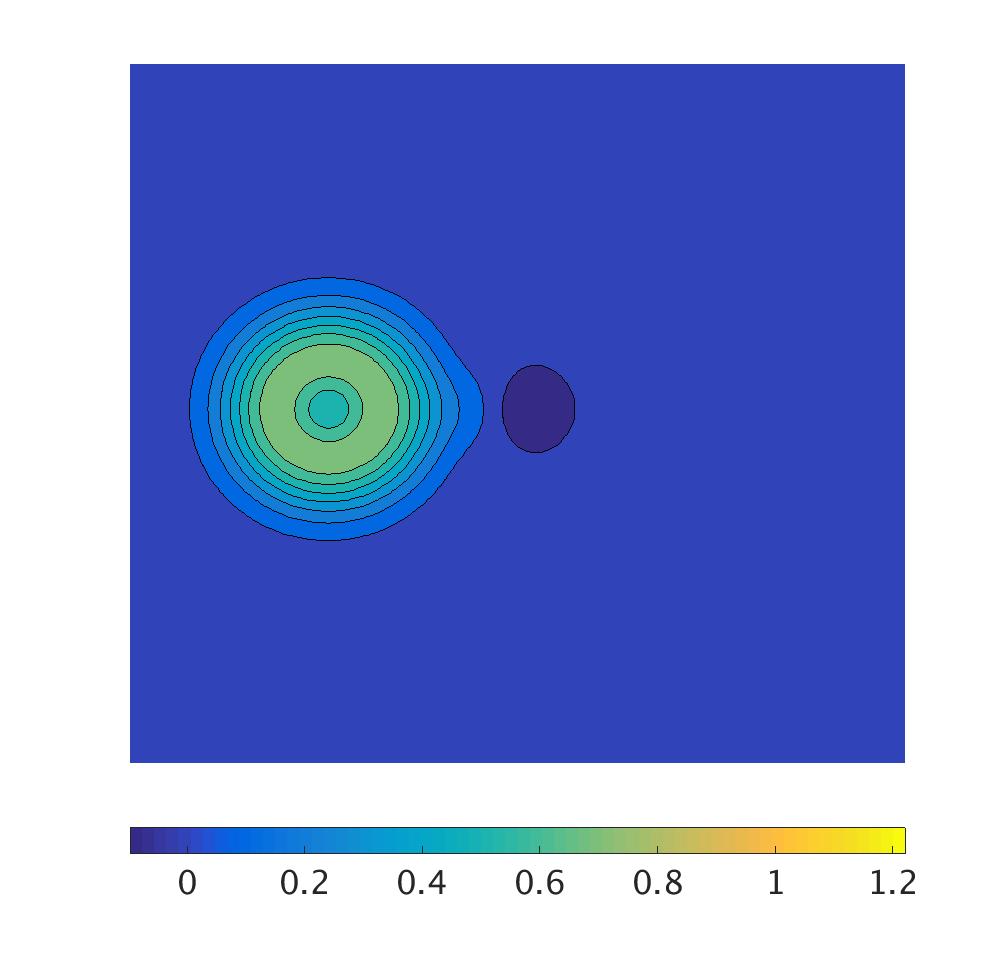}}
\subfloat[\, ]{\includegraphics[width=4.5cm,totalheight=4.8cm]{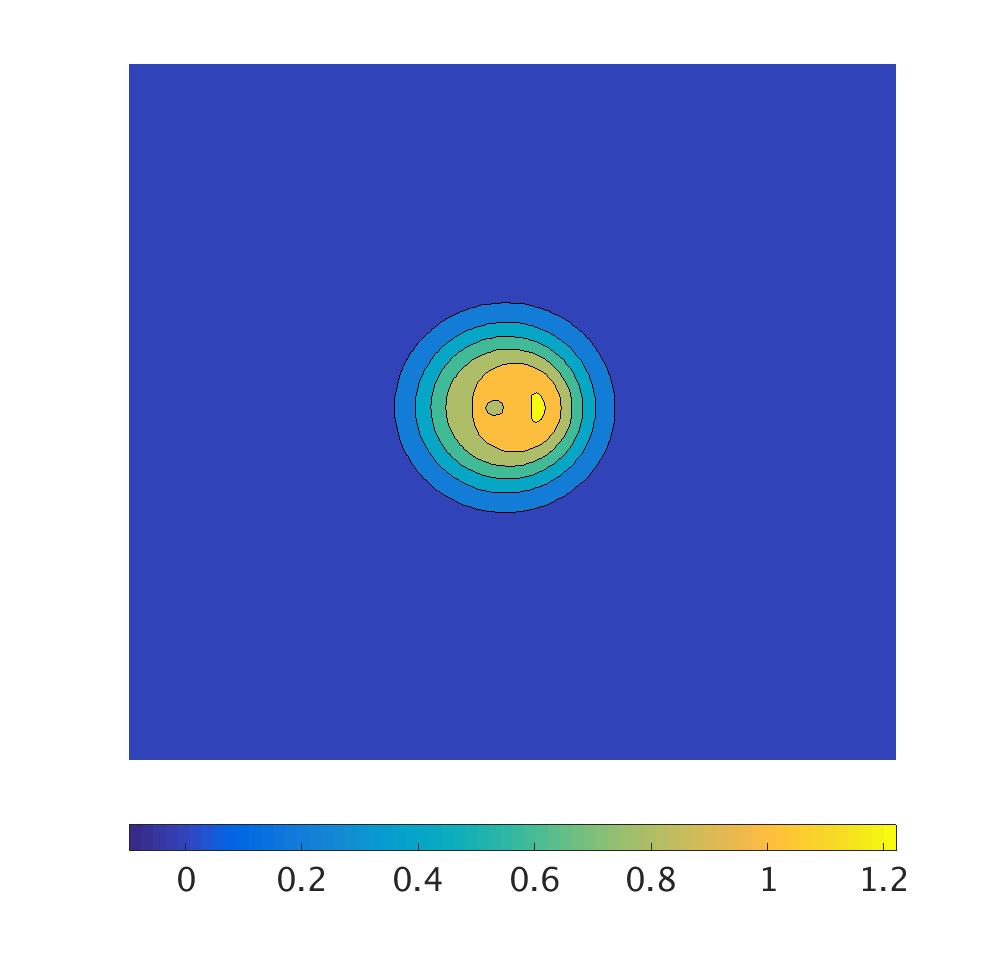}} 
\subfloat[\, ]{\includegraphics[width=4.5cm,totalheight=4.8cm]{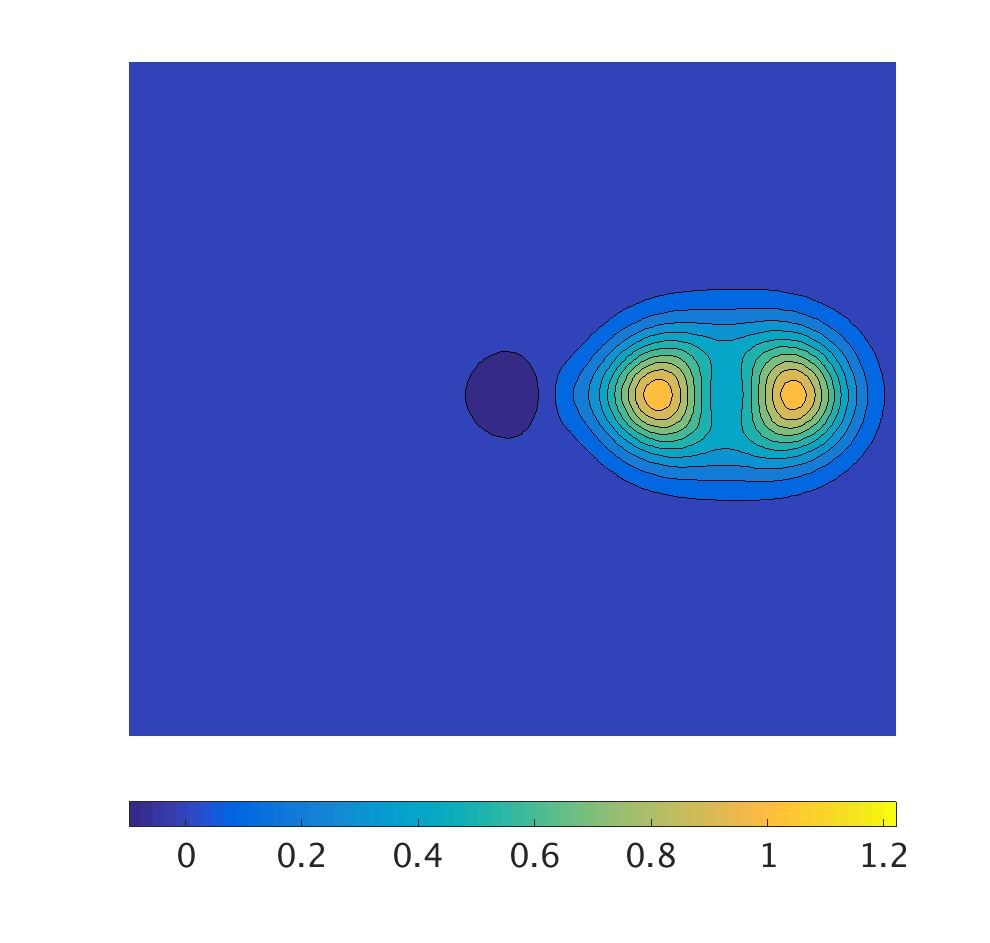}} \\
\subfloat[\, ]{\includegraphics[width=4.5cm,totalheight=4.8cm]{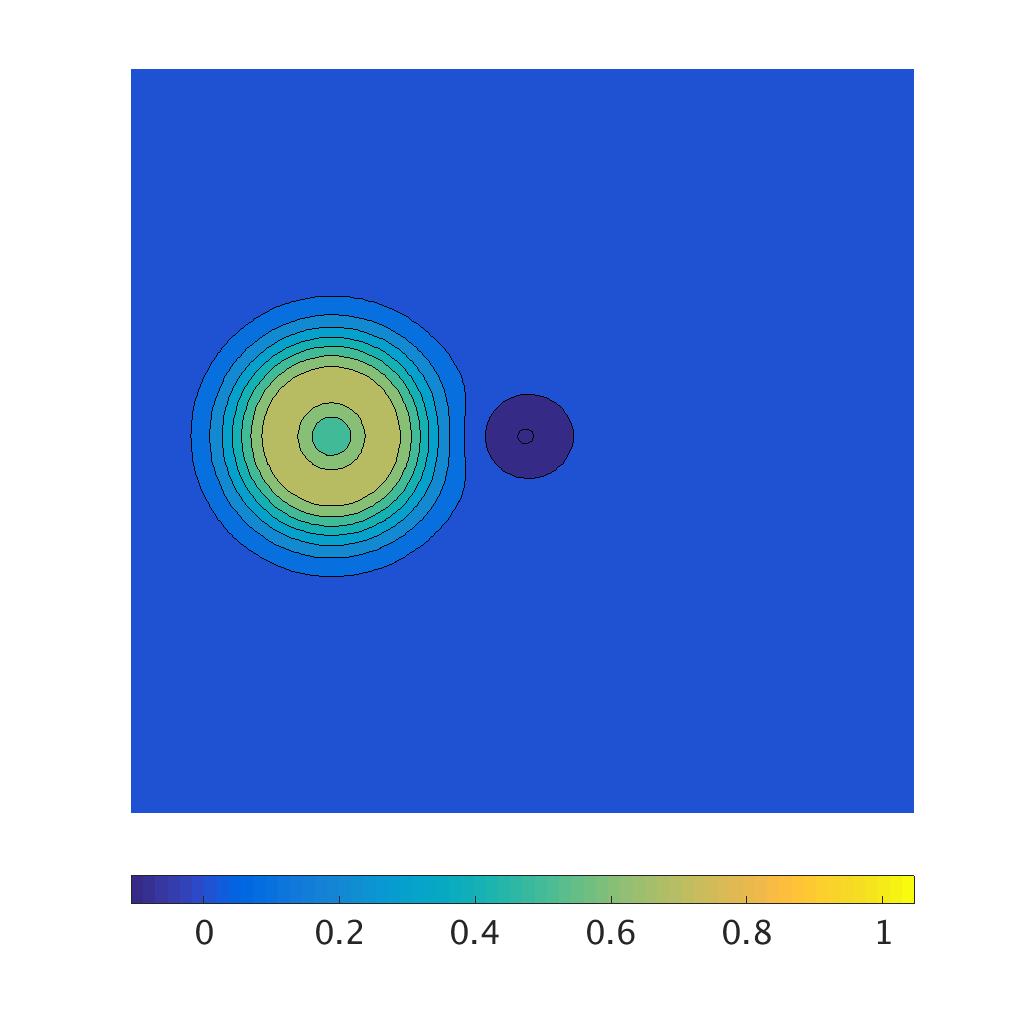}}
\subfloat[\, ]{\includegraphics[width=4.5cm,totalheight=4.8cm]{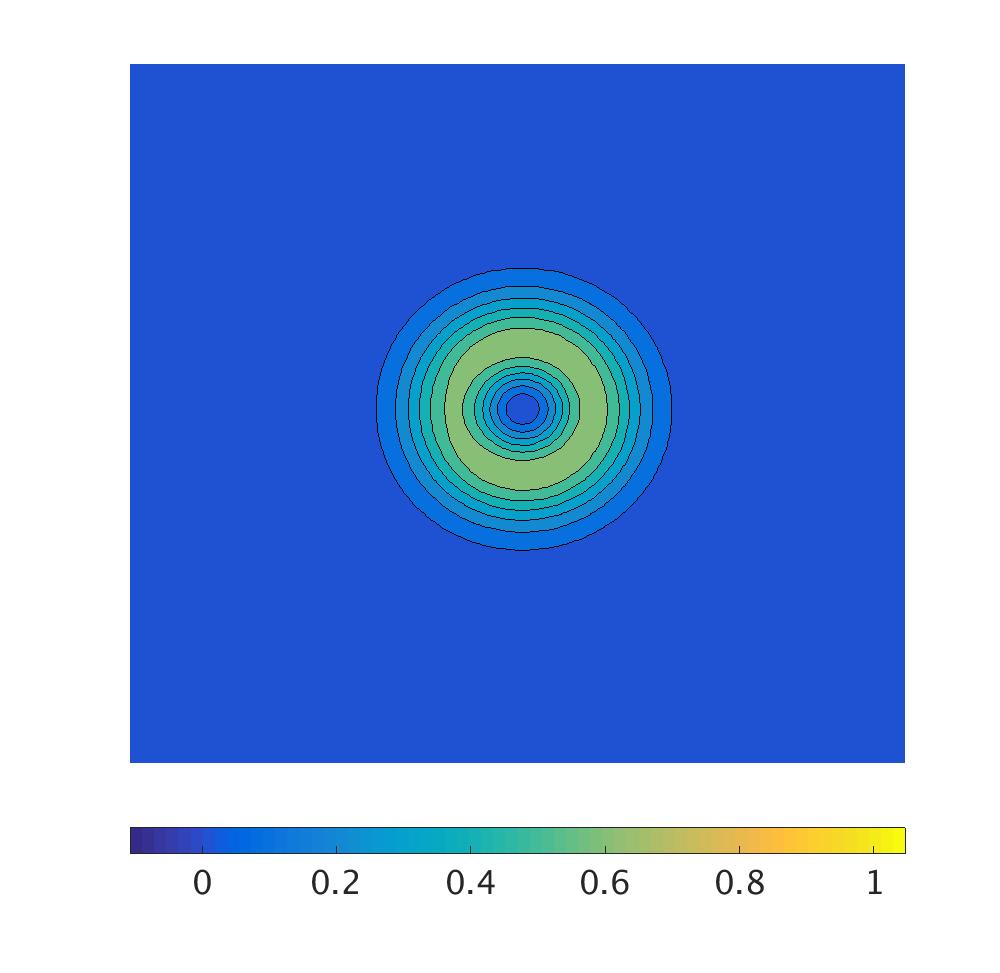}} 
\subfloat[\, ]{\includegraphics[width=4.5cm,totalheight=4.8cm]{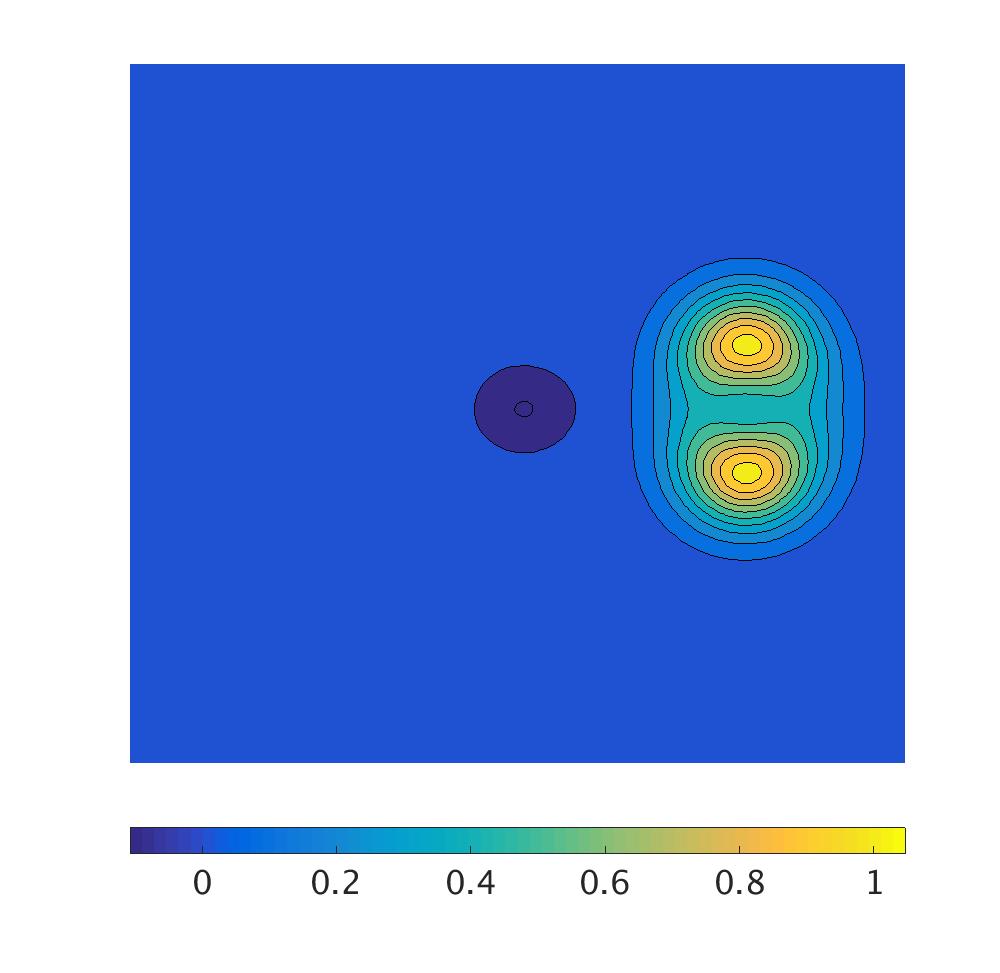}} 
\caption{Snapshots of the energy density during the scattering of a ring-shaped 2-vortex at critical coupling with impurities (a)-(c):~$\sigma(r) = e^{-r^2}$ and (d)-(f):~$\sigma(r) = -e^{-r^2}$ for initial velocity $v=0.3$.
}
\label{fig:vimp_N2snapshot}
\end{center}
\end{figure}


We now consider the scattering of a ring-shaped 2-vortex at critical coupling with the same two impurities. In Fig.~\ref{fig:vimp_N2snapshot} we display snapshots of the energy density at different times during the head-on scattering of a ring-shaped vortex with initial velocity $v=0.3$ and impurities (a)-(c):~$\sigma(r) = e^{-r^2}$ and (d)-(f):~$\sigma(r) = -e^{-r^2}$. In both cases, we initially see the vortex as a ring of energy density and the impurity as a region of negative energy density at the origin. As the vortex passes through the impurity for $c>0$, seen in Fig.~\ref{fig:vimp_N2snapshot}(b), a lump of energy taller than the initial ring-shaped vortex is formed at the origin. When the vortices emerge on the other side of the impurity in Fig.~\ref{fig:vimp_N2snapshot}(c), we see the ring break up into two vortices along the $x$-axis. Note that for $c>1,$ the impurity is attracting, so this motion can be understood as follows. The impurity pulls at the first vortex to spilt the ring-shaped $N=2$ vortex apart along the $x$-axis. While the second vortex is also attracted by the impurity it is simultaneously repelled by the first vortex which leads to the separation along the $x$-axis.
We observe a different behaviour for $c<0$. Fig.~\ref{fig:vimp_N2snapshot}(e) shows the ring-shaped vortex and impurity coincident. Here the energy forms a ring similar to an $N=3$ vortex solution. When the vortices emerge on the other side of the impurity, the ring has broken up into two vortices along the $y$-axis. This is shown in Fig.~\ref{fig:vimp_N2snapshot}(f). The vortices continue to move in the $x$-direction, but also separate to infinity in the $y$-direction. Note that for $c<0$ the impurity is repulsive and behaves more like a pinned vortex. Now, it is more natural for the ring-shaped 2-vortex to split along the $y$ direction, and both vortices are then repelled by the impurity. This dynamics shows similarities with the head-on collision of three vortices which leads to $\frac{\pi}{3}$ scattering, see \cite{MacKenzie:1995np} for a geometric explanation of $\frac{\pi}{N}$ scattering.

\begin{figure}[!htb]
\begin{center}
\includegraphics[totalheight=7.0cm]{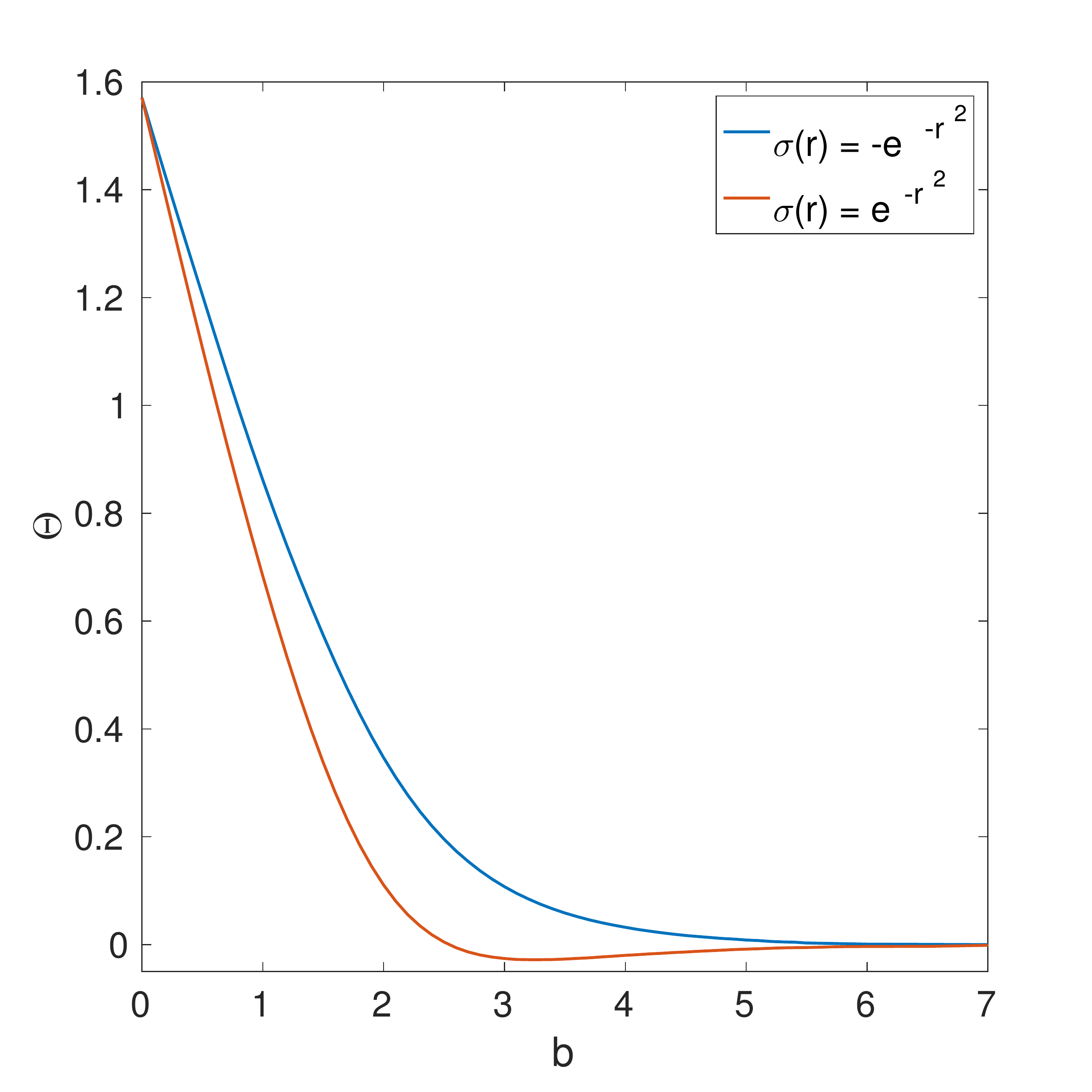}
%
\caption{ Scattering angle $\Theta$ in radians against impact parameter $b$ for the scattering of two vortices at critical coupling in the presence of impurities $\sigma(r)=e^{-r^2}$ (in red) and $\sigma(r)=-e^{-r^2}$ (in blue). }%
\label{fig:vimp_angle2N1}
\end{center}
\end{figure}

\begin{figure}[!htb]
\begin{center}
\subfloat[\, $\sigma(r) = e^{-r^2}$]{\includegraphics[totalheight=7.0cm]{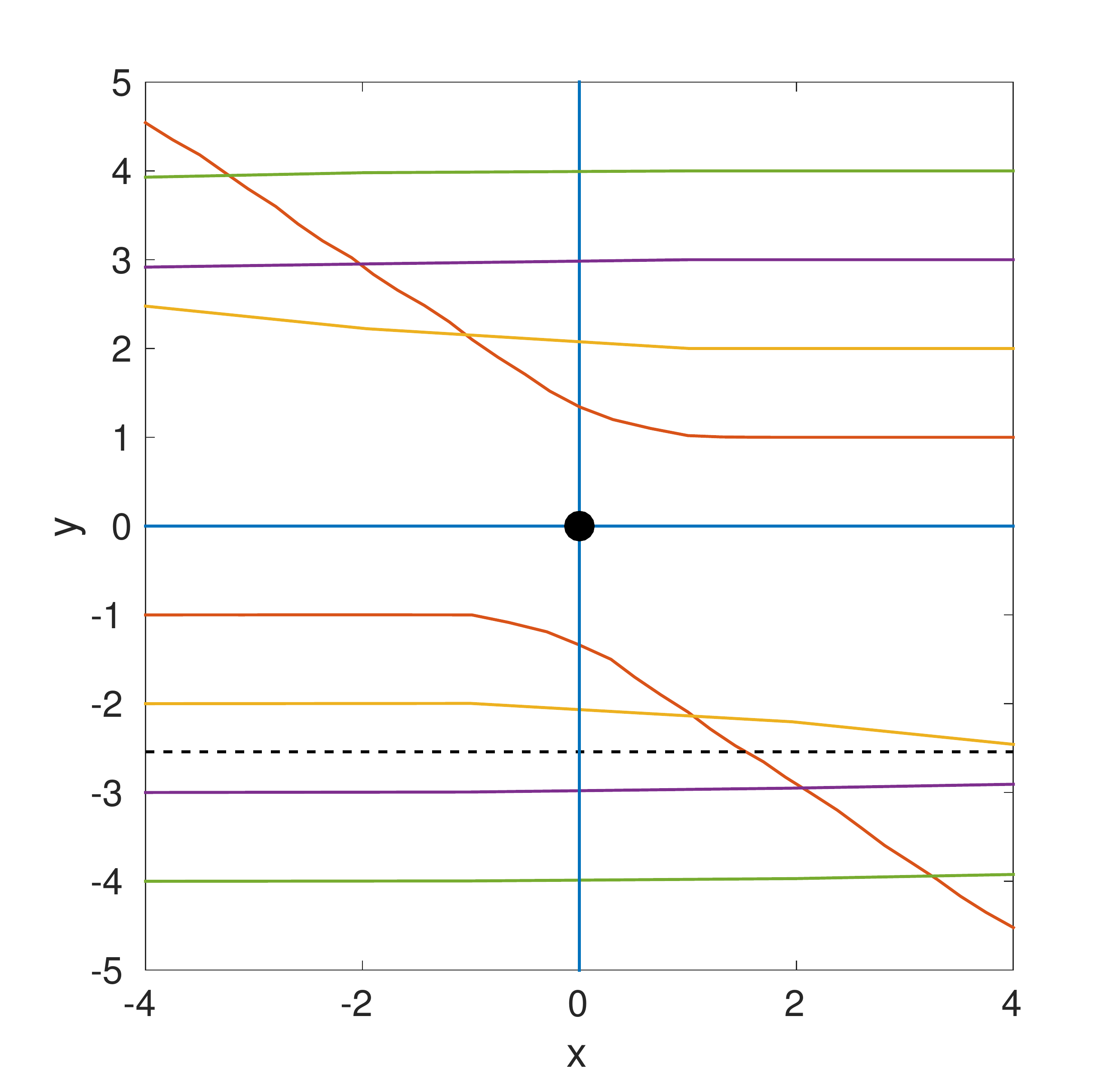}}
\subfloat[\, $\sigma(r) = -e^{-r^2}$]{\includegraphics[totalheight=7.0cm]{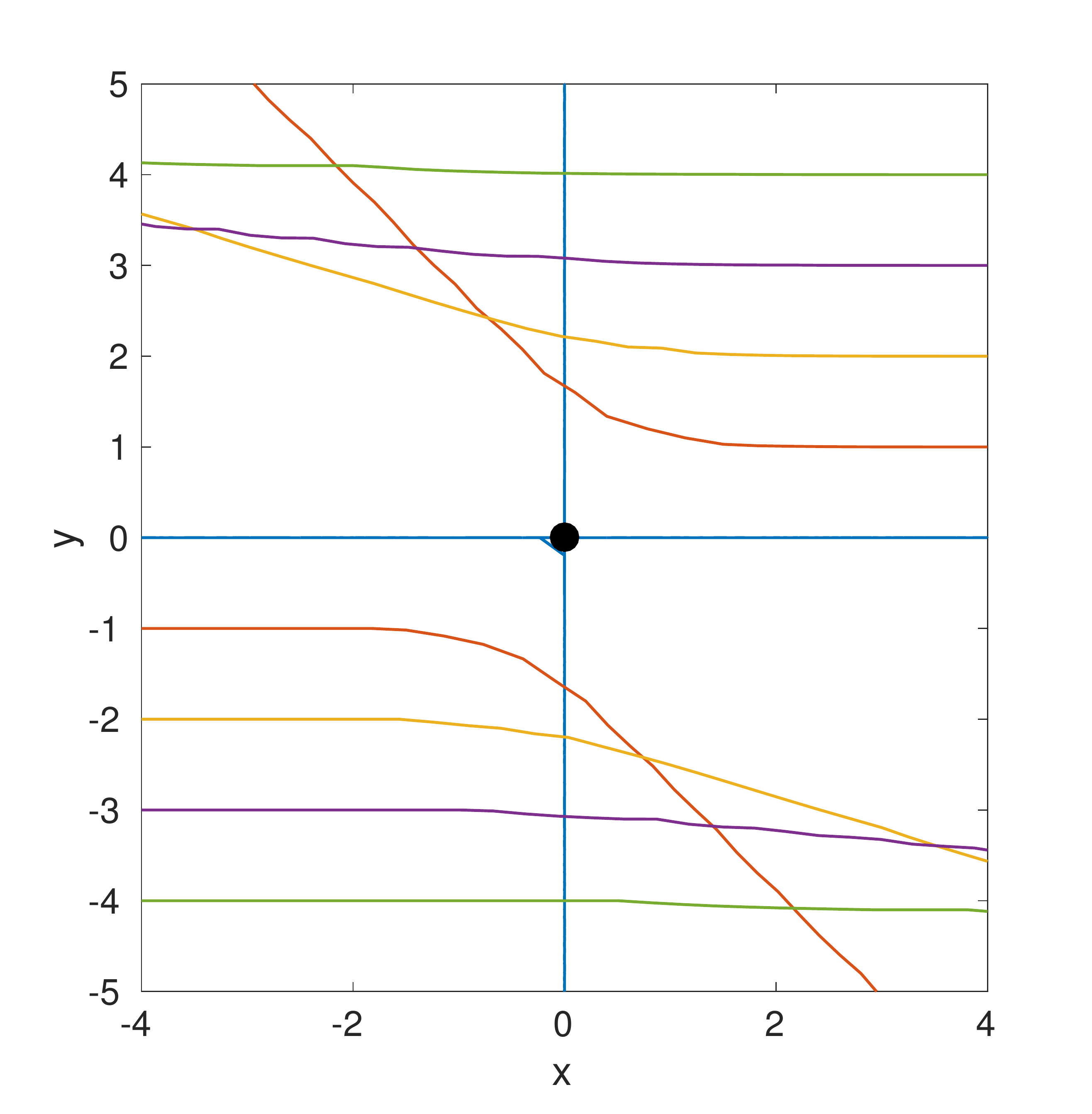}} 
\caption{Vortex trajectories during the scattering of two vortices at critical coupling in the presence of impurities (a)~$\sigma(r) = e^{-r^2}$ and (b)~$\sigma(r) = -e^{-r^2}$ for impact parameter values $b=0,1,2,3,4$ and initial velocity $v=0.3$. In (a), we also plot the trajectory for $b=2.54$ (where the scattering angle changes sign) as a black dashed line. The location of the impurity is indicated with a black dot, and the vortices are initially located at  $(\pm 4, \pm b)$.}
\label{fig:vimp_2N1pos}
\end{center}
\end{figure}

Finally, we investigate the scattering of two $N=1$ vortices at critical coupling in the presence of a magnetic impurity. In these simulations, we begin with a static impurity at the origin and two vortices located at $(\pm4,\pm b)$, where $b$ is the impact parameter between each vortex and the impurity, as illustrated in Fig.~\ref{fig:vimp_angle_diag}. The impact parameter between the two vortices is $2b$. We boost the vortices towards each other in the $x$-direction with initial velocity $v=0.3$. 

In Fig.~\ref{fig:vimp_angle2N1} we plot the scattering angle as a function of impact parameter for two vortices scattering in the presence of impurities  $\sigma(r)=e^{-r^2}$ and $\sigma(r)=-e^{-r^2}$. For impact parameter $b=0$, the vortices scatter at right angles, as is the case in a head-on collision of two vortices in the absence of an impurity. As the impact parameter increases, the scattering angle decreases. For  $\sigma(r)=-e^{-r^2}$, the vortex trajectories bend away from each other and the impurity. This can be seen in Fig.~\ref{fig:vimp_2N1pos}(b). For $\sigma(r)=e^{-r^2}$, the direction of the vortex trajectories changes after a certain value of the impact parameter. We see in Fig.~\ref{fig:vimp_2N1pos}(a) that the trajectories for $b=1,2$ bend away from the impurity, but for $b=3,4$ bend towards it. The critical value where the scattering angle crosses zero is $b=2.54$, and we show this trajectory as a black dashed line. In Fig.~\ref{fig:vimp_angle2N1}, this change in direction corresponds to the change in sign of the scattering angle $\Theta$. Initially the repulsion between the vortices is more significant than the attraction between each vortex and the impurity. Once the vortices are sufficiently separated, the attraction to the impurity becomes the strongest effect.

\begin{figure}[!htb]
\begin{center}
\subfloat[\, ]{\includegraphics[width=4.5cm,totalheight=4.8cm]{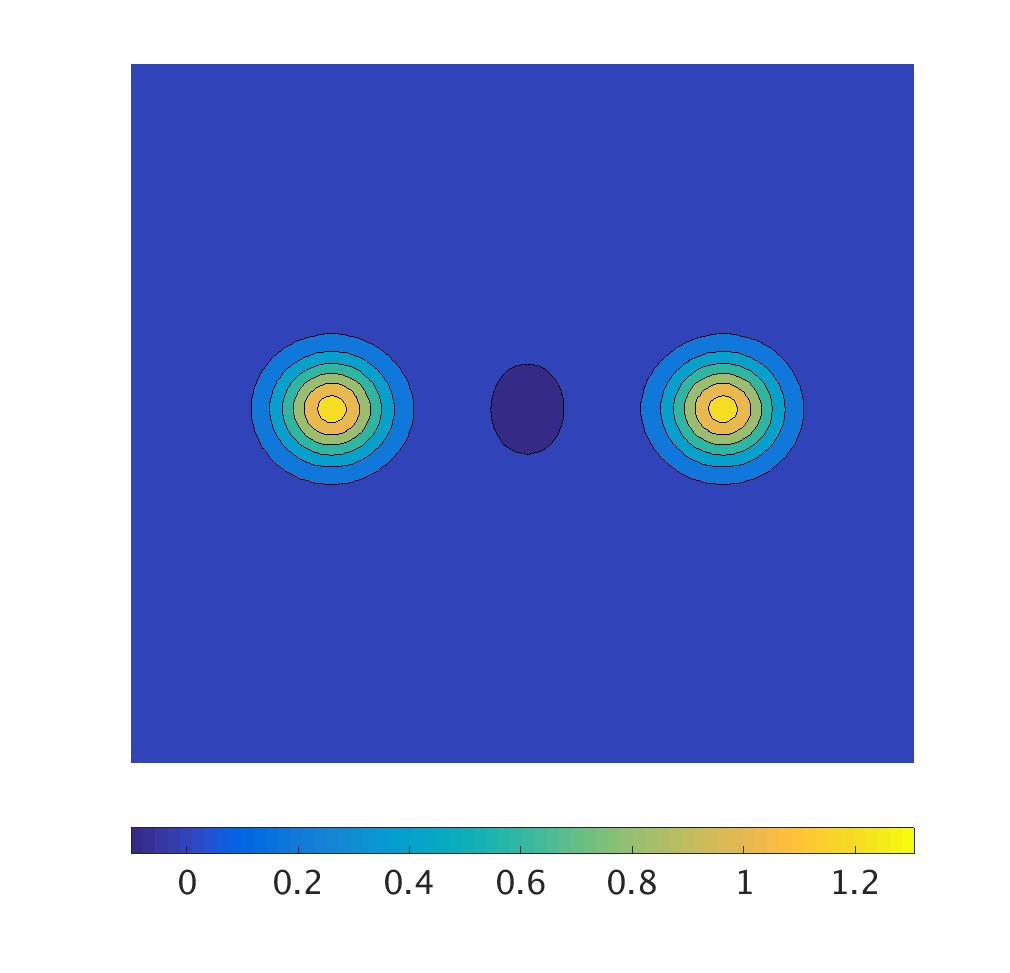}}
\subfloat[\, ]{\includegraphics[width=4.5cm,totalheight=4.8cm]{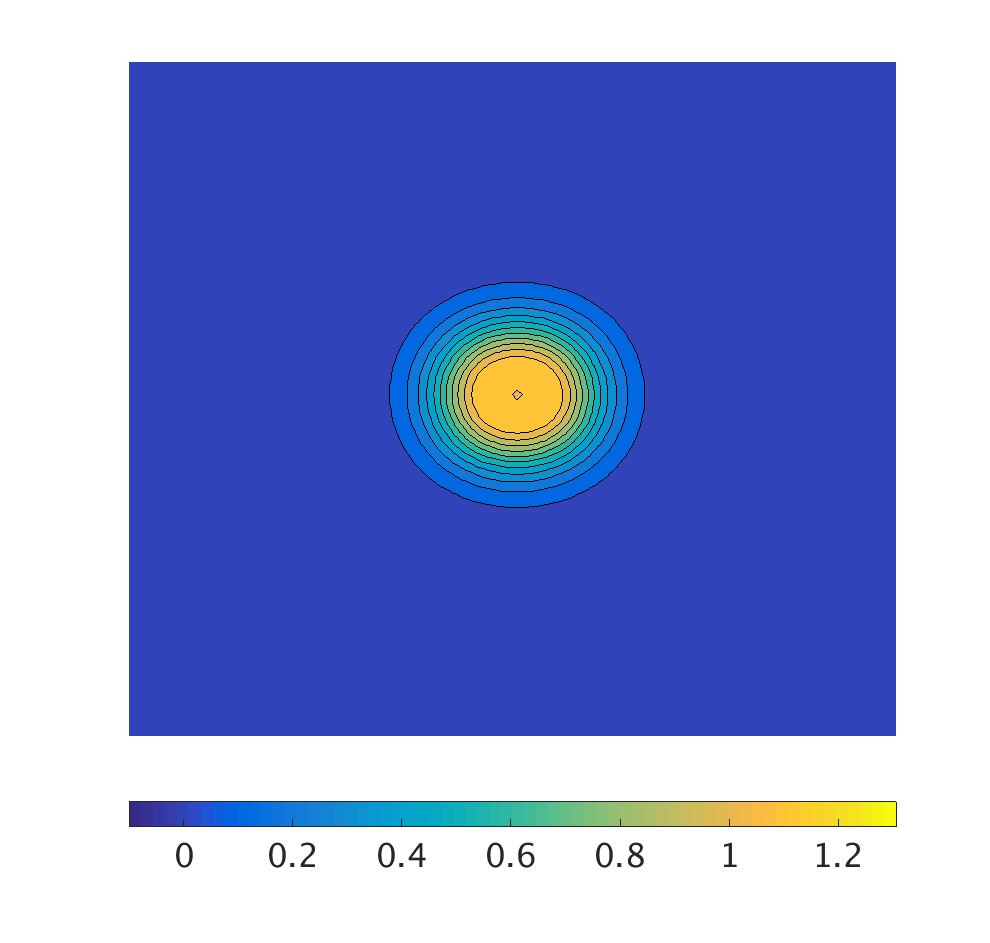}} 
\subfloat[\, ]{\includegraphics[width=4.5cm,totalheight=4.8cm]{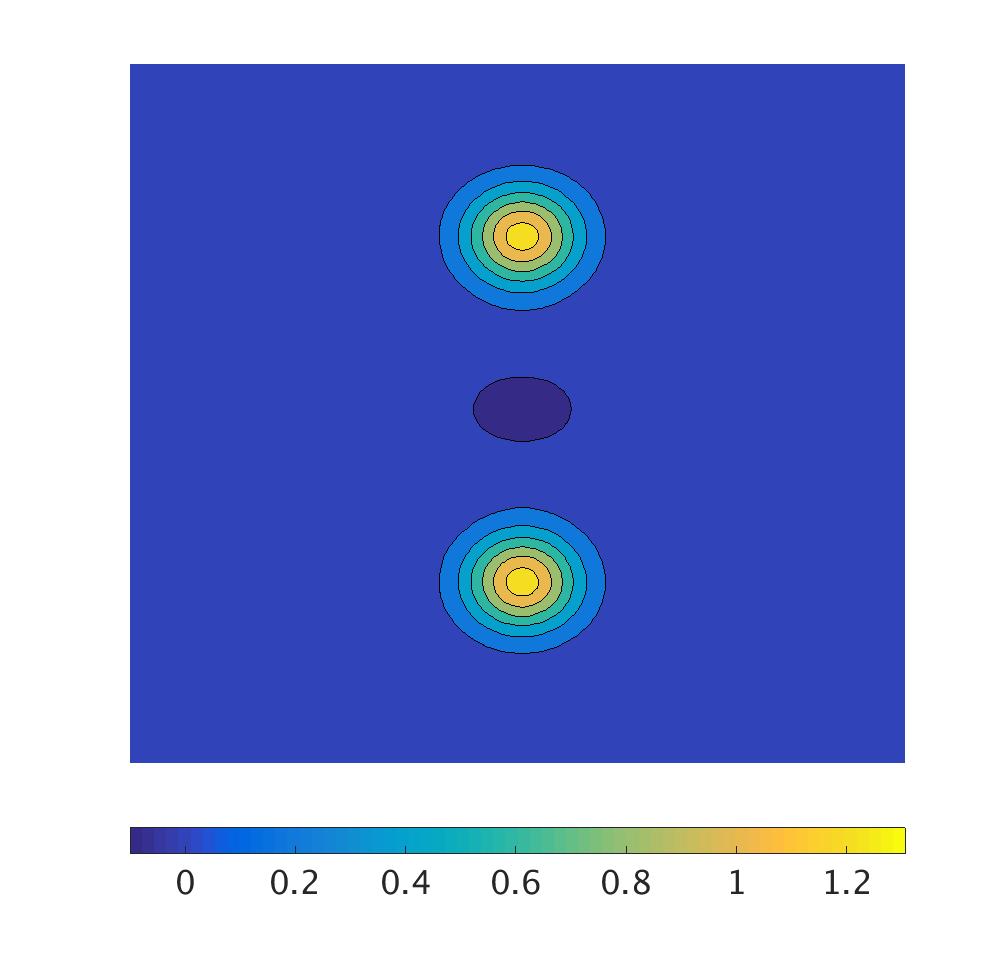}} \\
\subfloat[\, ]{\includegraphics[width=4.5cm,totalheight=4.8cm]{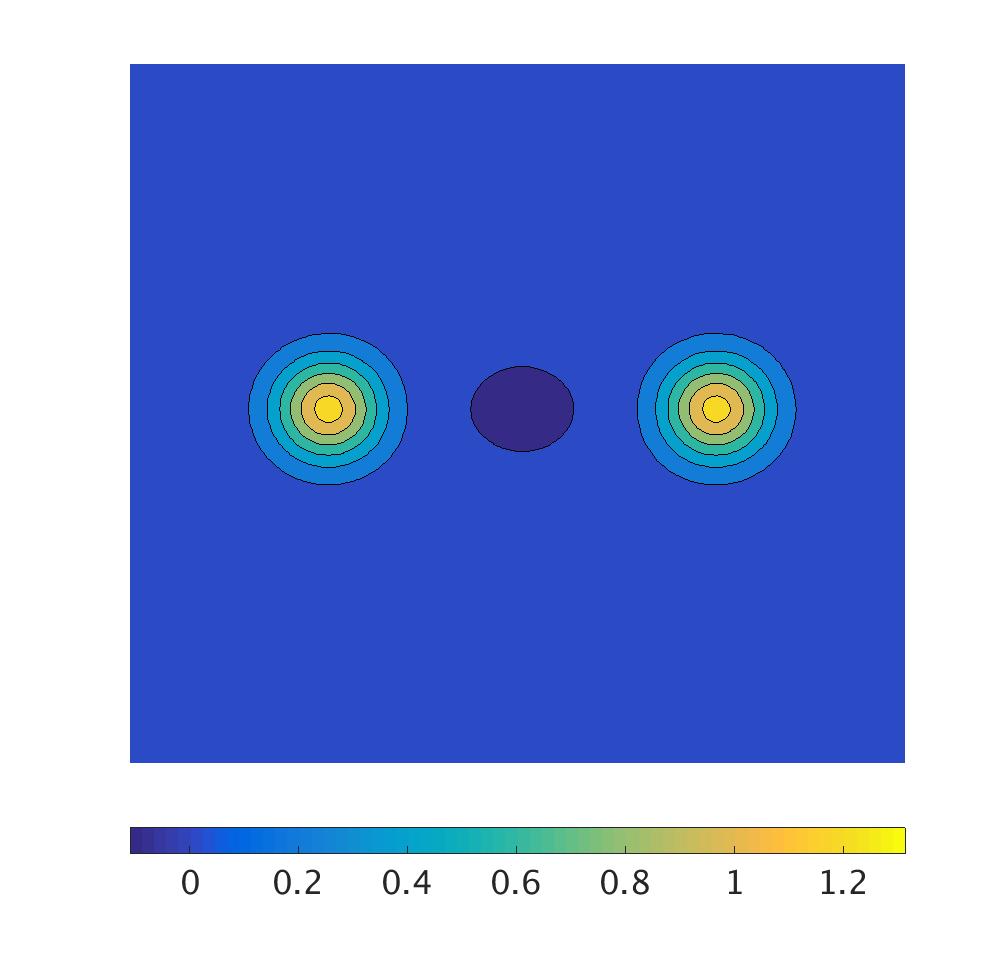}}
\subfloat[\, ]{\includegraphics[width=4.5cm,totalheight=4.8cm]{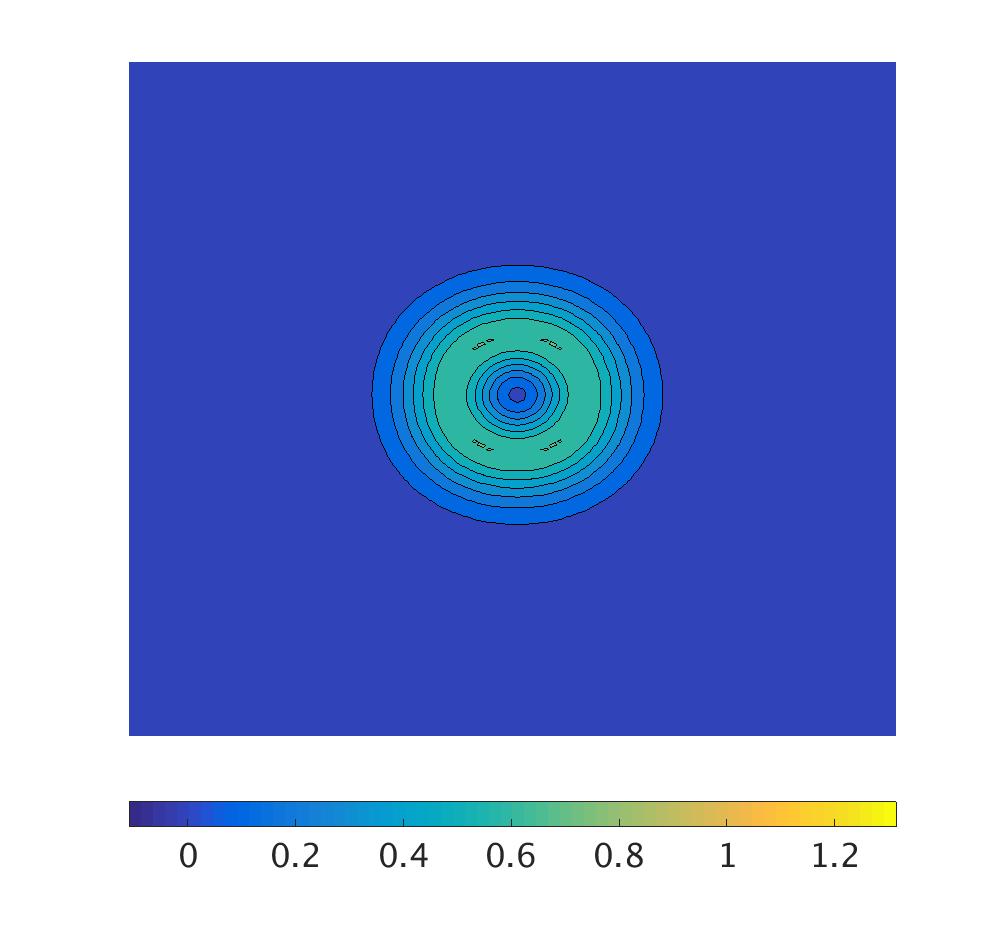}} 
\subfloat[\, ]{\includegraphics[width=4.5cm,totalheight=4.8cm]{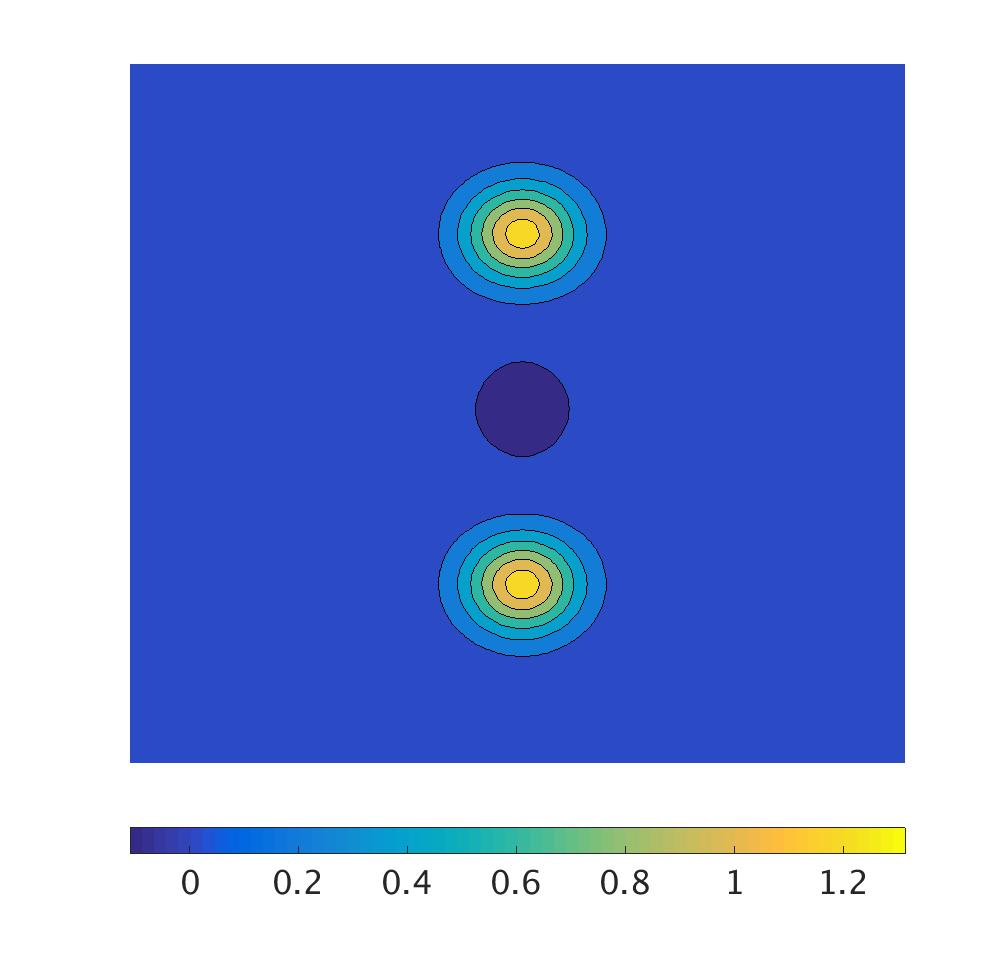}} 
\caption{Snapshots of the energy density during the scattering of two $N=1$ vortices at critical coupling through impurities (a)-(c):~$\sigma(r) = e^{-r^2}$ and (d)-(f):~$\sigma(r) = -e^{-r^2}$ for initial velocity $v=0.3$. }
\label{fig:vimp_2N1snapshot}
\end{center}
\end{figure}

Snapshots of the energy density during the head-on collision of two vortices through an impurity located at the origin are given in Fig.~\ref{fig:vimp_2N1snapshot}. The initial velocity given to the vortices is $v=0.3$, and the impurities considered are (a)-(c):~$\sigma(r) = e^{-r^2}$ and (d)-(f):~$\sigma(r) = -e^{-r^2}$.
Figs.~\ref{fig:vimp_2N1snapshot}(a) and (d) show the initial configurations of two vortices on either side of an impurity located at the origin. For $c>0$, we see in Fig.~\ref{fig:vimp_2N1snapshot}(b) that the vortices and impurity all meet at the origin, forming a large lump of energy density. The same situation for $c<0$ is seen in Fig.~\ref{fig:vimp_2N1snapshot}(e). Here the energy forms a ring surrounding the impurity. For both impurities, the overall result is that the vortices scatter at right angles, and we see in  Figs.~\ref{fig:vimp_2N1snapshot}(c) and~(e) that the vortices emerge and travel to infinity along the $y$-axis.

\subsection{Vortex scattering away from critical coupling}
\label{lambdanotzero}

In this section, we briefly discuss vortex scattering off impurities for $\lambda \neq 1.$ In this case, the dynamics of vortices can be approximated as moduli space dynamics with an induced potential. In particular, the trajectories are no longer just geodesics, so that trajectories depend significantly on the initial velocities. We restrict our investigation to the head-on collision of a single vortex and one of the impurities described in Fig. \ref{fig:vimp_en_vs_sep}. Hence, in the following we set $d=1$ and the coupling constant takes the values $\lambda = 0.5$ and $\lambda = 1.5.$

\begin{figure}[!htb]
\begin{center}
  \subfloat[\, $\frac{E}{\pi}$ for $c=4$]{\includegraphics[totalheight=4.9cm]
    {en_vs_sep_c4_long.pdf}}
\subfloat[\, $c=4$ and $\lambda=1.5$]{\includegraphics[totalheight=4.9cm]
  {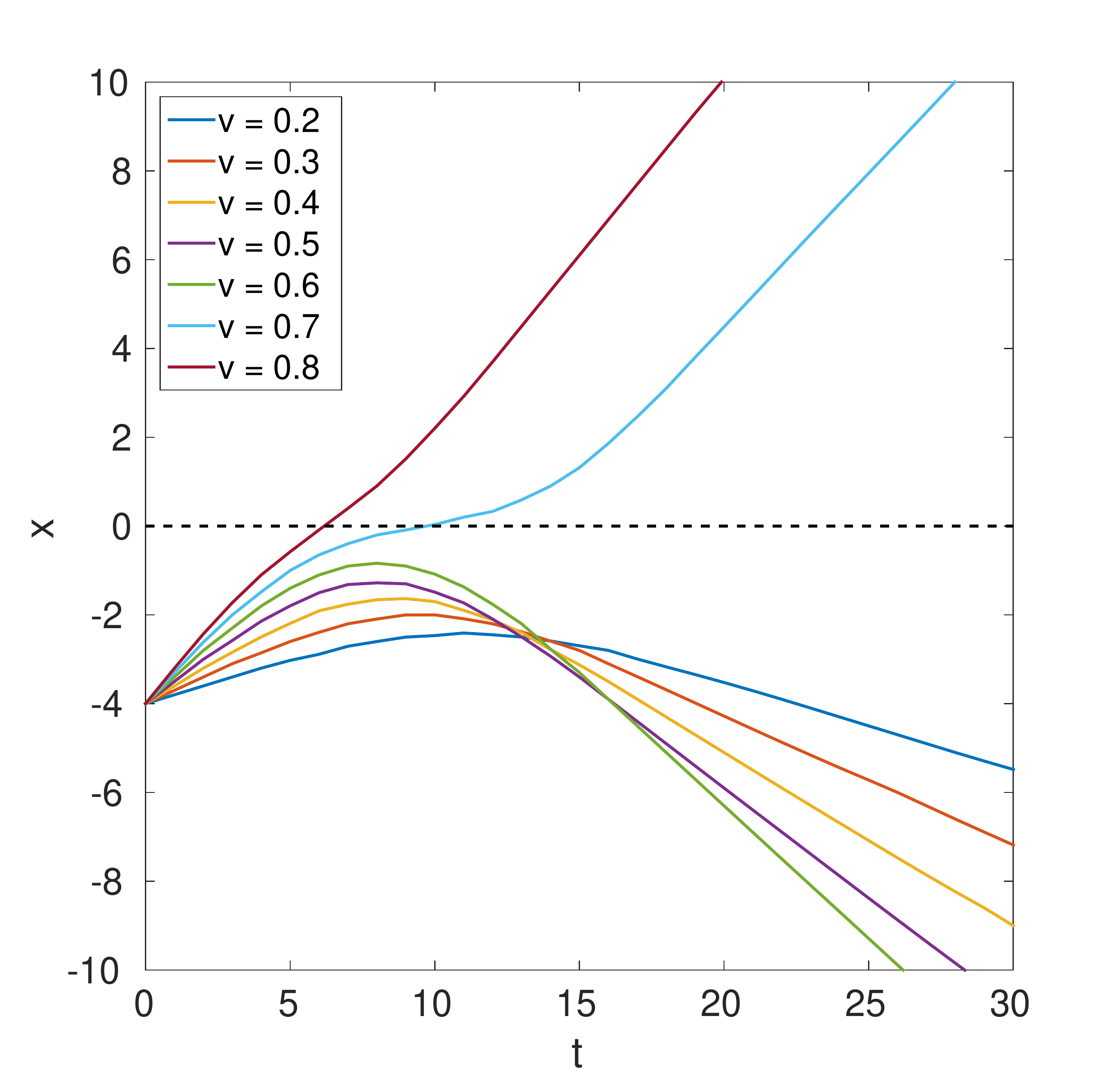}}
\subfloat[\, $c=4$ and $\lambda = 0.5$]{\includegraphics[totalheight=4.9cm]
  {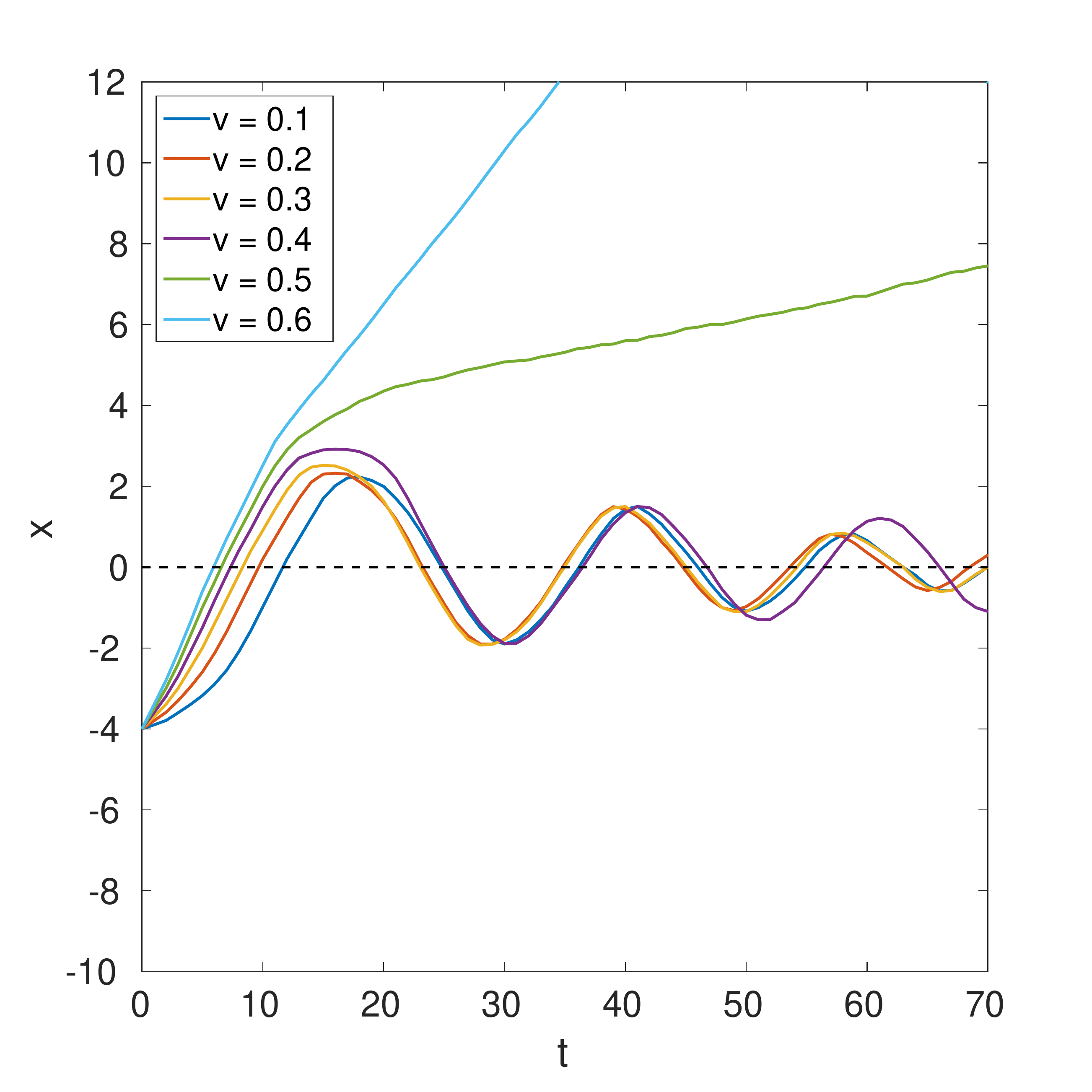}}
\\
 \subfloat[\, $\frac{E}{\pi}$ for $c=-4$]{\includegraphics[totalheight=4.9cm]
    {en_vs_sep_c-4_long.pdf}}
\subfloat[\, $c=-4$ and $\lambda=1.5$]{\includegraphics[totalheight=4.9cm]
  {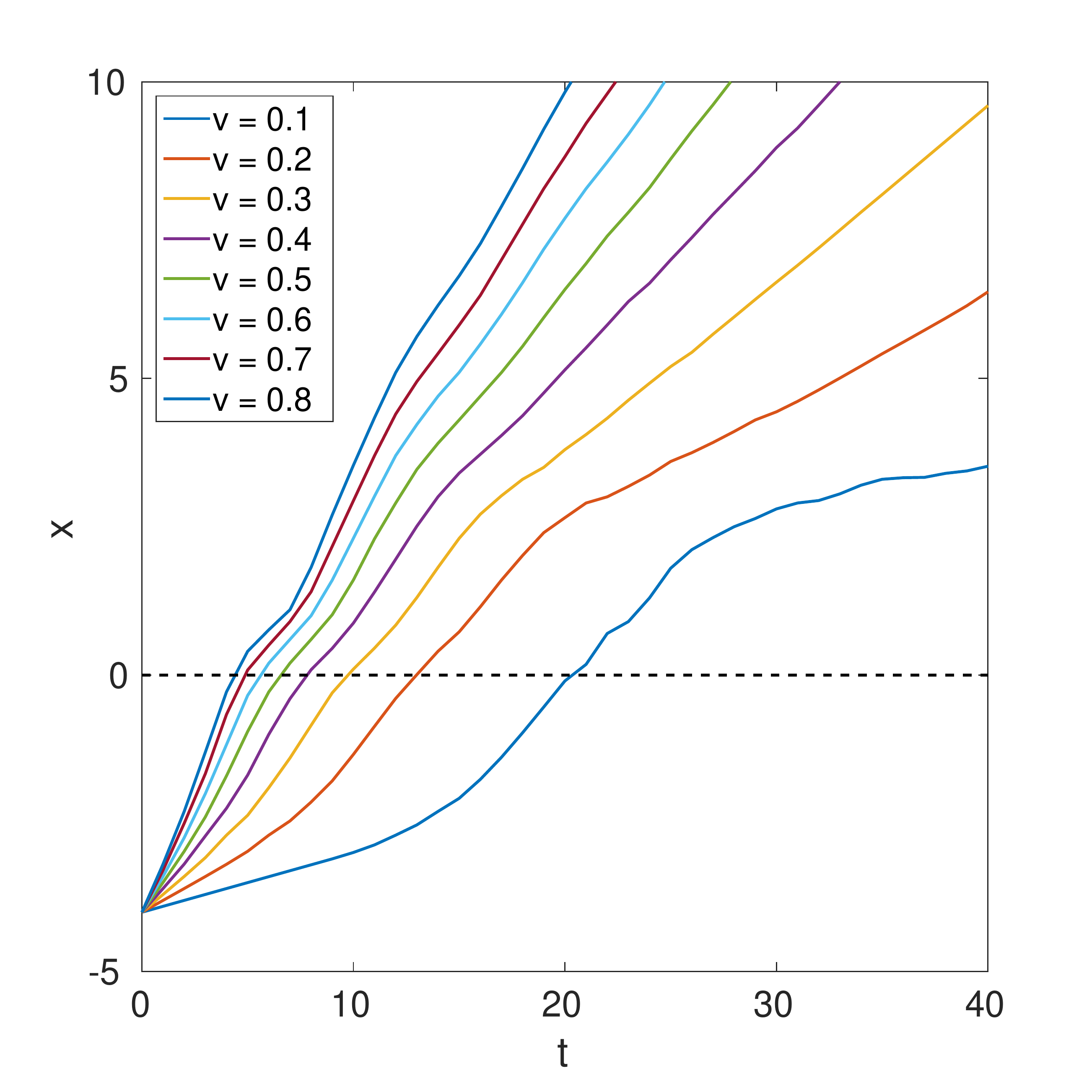}}
\subfloat[\, $c=-4$ and $\lambda = 0.5$]{\includegraphics[totalheight=4.9cm]
  {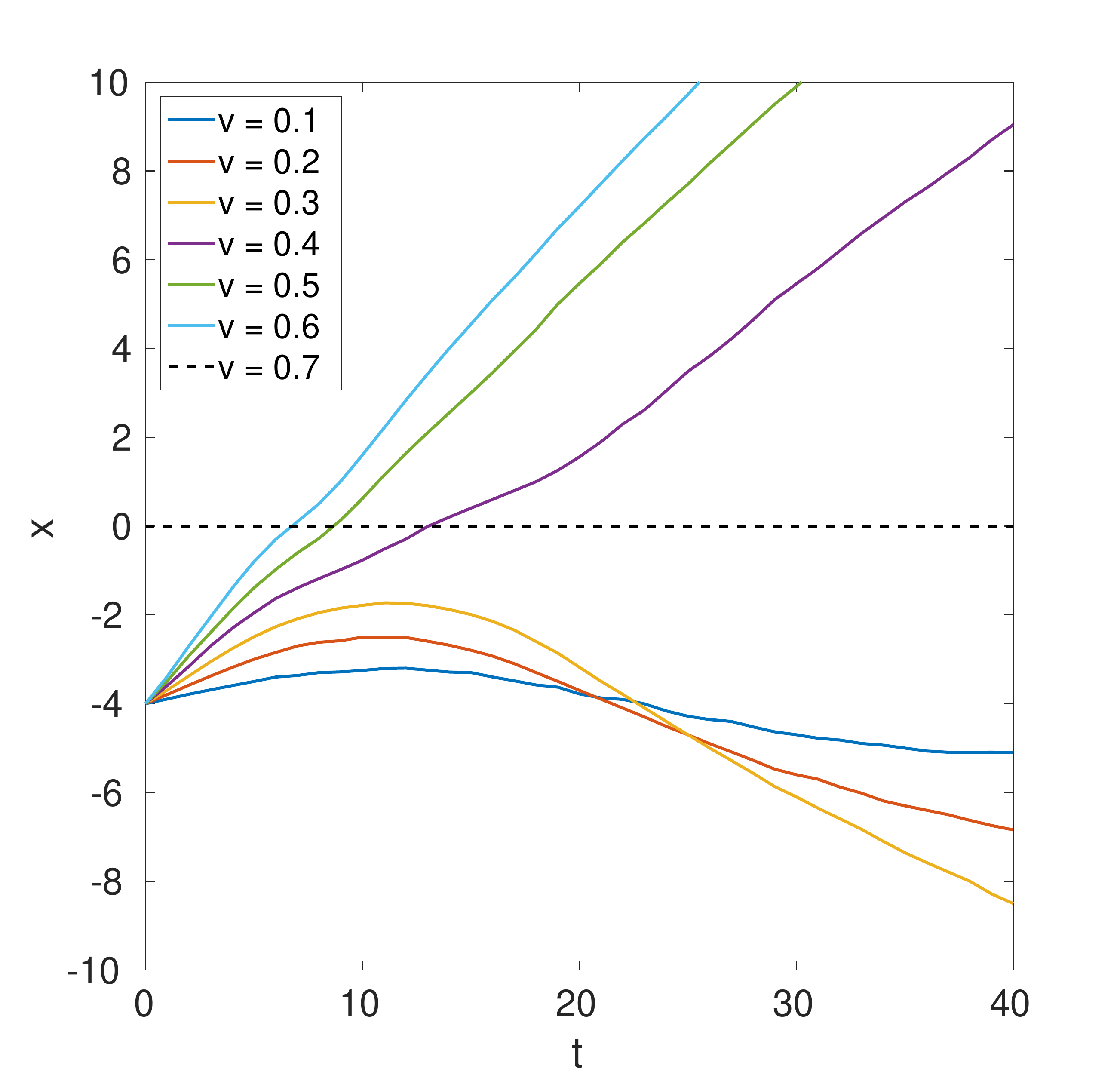}}\\
 \subfloat[\, $\frac{E}{\pi}$ for $c=-16$]{\includegraphics[totalheight=4.9cm]
    {en_vs_sep_c-16_ver2.pdf}}
\subfloat[\, $c=-16$ and $\lambda=1.5$]{\includegraphics[totalheight=4.9cm]
  {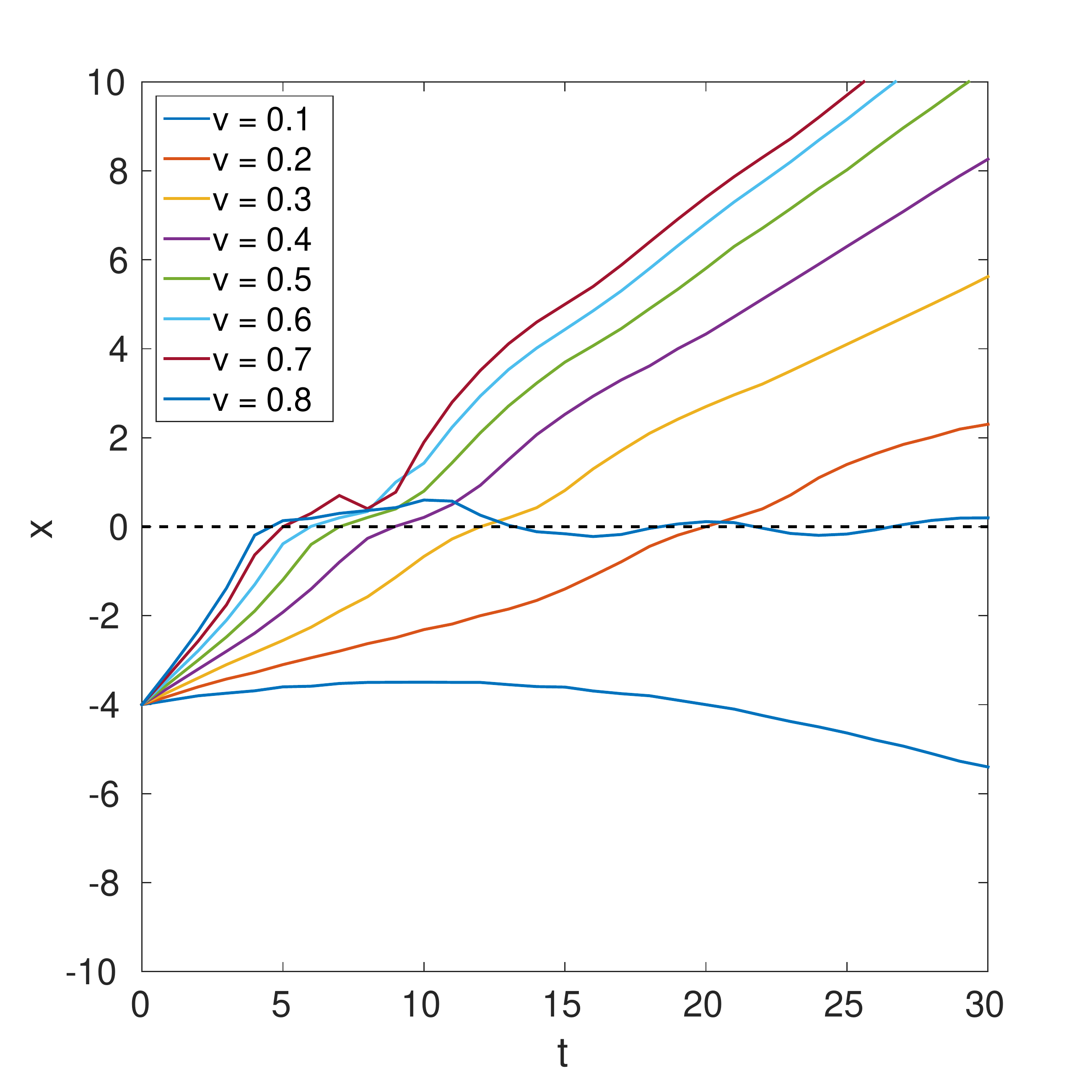}}
\subfloat[\, $c=-16$ and $\lambda = 0.5$]{\includegraphics[totalheight=4.9cm]
  {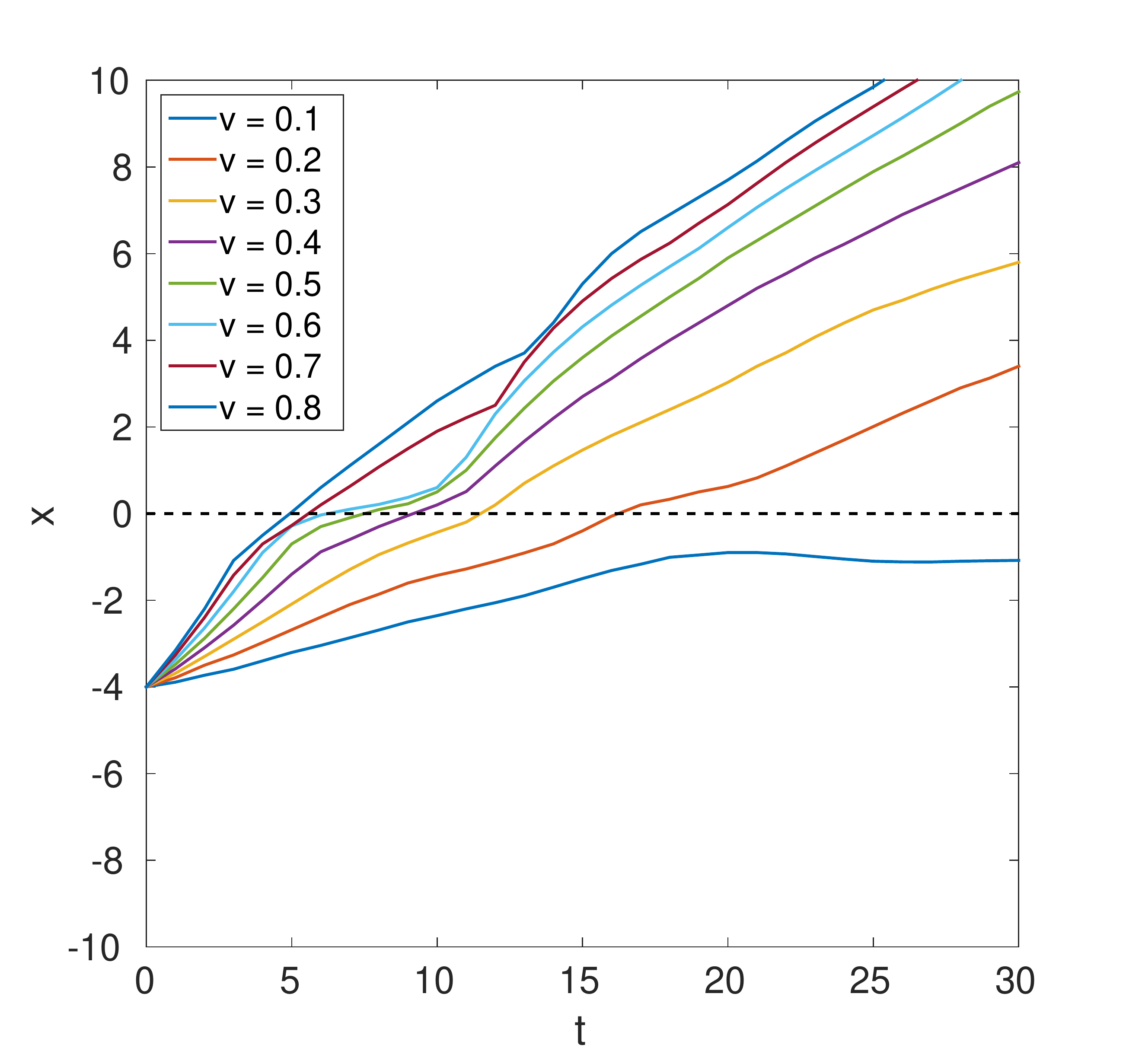}}
\caption{Scattering trajectories for $\lambda \neq 1.$ The potential energy $\frac{E}{\pi}$ is displayed in (a), (d) and (g). The corresponding vortex trajectories are displayed in the middle row for $\lambda =1.5$ and in the righthand row for $\lambda =0.5.$ A detailed discussion can be found in the main text.}
\label{fig:trajectories1}
\end{center}
\end{figure}

For $c=4$ and $\lambda = 1.5,$ the potential energy $\frac{E}{\pi}$ as a function of $s$ is displayed in Fig. \ref{fig:trajectories1} (a) top figure in red. It has a global maximum at the origin and a very shallow global minimum at $s \approx 4.$ Fig.  \ref{fig:trajectories1} (b) shows vortex trajectories with initial velocities $v=0.2, \dots, 0.8.$ For low initial velocity the vortex is reflected back elastically from the impurity. As the initial velocity increases the vortex be comes closer to the impurity before turning back. At $v=0.7$ the vortex crosses the impurity and moves away to infinity. The trajectory shows how the vortex slows down before reaching the impurity and then accelerates away from it. At $v=0.8$ the vortex crosses easily over the impurity, and its velocity is less affected by it. For very low initial velocities, we expect that the vortex could get trapped in the shallow minimum. 

The bottom figure in Fig. \ref{fig:trajectories1} (a) shows the potential energy $\frac{E}{\pi}$ in blue as a function of $s$ for $c=4$ and $\lambda = 0.5.$ There is a global minimum at the origin and a very shallow global maximum at  $s \approx 5.5.$ 
Fig.  \ref{fig:trajectories1} (c) shows vortex trajectories with initial velocities $v=0.2, \dots, 0.6.$ For low initial velocities the vortex gets trapped by the impurity and oscillates around the origin. For $v=0.5$ the vortex escapes but slows down significantly. For $v=0.6$ the vortex easily escapes the impurity. By scattering vortices from further away with very low initial velocity we expect that the vortex would be reflected by the impurity.

The top figure in Fig. \ref{fig:trajectories1} (d) shows the potential energy $\frac{E}{\pi}$ in red as a function of $s$ for $c=-4$ and $\lambda = 1.5.$ 
There is a global minimum at the origin and a global maximum at $s\approx 3.$ Fig.  \ref{fig:trajectories1} (e) shows vortex trajectories with initial velocities $v=0.1, \dots, 0.8.$ In all cases, the vortex passes the impurity. For low velocities the vortex accelerates towards the origin and then decelerates. For even lower initial velocities we expect that the vortex might get trapped by the impurity or reflected away by the maximum.

The bottom figure in Fig. \ref{fig:trajectories1} (d) shows the potential energy $\frac{E}{\pi}$ in blue as a function of $s$ for $c=-4$ and $\lambda = 1.5.$ 
There is a global maximum at the origin and a global minimum at $s\approx 4.$ Fig.  \ref{fig:trajectories1} (f) shows vortex trajectories with initial velocities $v=0.1, \dots, 0.7.$ For low velocities, the vortex is reflected back from the impurity. At $v=0.4$ the vortex passes the impurity, slowing down before reaching the centre of the impurity and then speeding up again. For higher velocities, the vortex also passes the impurity, but the velocity is less affected. For very low velocities, we expect the vortex to be trapped by the minimum at $s \approx 4.$

The top figure in Fig. \ref{fig:trajectories1} (g) shows the potential energy $\frac{E}{\pi}$ in red as a function of $s$ for $c=-16$ and $\lambda = 1.5.$ 
There is a local minimum at the origin and a global maximum at $s\approx 3.$ Fig.  \ref{fig:trajectories1} (h) shows vortex trajectories with initial velocities $v=0.1, \dots, 0.8.$ For initial velocity $v=0.1$ the vortex is reflected by the maximum. For larger velocities, the vortices pass the maximum, cross the impurity and escape to the other side. The only exception is $v=0.8$ which is trapped by the impurity and oscillates around the origin. This indicates that dependence on the initial velocity is quite subtle at higher velocities. We also expect that the vortex may get trapped by the impurity for initial velocities between $v =0.1$ and $v=0.2.$

The bottom figure in Fig. \ref{fig:trajectories1} (d) shows the potential energy $\frac{E}{\pi}$ in blue as a function of $s$ for $c=-16$ and $\lambda = 1.5.$ 
There is a local maximum at the origin and a global minimum at $s\approx 3.$ Fig.  \ref{fig:trajectories1} (i) shows vortex trajectories with initial velocities $v=0.1, \dots, 0.8.$ For initial velocity $v=0.1$ the vortex is trapped in the global minimum. For larger velocities, the vortex escapes the impurity.

In summary, Fig. \ref{fig:trajectories1} shows many interesting scattering trajectories for $\lambda \neq 1$ for different initial velocities. 
A more detailed discussion of vortex impurity scattering is beyond the scope of this paper. Solitons and impurities have been extensively studied from a theoretical point of view. To mention just one example, the interactions of kinks and impurities give rise to interesting trapping phenomena, double-bounce solutions and fractal windows \cite{Fei:1992dk, Goodman:2004}. Scattering of a Sine-Gordon kink with a kink trapped in an extended impurity has been studied in \cite{Goatham:2010dg} revealing various different outcomes such as double trapping, kink knock-out and double escape. Recently, a new way of introducing BPS impurities has been discovered in \cite{Adam:2018pvd,Adam:2019yst} which gives us a powerful tool to study solitons and impurities. We expect that a detailed study of vortex impurity dynamics will reveal similar phenomena.

Recent progress in creating vortices and impurities in experiments in controlled environments has led an increasing interest in the interaction of vortices and impurities. Vortex impurity pinning has been observed in condensed matter system \cite{Shapoval:2010}, Bose-Einstein condensates \cite{Tung:2006} and neutron stars \cite{Anderson:1975zze}. Relevant dynamics has been have been studied in \cite{Bulgac:2013nmn, Wlazlowski:2016yoe} using sophisticated methods and point-particle limits. In these system, the sign of the force can also be very sensitive to the precise nature of the impurity which may be point defects or extended impurities. A review on pinning of superconducting vortices by periodic impurities can be found in \cite{Velez:2008}.


\section{Conclusions}

We have numerically investigated the dynamics of vortices in the presence of magnetic impurities of the form $\sigma(r)=ce^{-dr^2}$. We began by introducing the model and the Bogomolny bound satisfied by vortices at critical coupling. We reproduced the vortex configurations at critical coupling obtained in Ref.~\cite{Cockburn:2015huk} by a different method. Our method also allows us to solve for vortices away from critical coupling, and we presented vacuum solutions for $\lambda=0.5$ and $\lambda=1.5$. We discovered that for $c<0$ a delta function impurity behaves like another vortex regardless of the coupling constant $\lambda.$ We illustrated this numerically by comparing vacuum profile functions for impurities approaching a delta function to the ordinary $N=1$ vortex profile functions. We also showed that the differential equations of an $N=n+m$ vortex are related to the differential equations of an $n$ vortex and an impurity of charge $m$ by a singular gauge transformation.

We determined how the attraction or repulsion between a vortex and an impurity depends on the coupling constant $\lambda$ and the impurity parameters $c$ and $d$ by considering the difference in energy between a well-separated vortex and impurity and a coincident vortex and impurity. We found that there are three different regimes: (i) $c>0$ where an impurity will attract a vortex for $\lambda<1$ and repel it for $\lambda>1$; (ii) $c<0$ with the impurity repelling a vortex for $\lambda<1$ and attracting it for $\lambda>1$; and (iii) $c<0$ with the impurity attracting a vortex for $\lambda<1$ and repelling it for $\lambda>1$. We calculated the critical line separating the two different types of behaviour for $c<0$ and compared this to the line $c=-4d$ along which impurities approach a delta function. We found that for sufficiently large $d$ these impurities should fall into category (iii) and thus do behave like vortices. We also investigated the attraction or repulsion between an impurity and a vortex by using the vortex asymptotics at large $r$, but found that this only agreed with the predictions of the energy calculation for case (iii). The disagreement between the two predictions was resolved by considering the energy as a function of the separation $s$ between the vortex and impurity. This revealed that whether a vortex and impurity attracts  also depends on the separation $s$, which explained the discrepancies. 
  
The final section was concerned with the scattering of vortices with magnetic impurities initially focusing on critical coupling. Our aim was to provide a numerical study of the scattering processes, similar to those already conducted for vortices in the absence of impurities \cite{Shellard:1988zx,Rebbi:1991wk,Myers:1991yh,Thatcher:1997da}. We first considered the scattering of a single vortex with an impurity for different impact parameters. In a head-on collision, the vortex will travel through the impurity and continue on its original trajectory, though the details differ depending on the sign of $c$. The scattering angle increases with impact parameter until it attains a maximum value (at impact parameter $b\approx1.5$ for the two impurities we considered), after which it begins decreasing to zero. When $c>0$, the vortex trajectory bends towards the impurity and when $c<0$ it bends away.  The general shape of the scattering angle plot was found to be  similar to a scattering angle plot given in Ref.~\cite{Cockburn:2015huk} for vortices and impurities in hyperbolic space. We also considered the head-on scattering of a ring-shaped 2-vortex with an impurity and found that the ring breaks up into two single  vortices as it passes through the impurity. For $c>0$, the ring breaks up along the $x$-axis, and for $c<0$ it breaks up along the $y$-axis.

Furthermore, we considered the scattering of two vortices in the presence of an impurity. In a head-on collision, we found that the vortices pass through the impurity and scatter at right angles, as they would in the absence of an impurity, see for example Refs.~\cite{Shellard:1988zx,Rebbi:1991wk}. The details of the scattering process are slightly different depending on the sign of $c$. As the impact parameter increases, the scattering angle decreases. Initially the repulsion between two vortices has the strongest effect on the vortex trajectories, even in the presence of an impurity with $c>0$. However we found that there is a certain value of the impact parameter ($b\approx2.54$ for the impurity we considered) past which the relationship between the vortex and the impurity controls the direction of the vortex trajectories. So for an impurity which attracts a vortex, the trajectories will bend towards the impurity, and for an impurity which repels a vortex, the trajectories will continue to bend away from the vortex. For large enough $b$, the vortex trajectories are no longer affected by each other, and the scattering angles are identical to those for a single vortex scattering with the impurity.  

Finally, we have studied head-on collisions of a single vortex and an impurity for $\lambda<1$ and $\lambda >1.$ We considered different values of $c$ to explore all the different regimes. Now, the trajectories are no longer geodesics, so we needed to include a variety of initial velocities. We found interesting trajectories such as vortices getting trapped by the impurity.

There remain many possibilities for further work on this subject.
When a ring-shaped 2-vortex scatters with an impurity, it breaks up into two vortices in a different way depending on the sign of $c$. The scattering of higher charge multi-vortices with an impurity should be investigated to see whether a similar behaviour is observed for them. Furthermore, the scattering of vortices with different impurities should be studied. In particular, it would be interesting to consider impurities which approach a delta function, as we have observed that these should behave like vortices.

Here we focused on relativistic dynamics of Abelian vortices which has applications in collisions of cosmic strings \cite{Vilenkin:2000jqa}. Vortices in real superconductors move according to first order dynamics. Manton proposed an elegant Schr\"odinger-Chern-Simons dynamics in \cite{Manton:1997tg} which is conservative and Galilean invariant and also has a description in terms of moduli space dynamics, see \cite{Demoulini:2009xp} for a rigorous justification. The moduli space approximation predicts that vortices close to critical coupling move around each other \cite{Manton:1997tg, Romao:2004df} which has been verified numerically in \cite{Krusch:2005wr}. It is not known yet how impurities will affect this dynamics.

So far, we have considered $\lambda \neq 1,$ which in the moduli space picture corresponds to inducing a potential on the moduli space of vortices. It would be interesting to explore how $\mu\neq 1$ will affect the dynamics of vortices. In that case, there will be an additional interaction between vortex and the impurity which is mediated by the magnetic field. As discussed at the end of section \ref{lambdanotzero} there are exciting experimental and theoretical developments concerning vortices and impurities in various systems. The review on vortices and periodic impurities \cite{Velez:2008} suggests that it would be very interesting to study the gauge Ginzburg-Landau vortices with Lagrangian \eqref{vimp_lag} with periodic impurities.

\acknowledgments

The authors would like to thank Mareike Haberichter and Tom Winyard for helpful discussions. S.K. is grateful to Muneto Nitta, Abera Muhamed, Yasha Shnir and Martin Speight for interesting suggestions. J.E.A. acknowledges the UK Engineering and Physical Sciences Research Council (Doctoral Training Grant Ref. EP/K50306X/1) and the University of Kent School of Mathematics, Statistics and Actuarial Science for a PhD studentship.


\begin{small}

\end{small}

\end{document}